\def\IEEEsubmission{0}

\def\realsNonnegative{\mathbb{R}_{0}^{+}}
\def\realsNonnegative{\mathbb{R}_{0}^{+}}
\def\complexNumbers{\mathbb{C}}
\def\realNumbers{\mathbb{R}}
\def\integers{\mathbb{Z}}
\def\integersPositive{\mathbb{Z}^{+}}
\def\integersNonnegative{\mathbb{Z}_{0}^{+}}

\def\constante{{\rm e}}
\def\constantj{{\rm j}}
\def\exponentialBase{\xi}

\def\indexU{u}
\def\indexV{v}

\def\varUpsample{U}

\def\lagForCorrelation{k}
\def\unitSequences{G}

\def\indexEleOfSeq{i}
\def\indexIteration{n}
\def\indexIterationdot{n}
\def\indexIterationANF{l}
\def\indexChoosen{l}
\def\indexFirstOrderMonomial{j}
\def\indexMonomial{k}
\def\orderMonomial[#1]{k_{#1}}
\def\coeffientsANF[#1]{c_{#1}}
\def\polyVariable{z}
\def\polyVariableIt{z}

\def\numberOfIterations{m}
\def\numberOfPointsForPSK{H}
\def\modulationSymbolF[#1]{m_{#1}}
\def\cardinalitySetOfOperators[#1]{{H}_{#1}}

\def\monomial[#1]{x_{#1}}

\def\lengthGaGb{N}

\def\scaleAexp[#1]{a_{#1}}
\def\scaleBexp[#1]{b_{#1}}
\def\scaleEexp[#1]{e_{#1}}
\def\angleexp[#1]{c_{#1}}
\def\angleexpAll[#1]{k_{#1}}
\def\angleexpAllN[#1]{q_{#1}}
\def\arbitraryPhaseKN{q'}
\def\angleScaleAexp[#1]{\dot{c}_{#1}}
\def\angleScaleBexp[#1]{\ddot{c}_{#1}}
\def\arbitraryScaleE{e'}

\def\arbitraryPhaseK{k'}
\def\arbitraryPhaseKPP{k''}
\def\separationGolay[#1]{d_{#1}}

\def\angleexpAllBit[#1]{\kappa_{#1}}
\def\arbitraryPhaseKBit{\kappa'}

\def\angleexpAllBitD[#1]{\hat{\kappa}_{#1}}
\def\arbitraryPhaseKBitD{\hat{\kappa}'}

\def\eleGa[#1]{{a}_{#1}}
\def\eleGb[#1]{{b}_{#1}}

\def\apac[#1][#2]{\rho_{#1}(#2)}
\def\apacPositive[#1][#2]{\rho^{+}_{#1}(#2)}
\def\binaryAsignment[#1][#2]{b_{#1}^{(#2)}}

\def\eleSeqf[#1]{{f}_{#1}}
\def\eleSeqg[#1]{{g}_{#1}}
\def\eleSeqcf[#1]{{c}_{f,#1}}
\def\eleSeqcg[#1]{{c}_{g,#1}}

\def\scaleA[#1]{\alpha_{#1}}
\def\scaleB[#1]{\beta_{#1}}
\def\angleGolay[#1]{\omega_{#1}}
\def\angleScaleA[#1][#2]{{\gamma}_{#1}^{#2}}
\def\angleScaleB[#1][#2]{{\eta}_{#1}^{#2}}
\def\permutationShift[#1]{{\psi_{#1}}}

\def\permutationMono[#1]{{\pi_{#1}}}
\def\permutationMonoD[#1]{{\hat{\pi}_{#1}}}
\def\seqPermutationShift{\bm{\psi}}
\def\seqPermutationCompShift{\bm{\pi}}
\def\seqPermutationCompShiftD{\bm{\hat{\pi}}}

\def\symbolDuration{T_{\rm s}}
\def\timeVar{t}

\def\Ptransmit{P_{\rm av}}

\def\separationIterative[#1]{{\tau}_{#1}}
\def\permutationShiftAux[#1]{{\phi_{#1}}}
\def\separationGolayAux[#1]{\tau'_{#1}}

\def\funczArbitrary[#1]{f(#1)}
\def\funcaArbitrary[#1]{f_1(#1)}
\def\funcbArbitrary[#1]{f_2(#1)}
\def\coefficientArbitrary[#1]{k_{#1}}

\def\sqrtofNumPointsInQuad{s}

\def\Squad{\mathbb{S}_{\numberOfPointsForPSK}}

\def\pointQAM{p}

\def\distanceToPoint[#1]{r_{#1}}
\def\ratioBetweenDistanceAndInner[#1]{r_{#1}} % change it from l to d
\def\angleBetweenPointAndXaxis[#1]{\theta_{#1}}
\def\angleBetweenPointAndXYDiagonal[#1]{\psi_{#1}}

\def\vecArrangement[#1]{\textbf{b}_{#1}}

\def\seqGa{\textit{\textbf{a}}}
\def\seqGb{\textit{\textbf{b}}}
\def\seqGaIt[#1]{\textit{\textbf{a}}^{(#1)}}
\def\seqGbIt[#1]{\textit{\textbf{b}}^{(#1)}}
\def\seqGc{\textit{\textbf{c}}}
\def\seqGd{\textit{\textbf{d}}}
\def\seqGf[#1]{\textit{\textbf{f}}_{#1}}
\def\seqGg[#1]{\textit{\textbf{g}}_{#1}}

\def\seqSub[#1]{\textit{\textbf{h}}_{#1}}

\def\seqGaP{A}
\def\seqGbP{B}
\def\seqGaItP[#1]{{A}^{(#1)}}
\def\seqGbItP[#1]{{B}^{(#1)}}
\def\seqGcP{C}
\def\seqGdP{D}
\def\seqSubP[#1]{{H}_{#1}}
\def\seqGfP[#1]{{{F}}_{#1}}
\def\seqGgP[#1]{{{G}}_{#1}}

\def\seqFirstOrderMonomial[#1]{\textit{\textbf{m}}_{#1}}
\def\seqx{\textit{\textbf{x}}}

\def\seqToBeModulated[#1]{\textit{\textbf{s}}_{#1}}

\def\flipConjugate[#1]{{{\tilde{#1}}}}
\def\expectationOperator[#1][#2]{{\mathbb{E}_{#2}}[#1]}
\def\operator[#1][#2]{\mathcal{O}_{#1}^{(#2)}}
\def\operatorDef{O}

\def\compositeOperatorF[#1][#2]{{\mathcal{F}}_{#1}{\{#2\}}}
\def\compositeOperatorG[#1][#2]{{\mathcal{G}}_{#1}{\{#2\}}}

\def\setOfOperators[#1]{{\mathfrak{J}}_{#1}}
\def\anOperatorBinary[#1]{O_{#1}}

\def\operatorBinary[#1][#2]{O_{#1}^{(#2)}}
\def\operatorSign[#1][#2]{{\rm S}_{#1}^{(#2)}}
\def\operatorScaleA[#1][#2]{{\rm{A}}_{#1}^{(#2)}}
\def\operatorScaleB[#1][#2]{{\rm{B}}_{#1}^{(#2)}}
\def\operatorAngle[#1][#2]{\Omega_{#1}^{(#2)}}
\def\operatorSeparation[#1][#2]{\Delta_{#1}^{(#2)}}
\def\operatorOrderA[#1][#2]{\dot{\rm O}_{#1}^{(#2)}}
\def\operatorOrderB[#1][#2]{\ddot{\rm O}_{#1}^{(#2)}}
\def\operatorAngleScaleA[#1][#2]{\dot{\Omega}_{#1}^{(#2)}}
\def\operatorAngleScaleB[#1][#2]{\ddot{\Omega}_{#1}^{(#2)}}
\def\operatorAngleConjScaleA[#1][#2]{\dot{\Upsilon}_{#1}^{(#2)}}
\def\operatorAngleConjScaleB[#1][#2]{\ddot{\Upsilon}_{#1}^{(#2)}}

\def\functionf[#1]{P^{(#1)}{(\polyVariable)}}
\def\functiong[#1]{Q^{(#1)}{(\polyVariable)}}
\def\functionh{R(\polyVariable)}
\def\functionfdot[#1]{{P}_{\indexIterationANF}^{(#1)}{(\polyVariable)}} %removed bars
\def\functiongdot[#1]{{Q}_{\indexIterationANF}^{(#1)}{(\polyVariable)}} % removed bars

\def\funcfForCommonAmplitude{c_{\rm r}}
\def\funcfForPartAAmplitude{f_{\rm r}}
\def\funcfForPartBAmplitude{g_{\rm r}}
\def\funcfForCommonPhase{c_{\rm i}}
\def\funcfForCommonPhaseA{f_{\rm i}}
\def\funcfForCommonPhaseB{g_{\rm i}}

\def\funcfForCommonShift{f_{\rm s}}
\def\funcfForCommonOrder{S}
\def\funcfForFinalPhase{f_{\rm i}}
\def\funcfForFinalAmplitude{f_{\rm r}}
\def\funcgForFinalPhase{g_{\rm i}}
\def\funcgForFinalAmplitude{g_{\rm r}}

\def\funcfForCommonShiftDec{\check{f}_{\rm s}}
\def\funcfForFinalPhaseDec{\check{f}_{\rm i}}
\def\funcfForFinalAmplitudeDec{\check{f}_{\rm r}}
\def\funcgForFinalPhaseDec{\check{g}_{\rm i}}
\def\funcgForFinalAmplitudeDec{\check{g}_{\rm r}}
\def\funcfForCommonOrderDec{\check{S}}

\def\funcfForANF{f}
\def\funcfForANFdec{\check{f}}
\def\funcEnum{{e}}
\def\funcEnumInv{{e^{-1}}}

\def\varMonomial{{i}}

\def\funcgForANF{g}
\def\funcGfForANF[#1]{f_{#1}}
\def\funcGgForANF[#1]{g_{#1}}
\def\polySeq[#1][#2]{{#1}(#2)}

\def\funcGfForANFdec[#1]{\check{f}_{#1}}
\def\funcGgForANFdec[#1]{\check{g}_{#1}}

\def\OFDMinTime[#1][#2]{s_{#1}(#2)}

\def\shortPlus{\scalebox{0.75}[1.0]{\( + \)}}
\def\shortMinus{\scalebox{0.75}[1.0]{\( - \)}}
\def\spectralEfficient{\rho}
\def\code{\mathcal{C}}

\def\numberOfPointsOnConstellation{S}
\def\constellationPoints{\mathbb{S}_{\rm esc}}

\def\phaseOffset{\Delta'}
\def\phaseOffsetl[#1]{\Delta_{#1}}
\def\phaseOffsetll[#1]{\Delta_{#1}}
\def\amplitudeOffset[#1]{\epsilon'_{#1}}
\def\amplitudeOffsetl[#1]{\epsilon_{#1}}

\def\numberOfPointsNotEquidistance{A}
\def\numberOfPointsEquidistanceButDifferentAbsAngle{B}

\newcommand\mydots{\hbox to 1em{.\hss.\hss.}}

\newcommand{\expnumber}[2]{{#1}\mathrm{e}{#2}}

\if\IEEEsubmission1
	\documentclass[journal,12pt,onecolumn,draftclsnofoot,]{IEEEtran} \def\baselineSize{1.5}
\else
	\documentclass[journal]{IEEEtran}\def\baselineSize{1}
\fi

\usepackage{multicol}
\usepackage{lipsum}
\usepackage{mathtools}
\usepackage{amsthm}
\usepackage{acronym}
\usepackage[bookmarksopen=true]{hyperref}

\usepackage{xcolor}
\usepackage{amsfonts}
\usepackage{acronym}
\usepackage{cite}
\usepackage{bm}
\usepackage{amsmath,amssymb}
\usepackage{graphicx}
\usepackage[english]{babel}
\usepackage[caption=false,font=footnotesize,skip=0pt]{subfig}
\usepackage[geometry]{ifsym}
\usepackage{array}
\usepackage[utf8]{inputenc}
\usepackage[T1]{fontenc}
\usepackage{array}
\usepackage{algpseudocode}

\usepackage[ruled,vlined]{algorithm2e}
\SetAlCapFnt{\small}

\usepackage{stfloats}

\makeatletter
\def\BState{\State\hskip-\ALG@thistlm}
\makeatother

\newcolumntype{L}[1]{>{\raggedright\let\newline\\\arraybackslash\hspace{0pt}}m{#1}}
\newcolumntype{C}[1]{>{\centering\let\newline\\\arraybackslash\hspace{0pt}}m{#1}}
\newcolumntype{R}[1]{>{\raggedleft\let\newline\\\arraybackslash\hspace{0pt}}m{#1}}

\DeclarePairedDelimiter\ceil{\lceil}{\rceil}
\DeclarePairedDelimiter\floor{\lfloor}{\rfloor}

\usepackage{cuted}
\makeatletter
\newif\ifAC@uppercase@first%
\def\Aclp#1{\AC@uppercase@firsttrue\aclp{#1}\AC@uppercase@firstfalse}%
\def\AC@aclp#1{%
	\ifcsname fn@#1@PL\endcsname%
	\ifAC@uppercase@first%
	\expandafter\expandafter\expandafter\MakeUppercase\csname fn@#1@PL\endcsname%
	\else%
	\csname fn@#1@PL\endcsname%
	\fi%
	\else%
	\AC@acl{#1}s%
	\fi%
}%
\def\Acp#1{\AC@uppercase@firsttrue\acp{#1}\AC@uppercase@firstfalse}%
\def\AC@acp#1{%
	\ifcsname fn@#1@PL\endcsname%
	\ifAC@uppercase@first%
	\expandafter\expandafter\expandafter\MakeUppercase\csname fn@#1@PL\endcsname%
	\else%
	\csname fn@#1@PL\endcsname%
	\fi%
	\else%
	\AC@ac{#1}s%
	\fi%
}%
\def\Acfp#1{\AC@uppercase@firsttrue\acfp{#1}\AC@uppercase@firstfalse}%
\def\AC@acfp#1{%
	\ifcsname fn@#1@PL\endcsname%
	\ifAC@uppercase@first%
	\expandafter\expandafter\expandafter\MakeUppercase\csname fn@#1@PL\endcsname%
	\else%
	\csname fn@#1@PL\endcsname%
	\fi%
	\else%
	\AC@acf{#1}s%
	\fi%
}%
\def\Acsp#1{\AC@uppercase@firsttrue\acsp{#1}\AC@uppercase@firstfalse}%
\def\AC@acsp#1{%
	\ifcsname fn@#1@PL\endcsname%
	\ifAC@uppercase@first%
	\expandafter\expandafter\expandafter\MakeUppercase\csname fn@#1@PL\endcsname%
	\else%
	\csname fn@#1@PL\endcsname%
	\fi%
	\else%
	\AC@acs{#1}s%
	\fi%
}%
\edef\AC@uppercase@write{\string\ifAC@uppercase@first\string\expandafter\string\MakeUppercase\string\fi\space}%
\def\AC@acrodef#1[#2]#3{%
	\@bsphack%
	\protected@write\@auxout{}{%
		\string\newacro{#1}[#2]{\AC@uppercase@write #3}%
	}\@esphack%
}%
\def\Acl#1{\AC@uppercase@firsttrue\acl{#1}\AC@uppercase@firstfalse}
\def\Acf#1{\AC@uppercase@firsttrue\acf{#1}\AC@uppercase@firstfalse}
\def\Ac#1{\AC@uppercase@firsttrue\ac{#1}\AC@uppercase@firstfalse}
\def\Acs#1{\AC@uppercase@firsttrue\acs{#1}\AC@uppercase@firstfalse}
\newtheorem{theorem}{Theorem}
\newtheorem{definition}{Definition}
\newtheorem{lemma}{Lemma}
\newtheorem{corollary}{Corollary}
\newtheorem{example}{\color{black} Example} 

\acrodef{SIC}{successive interference cancellation}
\acrodef{AACF}{aperiodic auto-correlation function}
\acrodef{OFDM}{orthogonal frequency-division multiplexing}
\acrodef{DFT}{discrete Fourier transform}
\acrodef{DC}{direct current}
\acrodef{CS}{complementary sequence}
\acrodef{GCP}{Golay complementary pair}
\acrodef{ANF}{algebraic normal form}
\acrodef{PSK}{phase shift keying}
\acrodef{QAM}{quadrature amplitude modulation}
\acrodef{QPSK}{quadrature phase-shift keying}
\acrodef{GDJ}{Golay-Davis-Jedwab}
\acrodef{PMEPR}{peak-to-mean-envelope-power ratio}
\acrodef{FFT}{fast Fourier transform}
\acrodef{BER}{bit-error rate}
\acrodef{SNR}{signal-to-noise ratio}
\acrodef{4G}{Fourth Generation}
\acrodef{5G}{Fifth Generation}
\acrodef{NR}{New Radio}
\acrodef{LTE}{Long-Term Evolution}
\acrodef{PTS}{partial transmit sequences}
\acrodef{PSD}{power spectral density}
\acrodef{LDPC}{low-density parity check}
\acrodef{SE}{spectral efficiency}
\acrodef{eLAA}{enhanced licensed-assisted access}
\acrodef{NR-U}{NR in unlicensed bands}
\acrodef{RM}{Reed-Muller}
\acrodef{NOMA}{non-orthogonal multiple access}
\acrodef{FSK}{frequency-shift keying}
\acrodef{LDPC}{low-density parity check}
\acrodef{ML}{maximum-likelihood}
\acrodef{AWGN}{additive white Gaussian noise}
\acrodef{CFR}{channel frequency response}
\acrodef{CIR}{channel impulse response}
\acrodef{CP}{cyclic prefix}
\acrodef{MMSE}{minimum mean square error}
\acrodef{FDE}{frequency-domain equalization}
\acrodef{FHT}{fast Hadarmard transformation}
\acrodef{BPSK}{binary phase-shift keying}
\acrodef{PRB}{physical resource block}
\acrodef{PA}{power amplifier}
\acrodef{BLER}{block-error rate}
\acrodef{ESC}{equiangular sub-symmetric constellation}
\begin{document}
\title{ 
%A Generic Complementary Sequence Construction and Encoder/Decoder 
A Generic Complementary Sequence Construction and Associated Encoder/Decoder Design
%Encoder and Decoder based on A Generic Complementary Sequence Construction
}
%Non-contiguous Complementary Sequence Encoder
%A Generalized Complementary Sequence Encoder
\author{Alphan~\c{S}ahin,~\IEEEmembership{Member,~IEEE,}
        and~Rui~Yang,~\IEEEmembership{Member,~IEEE}
\thanks{Alphan~\c{S}ahin and Rui~Yang are with University of South Carolina, Columbia, SC and InterDigital, NY, respectively. E-mail: asahin@mailbox.sc.edu, rui.yang@interdigital.com
}}
\maketitle

\begin{abstract}
In this study, we propose a flexible construction of \acp{CS} that can contain zero-valued elements. 
%This construction relies on modifying a polynomial associated with a complex sequence. 
To derive the construction, we use Boolean functions to represent a polynomial generated with a recursion. 
%To derive the construction, we use a method that obtains the coefficients of a polynomial generated through a recursion with linear operators. 
By applying this representation to recursive \ac{CS} constructions, we show   the impact of  construction parameters such as sign, amplitude, phase rotation used in the recursion on the elements of the synthesized \ac{CS}. As a result, we extend Davis and Jedwab's \ac{CS} construction by obtaining  independent functions for the amplitude and  phase of each element of the \ac{CS}, and the seed sequence positions in the \ac{CS}. 
The proposed construction  shows that a set of distinct \acp{CS}  compatible with non-contiguous resource allocations for \ac{OFDM} and various constellations can be synthesized systematically. It also leads to a low \ac{PMEPR}  multiple accessing scheme  in the uplink and a low-complexity recursive decoder. We demonstrate the performance of the proposed encoder and decoder through comprehensive simulations.
%By utilizing this construction, an encoder using \acp{CS} with various constellations and the corresponding decoder are developed. 
\end{abstract}
\begin{IEEEkeywords}
	Complementary sequences, Gray code, peak-to-average-power ratio, pseudo-Boolean function, Reed-Muller code, non-contiguous resource allocation 
\end{IEEEkeywords}
\acresetall

\section{Introduction}

\Ac{OFDM} enables a radio to transmit multiple modulation symbols on orthogonal channels separated in the frequency domain. 
%It has been the dominant multiplexing scheme for today’s major wireless communication standards.
%Because of its high \ac{SE}, in conjunction with its compatibility with multiple access techniques, it has been the dominant waveform for today’s major wireless communication standards. 
%such as 3GPP \ac{LTE} \cite{lte_2018}, 3GPP \ac{NR} \cite{nr_2018}, and IEEE 802.11 Wi-Fi \cite{ieee_2016}. 
One drawback of \ac{OFDM} transmission is that it causes high \ac{PMEPR} for arbitrary modulation symbols. % A large power back-off may also be required to avoid the non-linear distortion at the power amplifier at the expense of a reduced coverage range. 
%Reducing the instantaneous peak power of an \ac{OFDM} symbol is not a trivial task. 
In the literature, there are numerous approaches to reduce the \ac{PMEPR}  \cite{Han_2005, Rahmatallah_2013}.
One of the trends is to design codes such that the \ac{PMEPR} of the resulting signal is bounded (see \cite{Wunder_2013} and the references therein). 
In this direction, \acp{CS} from the cosets of \ac{RM} codes can limit the \ac{PMEPR} of an \ac{OFDM} symbol to be less than or equal to 3 dB while still providing coding gain \cite{davis_1999}. 
However, these sequences have to be contiguous with \ac{PSK} alphabet and limited choice of lengths. %, which is difficult to be used in major wireless standards
%However, these sequences are typically contiguous, limited to \ac{PSK} alphabet, and it is difficult to make their lengths to be compatible with the resource allocations in major wireless standards. 
In this study, we address the issue of constructing  \acp{CS} compatible with a set of flexible resource allocation in the frequency domain for \ac{OFDM} while achieving expressions for not only phase but also amplitude of the elements of the \acp{CS}.
%Particularly, a non-contiguous resource allocation in an \ac{OFDM} symbol, which may be beneficial for low-latency and reliable communications or meeting regulatory  requirements for the communications in unlicensed bands \cite{ericUL}, makes  the \ac{PMEPR} minimization  a more challenging problem \cite{Liu_2018, Gokceli_2019}. 

\Acp{CS} were introduced by M. Golay in \cite{Golay_1961}. He established a framework to synthesize \acp{GCP} with binary alphabet. 
%He presented two general construction methods, known as {\em Golay's concatenation} and {\em Golay's interleaving}, by concatenating or  interleaving \acp{GCP} of length $\lengthGaGb$ and $\lengthGcGd$ to form a \ac{GCP} of length $2\lengthGaGb\lengthGcGd$.
His constructions were later extended by R. J. Turyn \cite{Turyn_1974} and  Sivasamy  \cite{Sivaswamy_1978}. In \cite{Budisin_1990}, Budi{\v s}in introduced the permutations of shifted phase-rotated sequences instead of simple concatenation seed \acp{CS} to obtain more \acp{CS}. In \cite{Budisin_1990_ml}, he proposed another construction that explains how the scaling factors can be taken into account in a recursive method to generate multi-level \acp{GCP}. In \cite{Garcia_2010_ml} and \cite{garcia_2013}, recursive constructions with phase factors and scaling factors were combined in a single recursion.
Nevertheless, these earlier studies did not propose direct constructions  for \acp{CS} or investigate \acp{CS} with certain alphabets.

Appealing applications of \acp{CS} on communication systems were first demonstrated in \cite{boyd_1986,Popovic_1991,Davis_1997,davis_1999,paterson_2000}. By generalizing an earlier work of Boyd \cite{boyd_1986}, Popovic showed that \acp{CS} were beneficial to control the peak instantaneous power of \ac{OFDM} signals \cite{Popovic_1991}. Davis and Jedwab proved that a large class of \acp{CS} over $\integers_{2^h}$ of length $2^\numberOfIterations$ can be obtained through generalized Boolean functions related to  \ac{RM} codes \cite{Davis_1997,davis_1999} for $h\in\integersPositive$.
%In  , they showed that the proposed method results in  $\numberOfIterations!/2\cdot2^{h(\numberOfIterations+1)}$ \acp{CS} which occur as the elements of the cosets of the first-order \ac{RM} code within the second-order \ac{RM} code, where $h\ge1$ is an integer. Since the Hamming distance between two sequences in the set is at least $2^{\numberOfIterations-2}$, they established a key connection between \acp{CS} and \ac{RM} codes  to achieve a coding scheme guaranteeing low \ac{PMEPR} for \ac{OFDM} symbols while providing good error correction capability. 
Davis and Jedwab's result was also generalized to a larger class of \acp{CS} over  $\integers_{\numberOfPointsForPSK}$ of length $2^\numberOfIterations$  where $\numberOfPointsForPSK$ is even positive integer in \cite{paterson_2000}. In the literature, the \acp{CS} generated through Davis and Jedwab's construction is sometimes referred to as {\em \ac{GDJ}} or {\em standard} sequences. 

The connection between \acp{CS} and \ac{RM} codes has been a cornerstone of various research directions. The first major  challenge has been the low code rate of the encoder proposed by Davis and Jedwab. Since the sequence length increases much faster than the number of distinct \acp{CS} for the encoder, the corresponding code rate  decreases for a larger number of uncoded bits. To address this issue, one major direction has been exploring  the \acp{CS} which cannot be generated through method in \cite{davis_1999}, i.e., {\em nonstandard} sequences. Until Li and Chu found 1024 quaternary nonstandard \acp{CS} of length 16 via a computer search in 2005 \cite{Li_Chu_2005}, there were no evidence on the existence of nonstandard \acp{CS} of length $2^\numberOfIterations$. In \cite{Li_2005}, Li and Kao showed that these sequences  contain third-order monomials and can be synthesized by concatenating or interleaving of quaternary \ac{GCP} of length 8. In \cite{Fiedler_2006}, Fiedler and Jedwab provided another explanation how these sequences arise by showing that some of the standard \acp{CS} in different cosets have identical \ac{AACF}. In \cite{fiedler2008_multi}, they also provided a construction leading to the nonstandard \acp{CS} in \cite{Li_Chu_2005} by expressing \acp{CS} as a projection of a  multi-dimensional array.

To increase the number of distinct \acp{CS} for a given sequence length, another direction has been the \ac{CS} construction with larger alphabets such as $M$-\ac{QAM} constellations. In \cite{robing_2001}, R{\"o}{\ss}ing and V. Tarokh showed that 16-QAM sequences which result in low \ac{PMEPR} can be constructed through  a weighted sum of two \acp{CS} with the alphabet of \ac{QPSK} constellation. In \cite{Chong_2002,Chong_2003}, Chong et al. showed that a 16-\ac{QAM} \ac{CS} can be formed by introducing the "offset" sequences based on the Boolean functions to a "base" quaternary standard sequence  and there exist at least $(14+12\numberOfIterations)(\numberOfIterations!/2)4^{\numberOfIterations+1}$ distinct \acp{CS} of length $2^\numberOfIterations$ with the alphabet of 16-QAM constellation. In \cite{Li_2008}, Li revised Chong et al. and Lee and Golomb's theorems \cite{Lee_2006}  and introduced new offset pairs leading to  64-QAM \acp{CS} proposed in \cite{Chang_2010}. In \cite{Li_2010}, Li showed that there exist at least $[(\numberOfIterations+1)4^{2(q-1)}-(m+1)4^{(q-1)}+2^{q-1}](m!/2)4^{(m+1)}$ \acp{CS} for $4^q$-QAM sequences.  Although the offset method is helpful for synthesizing a large number of distinct \ac{QAM} \acp{CS}, it typically does not give concise constructions as Davis and Jedwab's method for the standard sequences \cite{Chong_2003, Li_2008}. %In \cite{Chong_2003}, %Chong et al. remarked that {\em "unfortunately, these sequences were difficult to classify based on their polynomial structure. The classification of these sequences and their realization as \ac{OFDM} \ac{QAM} codes with low encoding/decoding complexities remains an interesting open problem."} 
This issue is one of our motivations to synthesize \acp{CS}  by separating the functions for its amplitude and the phase terms. From this aspect, our study is in line with \cite{budisin_2018} and \cite{wang2020newQAM}. However, our work differs from \cite{budisin_2018} and \cite{wang2020newQAM} as %\cite{budisin_2018} focuses on only \ac{QAM} \acp{CS} and  synthesizes \acp{CS} by indexing the elements of the unitary matrices and using the properties of Gaussian integers.  In our study, 
we focus on a framework relying on linear operators,
%, which introduces not only individual polynomials for amplitude and phase, but also separate polynomials for seed sequences and the support of the synthesized \ac{CS}, systematically. 
which enables us to generate non-contiguous \acp{CS}, i.e., \ac{CS} with zero-valued elements,  the \acp{CS} of length non-power-of-two, and the \ac{CS} with various constellations. They also pave the way for developing an encoder and a decoder.

Another noteworthy direction for increasing the number of distinct sequences is to extend \acp{CS} to complementary sets or complete complementary codes at the expense of a larger  \ac{PMEPR} \cite{wang2020new,  Zhou_2020, Chen_2016}. For these extensions, paraunitary matrices with larger sizes are often utilized. Since we focus on \acp{CS} in this study, we refer the reader to \cite{wang2020new} and \cite{Zhou_2020} and the references therein for further details.

\subsection{Contributions and Organization}

{\textbf{A framework based on linear operators:} }The first contribution  is Lemma~\ref{th:framework} that re-expresses the result of a recursion evolved with two linear operators at each step with Boolean functions. It describes how the operators applied to the input polynomials at the $\indexIteration$th recursion step are distributed to the coefficients of the final polynomials.

{\textbf{A novel CS construction:} }
By applying Lemma~\ref{th:framework} to recursive \ac{CS} constructions \cite{Budisin_1990,Budisin_1990_ml,Garcia_2010_ml, garcia_2013,parker_2003}, we show the impact of each parameter in the recursion on the synthesized \ac{CS}. By removing the redundant parameters, we obtain a concise construction given in Theorem~\ref{th:reduced} as the main contribution of this study. 
%This construction provides new insights into \acp{GCP} that can be hard to infer from the previous methods. 
It extends James and Davis's result \cite{davis_1999} by providing independent pseudo-Boolean functions for the amplitude, phase, seed sequences, and the support of the \ac{CS}.

{\textbf{Distinct CSs with a wide range of constellations:} } We introduce a constellation family consisting of \ac{QAM}, \ac{PSK}, DVB-16APSK, DVB-32APSK, and a modified version IEEE 802.11ay 64-NUC, called {\em\ac{ESC}}. We show that distinct \acp{CS} can be enumerated systematically for a given \ac{ESC}. We also discuss how to generate  non-contiguous \acp{CS} where the non-zero-valued elements are in \ac{ESC}.

{\textbf{Encoder/decoder for \acp{CS} with an \ac{ESC}:} } %In the literature, there are few studies demonstrate the performance of \ac{CS} with various constellation. 
%In this study, 
We develop a generic encoder and \ac{ML}-based decoder for \acp{CS} with \ac{ESC}. We demonstrate its applicability for a practical wireless communication system through a low-\ac{PMEPR} multiple accessing scheme in the uplink  with non-contiguous resource allocation for \ac{OFDM}.

The rest of the paper is organized as follows. In Section \ref{sec:pre}, preliminaries are provided.
% and set the notation used in the rest of the sections. 
In Section \ref{sec:encoding-with-an-iterative-process}, we establish Lemma~\ref{th:framework}.  In Section \ref{sec:derivation-of-complementary-sequence-encoder}, Theorem~\ref{th:reduced} is discussed. In Section~\ref{sec:constellation}, we enumerate \acp{CS} for an arbitrary \ac{ESC}. In Section~\ref{sec:encDec}, we discuss the  encoder and the decoder.
In Section \ref{sec:numerical},  we provide numeral results. We conclude the paper in Section \ref{sec:conclusion}.

{\em Notation:} The sets of complex numbers, real numbers, non-negative real numbers, integers,  non-negative integers,  positive integers, and integers modulo $\numberOfPointsForPSK$ are denoted by $\complexNumbers$,  $\realNumbers$, $\realsNonnegative$,  $\integers$, $\integersNonnegative$, $\integersPositive$, and $\integers_\numberOfPointsForPSK$, respectively. The set of $\numberOfIterations$-dimensional integers where each element is in $\integers_\numberOfPointsForPSK$  is denoted by $\integers^\numberOfIterations_\numberOfPointsForPSK$.  
The transpose, Hermitian, the complex conjugation,  the modulo 2, and the assignments are denoted by $(\cdot)^{\rm T}$, $(\cdot)^{\rm H}$, $(\cdot)^*$, $(\cdot)_2$, and  $\leftarrow$, respectively.  The constant $\constantj$ is $\sqrt{-1}$.

\section{Preliminaries and Definitions}
\label{sec:pre}
\subsection{Sequences, OFDM, and PMEPR}
\label{subsec:poly_sequence}
Let $\seqGa=(\eleGa[{\indexEleOfSeq}])_{i=0}^{\lengthGaGb-1}\triangleq(\eleGa[0],\eleGa[1],\dots, \eleGa[\lengthGaGb-1])$ be a sequence  of length $\lengthGaGb$,  where $\eleGa[{\indexEleOfSeq}]\in\complexNumbers$ and $\eleGa[\lengthGaGb-1]\neq0$. We associate the sequence $\seqGa$ with the polynomial
$\polySeq[\seqGaP][\polyVariable] = \eleGa[\lengthGaGb-1]\polyVariable^{\lengthGaGb-1} + \eleGa[\lengthGaGb-2]\polyVariable^{\lengthGaGb-2}+ \dots + \eleGa[0]$
in indeterminate $\polyVariable$.  When it is clear from the context,  $\polySeq[\seqGaP][\polyVariable]$ is a polynomial function.

Let $\OFDMinTime[\seqGa][\timeVar]=\sum_{\indexEleOfSeq=0}^{\lengthGaGb-1}\eleGa[{\indexEleOfSeq}]\constante^{\constantj2\pi\indexEleOfSeq\frac{\timeVar}{\symbolDuration}}$ for $\timeVar\in[0,\symbolDuration)$ be the  continuous-time baseband \ac{OFDM} symbol generated from the sequence $\seqGa$  with the symbol duration $\symbolDuration$. Since $\OFDMinTime[\seqGa][\timeVar] = \polySeq[\seqGaP][{\constante^{\constantj\frac{2\pi\timeVar}{\symbolDuration}}}]$,  
the {instantaneous envelope power} of the \ac{OFDM} symbol can be examined by evaluating the polynomial $\polySeq[\seqGaP][\polyVariable]$ at $|\polyVariable|=1$. When $\polyVariable$ is restricted  to be on the unit circle in the complex plane, $|\polySeq[\seqGaP][\polyVariable]|^2$ corresponds to $\sum_{\lagForCorrelation=-N+1}^{\lengthGaGb-1}\apac[\seqGa][\lagForCorrelation]\polyVariable^{\lagForCorrelation}$, where
$\apac[\seqGa][\lagForCorrelation]$ is the \ac{AACF} of the sequence $\seqGa$ given by
\begin{align}
	\apac[\seqGa][\lagForCorrelation]\triangleq
	\begin{cases}
		\sum_{\indexEleOfSeq=0}^{\lengthGaGb-\lagForCorrelation-1} \eleGa[\indexEleOfSeq]^*\eleGa[\indexEleOfSeq+\lagForCorrelation], & 0\le\lagForCorrelation\le\lengthGaGb-1\\
		\sum_{\indexEleOfSeq=0}^{\lengthGaGb+\lagForCorrelation-1} \eleGa[\indexEleOfSeq]\eleGa[\indexEleOfSeq-\lagForCorrelation]^*, & -\lengthGaGb+1\le\lagForCorrelation<0\\
		0,& \text{otherwise}
	\end{cases}~.
\end{align}
%, i.e., the Z-transform of the \ac{AACF} of the sequence $\seqGa$. 
Hence, an upper bound for the instantaneous power can be obtained as
$
\max_{\timeVar\in[0, \symbolDuration)}|\OFDMinTime[\seqGa][\timeVar]|^2=\max_{|\polyVariable|=1}|\polySeq[\seqGaP][\polyVariable]|^2 \le\apac[\seqGa][0] + 
		2\sum_{\lagForCorrelation=1}^{\lengthGaGb-1}|\apac[\seqGa][\lagForCorrelation]|
$, which indicates that a sequence with a smaller $|\apac[\seqGa][\lagForCorrelation]|$ for $\lagForCorrelation\neq0$  also provides less power fluctuations  when it is used for \ac{OFDM} transmission.

Let $\code$ denote a code of length $\lengthGaGb$. Let $\seqGc\in\complexNumbers^{\lengthGaGb}$ be an admissible sequence, i.e., a codeword, in $\code$. In this study, we define the \ac{PMEPR} of the codeword $\seqGc$ as
$\max_{\timeVar\in[0, \symbolDuration)}|\OFDMinTime[\seqGc][\timeVar]|^2/{\Ptransmit}
$, where $\Ptransmit=\expectationOperator[{\apac[\seqGc][0]}][\seqGc]$ is a constant that depends on the code. %While $\PAPR$ measures the fluctuations within an \ac{OFDM} symbol, \ac{PMEPR} considers the entire signal generated with many \ac{OFDM} symbols.  

\subsection{Sequence Synthesis}
\label{subsec:algebraic_sequence}
A generalized Boolean function is a function $\funcfForANF$ that maps from $\integers^\numberOfIterations_2=\{(\monomial[1],\monomial[2],\dots, \monomial[\numberOfIterations])|\forall \monomial[\indexFirstOrderMonomial]\in\integers_2\}$ to $\integers_\numberOfPointsForPSK$ as $\funcfForANF:\integers^\numberOfIterations_2\rightarrow\integers_\numberOfPointsForPSK$, where $\numberOfPointsForPSK$ is an integer. It can be uniquely expressed as a linear combination of the monomials over $\integers_\numberOfPointsForPSK$, i.e.,
\begin{align}
	\hspace{-2mm}\funcfForANF(\seqx)= \sum_{\indexMonomial=0}^{2^\numberOfIterations-1} \coeffientsANF[\indexMonomial]\prod_{\indexFirstOrderMonomial=1}^{\numberOfIterations} \monomial[\indexFirstOrderMonomial]^{\orderMonomial[\indexFirstOrderMonomial]} = \coeffientsANF[0]1+ \dots+ \coeffientsANF[2^\numberOfIterations-1]\monomial[1]\monomial[2]\mydots\monomial[\numberOfIterations]~,
	\label{eq:ANF}
\end{align} 
where $\seqx\triangleq(\monomial[1],\monomial[2],\dots, \monomial[\numberOfIterations])$ and $\coeffientsANF[\indexMonomial]\in \integers_\numberOfPointsForPSK$ for $\indexMonomial = \sum_{\indexFirstOrderMonomial=1}^{\numberOfIterations}\orderMonomial[\indexFirstOrderMonomial]2^{\numberOfIterations-\indexFirstOrderMonomial}$ and $\orderMonomial[\indexFirstOrderMonomial]\in\integers_2$ (i.e., the coefficient of ($\indexMonomial+1$)th monomial $\monomial[1]^{\orderMonomial[1]}\monomial[2]^{\orderMonomial[2]}\mydots\monomial[\numberOfIterations]^{\orderMonomial[\numberOfIterations]}$  belongs to $\integers_\numberOfPointsForPSK$). The expression given in \eqref{eq:ANF} is called \ac{ANF} of  $\funcfForANF(\seqx)$ \cite{Fiedler_2008}.  

The co-domain of  the function $\funcfForANF$  can be extended to $\realNumbers$ and $\integersNonnegative$. In the former case, $\funcfForANF:\integers^\numberOfIterations_2\rightarrow\realNumbers$ is a pseudo-Boolean function and the sequences generated through the monomials construct a basis in $\realNumbers^{2^\numberOfIterations}$ and $\coeffientsANF[\indexMonomial]\in \realNumbers$. In the latter case, i.e., $\funcfForANF:\integers^\numberOfIterations_2\rightarrow\integersNonnegative$, the monomial coefficients are in $\integersNonnegative$. In this study, we use $\check{\cdot}$ on top of a function to denote a function that maps the non-negative integers to the co-domain of $\funcfForANF$ as $\funcfForANFdec(\varMonomial)\triangleq\funcfForANF\circ\funcEnumInv(\varMonomial)$, where $\varMonomial=\funcEnum(\seqx) \triangleq   \sum_{\indexFirstOrderMonomial=1}^{\numberOfIterations}\monomial[\indexFirstOrderMonomial]2^{\numberOfIterations-\indexFirstOrderMonomial}$, i.e.,  a decimal representation of the binary number constructed using all elements in the sequence $\seqx$, where the most significant bit is $\monomial[1]$.

Let $\funcfForFinalAmplitude:\integers^\numberOfIterations_2\rightarrow\realNumbers$, $\funcfForFinalPhase:\integers^\numberOfIterations_2\rightarrow\realNumbers$, and $\funcfForCommonShift:\integers^\numberOfIterations_2\rightarrow\integersNonnegative$. In this study, we generate a complex sequence $\seqGa=(\eleGa[{\indexEleOfSeq}])_{i=0}^{2^{\numberOfIterations}-1}$  by listing  $\eleGa[{\indexEleOfSeq}]=\exponentialBase^{\funcfForFinalAmplitudeDec(\varMonomial)   + \constantj \funcfForFinalPhaseDec(\varMonomial)}$ for $\exponentialBase\triangleq\constante^{\frac{2\pi}{\numberOfPointsForPSK}}$. We  utilize the function $\funcfForCommonShift(\seqx)$ for generating complex sequences with zero-valued elements by modifying the associated polynomial of $\seqGa=(\eleGa[{\indexEleOfSeq}])_{i=0}^{2^{\numberOfIterations}-1}$ as
\begin{align}
\polySeq[\seqGcP][\polyVariable] =&\sum_{\varMonomial=0}^{2^\numberOfIterations-1} \eleGa[\varMonomial]\polyVariable^{\funcfForCommonShiftDec(\varMonomial)+\varMonomial}\nonumber~.
\end{align}
Therefore, the sequence $\seqGc$ can be characterized with $\funcfForFinalAmplitude(\seqx)$, $\funcfForFinalPhase(\seqx)$, and $\funcfForCommonShift(\seqx)$.

Let $\funcfForANF:\integers^\numberOfIterations_2\rightarrow\integers_2$ be a Boolean function and $\seqGa$ and $\seqGb$ be seed sequences of length $\lengthGaGb$. The polynomial
$\funcfForCommonOrder(\seqx,\polyVariable)=\polySeq[{\seqGaP}][\polyVariable](1+\funcfForANF(\seqx))_2+\polySeq[{\seqGbP}][\polyVariable]\funcfForANF(\seqx)$ is then equal to $\polySeq[{\seqGaP}][\polyVariable]$ for $\funcfForANF(\seqx)=0$ and $\polySeq[{\seqGbP}][\polyVariable]$ for $\funcfForANF(\seqx)=1$. For  $\varUpsample\in\integersNonnegative$, consider the polynomial given by
\begin{align}
	\polySeq[{\seqGcP}][\polyVariable] = 
	\sum_{\varMonomial=0}^{2^\numberOfIterations-1}
	\funcfForCommonOrderDec(\varMonomial,\polyVariable)
	\eleGa[\varMonomial]
	\polyVariable^{\funcfForCommonShiftDec(\varMonomial)+\varUpsample\varMonomial}.\nonumber 
\end{align} For $\varUpsample=\lengthGaGb$ and $\funcfForCommonShiftDec(\varMonomial)=0$ for all $\varMonomial$, the polynomial indicates that the $2^\numberOfIterations$  sequences (i.e., $\polySeq[\seqGaP][\polyVariable]\exponentialBase^{\funcfForFinalAmplitudeDec(\varMonomial)   + \constantj \funcfForFinalPhaseDec(\varMonomial)}$ or $\polySeq[\seqGbP][\polyVariable]\exponentialBase^{\funcfForFinalAmplitudeDec(\varMonomial)   + \constantj \funcfForFinalPhaseDec(\varMonomial)}$ based on $\funcfForANFdec(\varMonomial)$) are concatenated. %, which is similar to the projection operation for multi-dimensional arrays introduced in \cite{fiedler2008_multi}. 
While  the seed sequences are separated by $\varUpsample-\lengthGaGb$ zero-valued elements for $\varUpsample>\lengthGaGb$, the $\lengthGaGb-\varUpsample$ elements of the two adjacent seed sequences are overlapped for $\varUpsample<\lengthGaGb$. %In this study, we set $\varUpsample\ge\lengthGaGb$ unless otherwise stated.

\subsection{Constellations}
Let $\pointQAM_\indexU\in\complexNumbers$  for $\indexU\in\{1,2,\mydots,\numberOfPointsOnConstellation\}$ and $\numberOfPointsOnConstellation\in\integersPositive$. Let  $\distanceToPoint[\indexU]$ and $\angleBetweenPointAndXYDiagonal[\indexU]$ denote 
the amplitude and the phase of $\pointQAM_\indexU$ subtracted by $\pi/4$ radian, respectively.
%the distance between $\pointQAM_\indexU$ and the origin and the angle between the vector that passes through $\pointQAM_\indexU$ and the $xy$-diagonal axis, respectively. 
We  define $\Squad$  as a set of distinct $\numberOfPointsOnConstellation$ complex numbers such that $-\constantj\pointQAM_\indexU^*\in\Squad$ if $|\angleBetweenPointAndXYDiagonal[\indexU]|<\pi/\numberOfPointsForPSK$ for all $\pointQAM_\indexU\in\Squad$ and $-\pi/\numberOfPointsForPSK< \angleBetweenPointAndXYDiagonal[\indexU]\le\pi/\numberOfPointsForPSK$. In other words, the points in $\Squad$ are symmetric with respect to the $xy$-diagonal of the complex plane (i.e., sub-symmetricity), except the points on the line $y=\tan(\pi/\numberOfPointsForPSK+\pi/4)x$. By rotating all the points in $\Squad$  by an integer multiple of $2\pi/\numberOfPointsForPSK$  (i.e., equiangularity), we  define  \ac{ESC} as follows:
\begin{definition} Equiangular sub-symmetric constellation is $\constellationPoints=\{\pointQAM_\indexU\times\constante^{\constantj2\pi k/H }|\forall\pointQAM_\indexU\in\Squad \text{ {\rm and} } \forall k\in\integers_{\numberOfPointsForPSK}\}$.%, where the number of elements in $\constellationPoints$ is $\numberOfPointsForPSK\numberOfPointsOnConstellation$. 
\end{definition}

\if\IEEEsubmission0
\begin{figure}
	\centering
	\subfloat[$4\sqrtofNumPointsInQuad^2$-QAM ($\sqrtofNumPointsInQuad=4,\numberOfPointsOnConstellation=16,\numberOfPointsForPSK=4$).]{\includegraphics[width =1.65in]{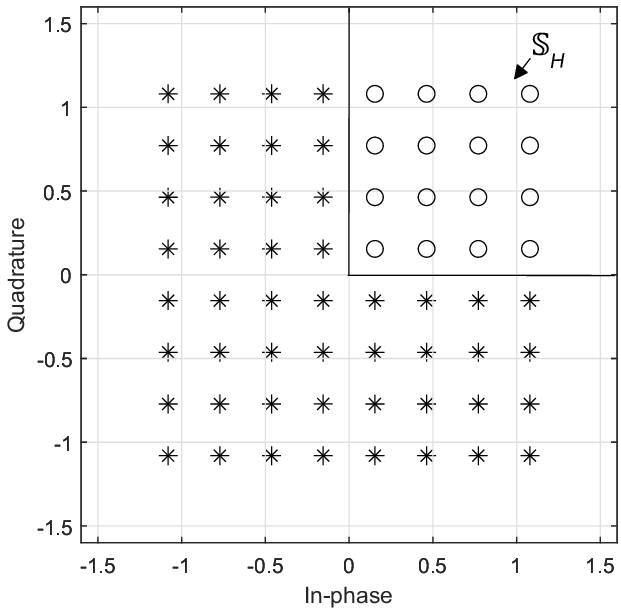}
		\label{subfig:qam}}~
	\subfloat[$\numberOfPointsForPSK\numberOfPointsOnConstellation$-Star ($\numberOfPointsOnConstellation=4,\numberOfPointsForPSK=8$).]{\includegraphics[width =1.65in]{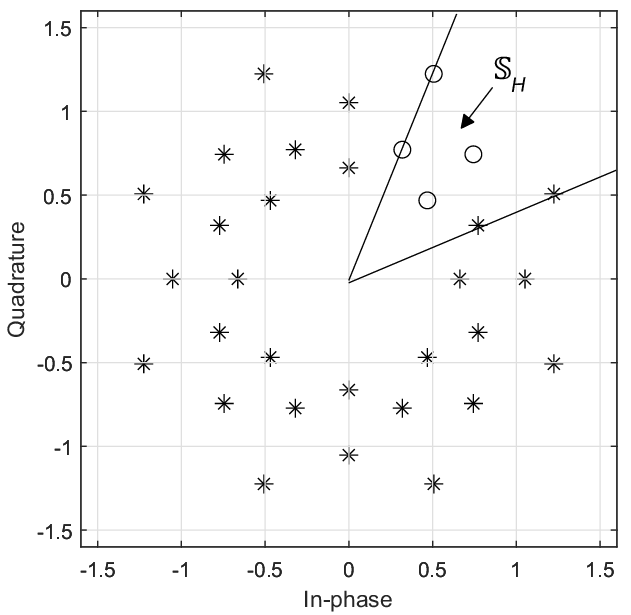}
		\label{subfig:star}}
	\\
	\subfloat[DVB-32APSK ($\numberOfPointsOnConstellation=8,\numberOfPointsForPSK=4$).]{\includegraphics[width =1.65in]{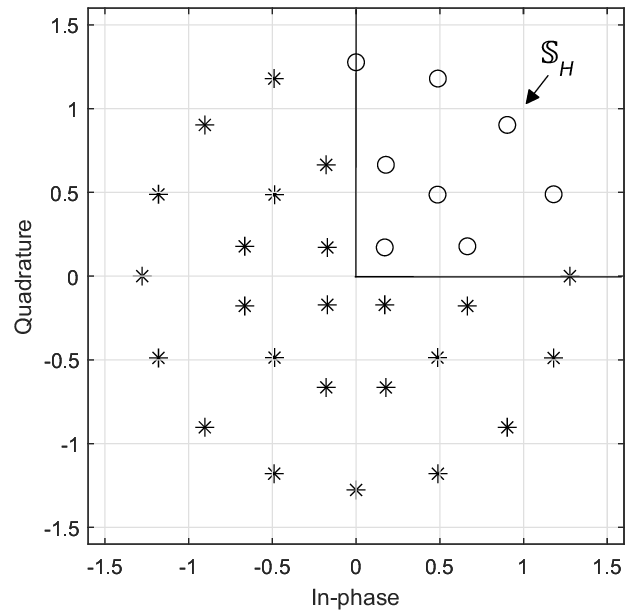}
		\label{subfig:apsk}}~	
	\subfloat[Modified 64-NUC ($\numberOfPointsOnConstellation=16,\numberOfPointsForPSK=4$).]{\includegraphics[width =1.65in]{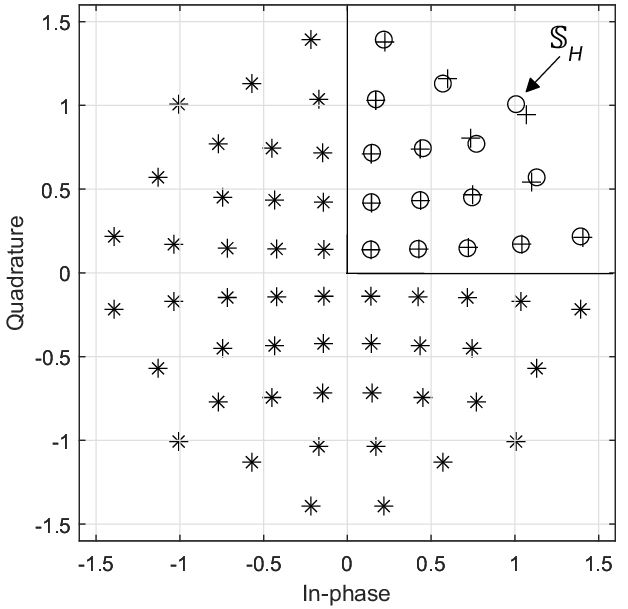}
		\label{subfig:nuc}}
	\caption{Example \acp{ESC} ($*$: points in $\constellationPoints$, o: points in $\Squad$, +: original 64-NUC constellation on the first quadrant).}
	\label{fig:escs}
	\vspace{-2mm}
\end{figure}
\else
\begin{figure}
	\centering
	\subfloat[$4\sqrtofNumPointsInQuad^2$-QAM ($\sqrtofNumPointsInQuad=4,\numberOfPointsOnConstellation=16,\numberOfPointsForPSK=4$).]{\includegraphics[width =1.6in]{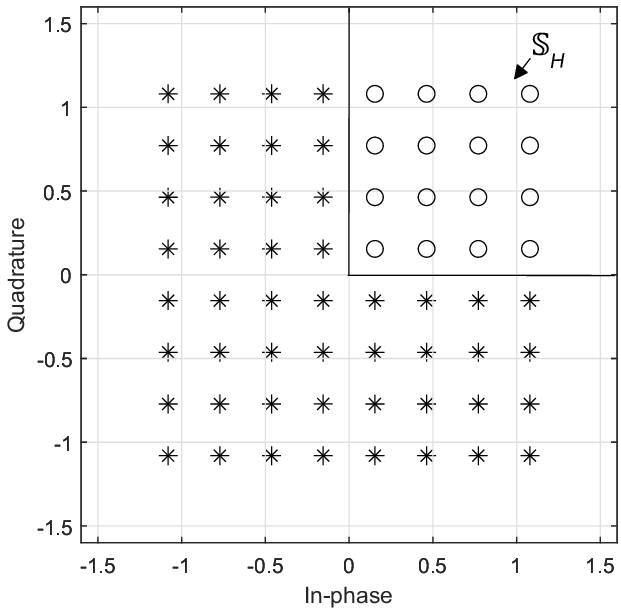}
		\label{subfig:qam}}
	\subfloat[$\numberOfPointsForPSK\numberOfPointsOnConstellation$-Star ($\numberOfPointsOnConstellation=4,\numberOfPointsForPSK=8$).]{\includegraphics[width =1.6in]{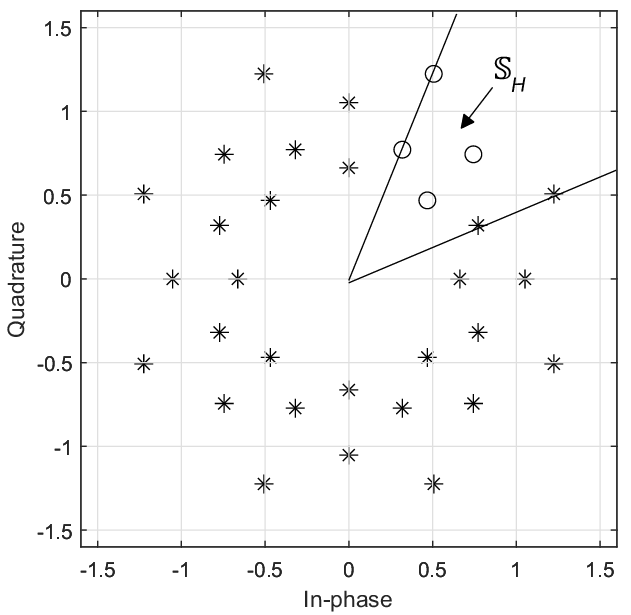}
		\label{subfig:star}}
	\subfloat[DVB-32APSK ($\numberOfPointsOnConstellation=8,\numberOfPointsForPSK=4$).]{\includegraphics[width =1.6in]{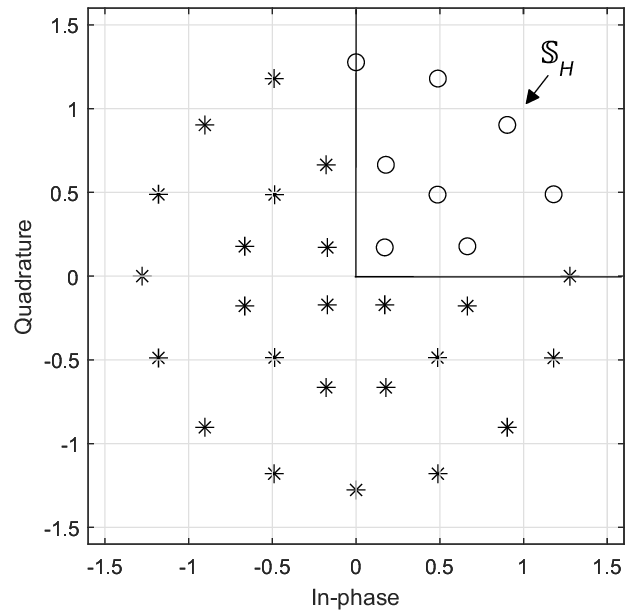}
		\label{subfig:apsk}}	
	\subfloat[Modified 64-NUC ($\numberOfPointsOnConstellation=16,\numberOfPointsForPSK=4$).]{\includegraphics[width =1.6in]{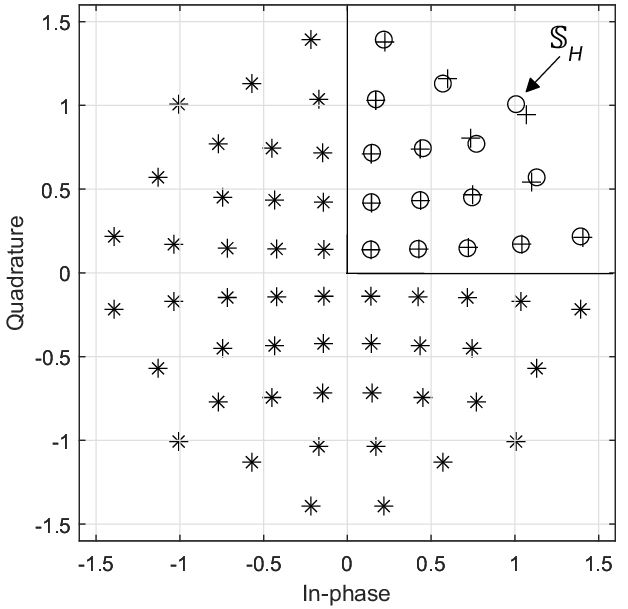}
		\label{subfig:nuc}}
	\caption{Example \acp{ESC} ($*$: points in $\constellationPoints$, o: points in $\Squad$, +: original 64-NUC constellation on the first quadrant).}
	\label{fig:escs}
	\vspace{-2mm}
\end{figure}
\fi
\ac{QAM} and \ac{PSK} are \acp{ESC}. For example,  let $4\sqrtofNumPointsInQuad^2$-QAM constellation be a set of points given by $\{(2\indexU-1)+\constantj(2\indexV-1)|\indexU,\indexV\in\integers,-\sqrtofNumPointsInQuad<\indexU\le\sqrtofNumPointsInQuad,-\sqrtofNumPointsInQuad<\indexV\le\sqrtofNumPointsInQuad\}$ for $\sqrtofNumPointsInQuad\in\integersPositive$.  A $4\sqrtofNumPointsInQuad^2$-QAM constellation is an \ac{ESC} for $\numberOfPointsForPSK=4$ and the corresponding $\Squad$ is the set of $\numberOfPointsOnConstellation=\sqrtofNumPointsInQuad^2$ points on the first quadrant as shown in \figurename~\ref{fig:escs}\subref{subfig:qam}  for $\sqrtofNumPointsInQuad=4$.
$\numberOfPointsForPSK\numberOfPointsOnConstellation$-Star is another \ac{ESC} defined as the points in $\Squad$ satisfying $\ratioBetweenDistanceAndInner[\indexU]<\ratioBetweenDistanceAndInner[\indexU+1]$ and $\angleBetweenPointAndXYDiagonal[\indexU]=\pi/\numberOfPointsForPSK\times((\indexU-1)\mod{2})$ for $\indexU=1,2,\mydots,\numberOfPointsOnConstellation$ as exemplified in  \figurename~\ref{fig:escs}\subref{subfig:star}.
DVB-16/32APSK  defined in \cite{etsi_2014_dvbs2} are also \acp{ESC} for $\numberOfPointsForPSK=4$. DVB-32APSK is shown in \figurename~\ref{fig:escs}\subref{subfig:apsk}. IEEE 802.11ay 64-NUC constellation \cite{ieee_11ay_d50} is not an \ac{ESC}. However, it can be modified to be an \ac{ESC} for $\numberOfPointsForPSK=4$ by switching the real and imaginary components and calculating the mean of two closest points in 2-D complex plane. The modified IEEE 802.11ay 64-NUC is provided in \figurename~\ref{fig:escs}\subref{subfig:nuc}.

In this study, we synthesize complex sequences where their non-zero-valued elements in an \ac{ESC} via  $\exponentialBase^{\funcfForFinalAmplitude(\seqx)   + \constantj \funcfForFinalPhase(\seqx)}$ for $\funcfForFinalAmplitude:\integers^\numberOfIterations_2\rightarrow\realNumbers$ and $\funcfForFinalPhase:\integers^\numberOfIterations_2\rightarrow\realNumbers$. The fundamental strategy that we use is to offset the monomial coefficients of $\funcfForFinalAmplitude(\seqx)$ and $\funcfForFinalPhase(\seqx)$ such that the elements are in the targeted \ac{ESC}. For instance, let $\funcfForFinalPhase(\seqx)=\arbitraryPhaseK+\angleexpAll[1]\monomial[1]$ and $\funcfForFinalAmplitude(\seqx)=\scaleEexp[1]\monomial[1]$ for $\numberOfIterations=2$, $\arbitraryPhaseK=\arbitraryPhaseKBit+\phaseOffsetl[1],\angleexpAll[1]\in\integers_{2},\arbitraryPhaseKBit\in\integers_{2}$, $\scaleEexp[1]\in\realNumbers$, and $\numberOfPointsForPSK=2$. For $\{\phaseOffsetl[1]=0,\scaleEexp[1]=0\}$, $\{\phaseOffsetl[1]=\pi/2,\scaleEexp[1]=0\}$, $\{\phaseOffsetl[1]=0,\scaleEexp[1]=\ln(3)/\pi\}$, and $\{\phaseOffsetl[1]=\pi/2,\scaleEexp[1]=\ln(3)/\pi\}$, the elements of the complex sequence are in $\{-1,1\}$, $\{-\constantj,\constantj\}$, $\{-3,-1,1,3\}$, and $\{-3j,-1j,1j,3j\}$, respectively, which forms an \ac{ESC} of $\numberOfPointsOnConstellation=8$ and $\numberOfPointsForPSK=2$.

\subsection{Golay Complementary Pair and Complementary Sequence}
The sequence pair  $(\seqGa,\seqGb)$ of length  $\lengthGaGb$ is  a \ac{GCP} if $\apac[\seqGa][\lagForCorrelation]+\apac[\seqGb][\lagForCorrelation] = 0$ for $\lagForCorrelation\neq0$. The sequences $\seqGa$ and $\seqGb$ are referred to as \acp{CS}. By using the definition of \ac{GCP}, a GCP $(\seqGa,\seqGb)$ satisfies $|\polySeq[\seqGaP][\polyVariable]|^2+|\polySeq[\seqGbP][\polyVariable]|^2 =\apac[\seqGa][0]+\apac[\seqGb][0]$ for  $|\polyVariable|=1$. The instantaneous peak power of the corresponding \ac{OFDM} signal generated through a \ac{CS} $\seqGa$ is bounded since
$\max_{\timeVar\in[0, \symbolDuration)}|\OFDMinTime[\seqGa][\timeVar]|^2 \le \apac[\seqGa][0]+\apac[\seqGb][0]$.

A general  well-known recursive \ac{GCP} construction can be given as follows:
\begin{theorem}[\cite{Budisin_1990,Budisin_1990_ml,Garcia_2010_ml,garcia_2013,parker_2003}]
	\label{th:golayIterative}
	Let $(\seqGa,\seqGb)$ be a \ac{GCP} of length $\lengthGaGb$. For $\scaleA[\indexIteration],\scaleB[\indexIteration]\in \realsNonnegative$,  $\separationIterative[\indexIteration],\varUpsample\in  \integersNonnegative$,  $\angleGolay[\indexIteration],\angleScaleA[\indexIteration][],\angleScaleB[\indexIteration][]\in \{u:u\in\complexNumbers, |u|=1\}$ for $\indexIteration=1,2,\mydots,\numberOfIterations$, and the sequence $\seqPermutationShift=(\permutationShift[{\indexIteration}])_{i=1}^{\numberOfIterations}$ defined by a permutation of $\{0,1,\dots,\numberOfIterations-1\}$,  assume that the following operations occur at the $\indexIteration$th step of a recursion:
	\begin{align}
	\polySeq[{\seqGaItP[\indexIteration]}][\polyVariable] =&\angleScaleA[\indexIteration][]\scaleA[\indexIteration] \polySeq[{\seqGaItP[\indexIteration-1]}][\polyVariable]  
	+ \angleScaleB[\indexIteration][]\scaleB[\indexIteration]\angleGolay[\indexIteration]\polySeq[{\seqGbItP[\indexIteration-1]}][\polyVariable] \polyVariable^{\separationIterative[\indexIteration]+\varUpsample2^\permutationShift[{\indexIteration}]}~, \nonumber\\
	\polySeq[{\seqGbItP[\indexIteration]}][\polyVariable] =& \angleScaleB[\indexIteration][*]\scaleB[\indexIteration]\polySeq[{\seqGaItP[\indexIteration-1]}][\polyVariable]  
	- \angleScaleA[\indexIteration][*] \scaleA[\indexIteration]\angleGolay[\indexIteration]\polySeq[{\seqGbItP[\indexIteration-1]}][\polyVariable] \polyVariable^{\separationIterative[\indexIteration]+\varUpsample2^\permutationShift[{\indexIteration}]}~, 
	\nonumber
	\end{align}
	where $\polySeq[{\seqGaItP[0]}][\polyVariable] = \polySeq[{\seqGaP}][\polyVariable]$, $\polySeq[{\seqGbItP[0]}][\polyVariable] =\polySeq[{\seqGbP}][\polyVariable]$. The sequences $\seqGaIt[\numberOfIterations]$ and $\seqGbIt[\numberOfIterations]$ associated with the polynomials $\polySeq[{\seqGaItP[\numberOfIterations]}][\polyVariable]$ and $\polySeq[{\seqGbItP[\numberOfIterations]}][\polyVariable]$, respectively, construct a GCP for $\numberOfIterations\ge 1$.
\end{theorem}
%For completeness, we provide the proof in Appendix~\ref{app:gcp}. 

Theorem~\ref{th:golayIterative} shows the Budi{\v s}in's recursive constructions in \cite[Eq. (4)]{Budisin_1990} and \cite[Eqs. (3-4)]{Budisin_1990_ml}, the Garc\'{i}a's constructions in \cite[Eq. (2)]{Garcia_2010_ml} and \cite[Eq. (5)]{garcia_2013},  a special case of Golay's concatenation in \cite[Eq. (10)]{parker_2003}.  
The expression in Theorem~1 can also be written in a matrix form as
\begin{align}
\begin{bmatrix}
	\polySeq[{\seqGaItP[\indexIteration]}][\polyVariable]\\
	\polySeq[{\seqGbItP[\indexIteration]}][\polyVariable]
\end{bmatrix}=
\begin{bmatrix}
	\angleScaleA[\indexIteration][]\scaleA[\indexIteration]&
 	\angleScaleB[\indexIteration][]\scaleB[\indexIteration]\angleGolay[\indexIteration]\polyVariable^{\separationIterative[\indexIteration]+\varUpsample2^\permutationShift[{\indexIteration}]}\\
 	\angleScaleB[\indexIteration][*]\scaleB[\indexIteration]&
 	-\angleScaleA[\indexIteration][*] \scaleA[\indexIteration]\angleGolay[\indexIteration]\polyVariable^{\separationIterative[\indexIteration]+\varUpsample2^\permutationShift[{\indexIteration}]}
\end{bmatrix}
\begin{bmatrix}
	\polySeq[{\seqGaItP[\indexIteration-1]}][\polyVariable]\\
	\polySeq[{\seqGbItP[\indexIteration-1]}][\polyVariable]
\end{bmatrix}~,\nonumber
\end{align}
which shows that \ac{CS} construction can involve 2-by-2 paraunitary matrices. The special cases of the paraunitary matrices are investigated in the literature. For example, \cite{budisin_2018} and \cite{wang2020new}  consider the case \{$\seqGa=\seqGb=1$, $\separationIterative[\indexIteration]=0$, $\varUpsample=1$ for $\indexIteration=1,2,\mydots,\numberOfIterations$\}. Both \cite{budisin_2018} and \cite{wang2020new} show that Davis and Jedwab's construction can also be derived as a special case of paraunitary matrices, which implies that Theorem~\ref{th:golayIterative} is general enough to construct standard \acp{CS}. 

To utilize \acp{CS} in communications, their distinctness must be ensured. However, the recursion in Theorem~\ref{th:golayIterative} (and the corresponding paraunitary matrix representation) by itself does not guarantee distinct \acp{CS} for different parameters \cite[Section IV]{budisin_2018} after $\numberOfIterations$ iterations:
\begin{example}
\rm 
Assume that $\seqGa=\seqGb=1$, $\varUpsample=1$, $\numberOfIterations=3$, 
$\seqPermutationShift=(1,2,3)$, $\separationIterative[\indexIteration]=0$, $\angleScaleA[\indexIteration][]=0$,  and $\angleScaleB[\indexIteration][]=0$ for $\indexIteration=1,2,3$.
Consider two different configurations: $\{(\scaleA[1],\scaleA[2],\scaleA[3])=(1,2,3)$, $(\scaleB[1],\scaleB[2],\scaleB[3])=(1/2,2,3)\}$ and $\{(\scaleA[1],\scaleA[2],\scaleA[3])=(3,1,2)$,$(\scaleB[1],\scaleB[2],\scaleB[3])=(3/2,1,2)\}$. Both configurations result in the same \ac{CS} $\seqGa^{(3)}=(6,3,3,-6,6,6,3,-3,6)$.
\end{example}
To resolve this issue, in \cite{budisin_2018}, the distinct \acp{CS} are generated by using the properties of Gaussian integers for \ac{QAM} alphabet. In this study, we eliminate the redundancy in Theorem 1 by using pseudo-Boolean functions, which leads to a generalization of Davis and Jedwab's construction.

%As we combine the recursions in \cite{Budisin_1990,Budisin_1990_ml,Garcia_2010_ml, garcia_2013}, we provide the corresponding proof in Appendix~\ref{app:gcp} for completeness. 
	
%The cases \{$\scaleA[\indexIteration]=1$, $\scaleB[\indexIteration]=1$, $\angleScaleA[\indexIteration][]=1$, $\angleScaleB[\indexIteration][]=1$,  $\separationIterative[\indexIteration]=0$, $\varUpsample=\lengthGaGb$ for $\indexIteration=1,2,\mydots,\numberOfIterations$\} and \{$\angleScaleA[\indexIteration][]=1$, $\angleScaleB[\indexIteration][]=1$,  $\angleGolay[\indexIteration]=1$, $\separationIterative[\indexIteration]=0$, $\varUpsample=\lengthGaGb$ for $\indexIteration=1,2,\mydots,\numberOfIterations$\} are the Budi{\v s}in's constructions in \cite[Eq. (4)]{Budisin_1990} and \cite[Eqs. (3-4)]{Budisin_1990_ml}, respectively. The Garc\'{i}a's constructions in \cite[Eq. (2)]{Garcia_2010_ml} and \cite[Eq. (5)]{garcia_2013} can also be obtained for the case \{$\angleScaleA[\indexIteration][]=1$, $\angleScaleB[\indexIteration][]=1$, $\angleGolay[\indexIteration]=1$,  $\separationIterative[\indexIteration]=0$, $\varUpsample=\lengthGaGb$ for $\indexIteration=1,2,\mydots,\numberOfIterations$\}. The case \{$\scaleA[\indexIteration]=1$, $\scaleB[\indexIteration]=1$, $\angleGolay[\indexIteration]=1$, $\separationIterative[\indexIteration]=\lengthGaGb$, $\varUpsample=0$ for $\indexIteration=1$\} also corresponds to a special Golay's concatenation in \cite[Eq. (10)]{parker_2003}. 

\subsection{Operators}
An operator  $\operatorDef$ transforms one polynomial to another polynomial. If  $\operatorDef\{a\polySeq[\seqGaP][\polyVariable]+b\polySeq[\seqGbP][\polyVariable]\}=a\operatorDef\{\polySeq[\seqGaP][\polyVariable]\}+b\operatorDef\{\polySeq[\seqGbP][\polyVariable]\}$ for all  $\polySeq[\seqGaP][\polyVariable]$ and $\polySeq[\seqGbP][\polyVariable]$ and $a,b\in\complexNumbers$, the operator  $\operatorDef$ is linear. Several examples of linear operators are $\operatorDef\{{\functionh}\}= \polyVariable^{\separationGolay[]}\functionh$, $\operatorDef\{{\functionh}\}= -\functionh$, $\operatorDef\{{\functionh}\}= 	\exponentialBase^{\scaleAexp[]}\functionh$, and $\operatorDef\{{\functionh}\}=\exponentialBase^{\constantj\scaleBexp[]}\functionh$ for $\separationGolay[]\in\integersNonnegative$ and $ \scaleAexp[], \scaleBexp[] \in \realNumbers $, where $\functionh$ is an arbitrary polynomial.

Let $\anOperatorBinary[0]$ and $\anOperatorBinary[1]$ be two linear operators and $\funcfForANF:\integers^\numberOfIterations_2\rightarrow\integers_2$ be a Boolean function. Consider the polynomial given by
$
%\begin{align}
	\functionf[\numberOfIterations] = \sum_{\varMonomial=0}^{2^\numberOfIterations-1} \anOperatorBinary[{\funcGfForANFdec[](\varMonomial)}]\{\functionh\}\polyVariableIt^{\varMonomial}
$. %\end{align}
The  Boolean function ${\funcGfForANFdec[](\varMonomial)}$ determines which one of the two operators affects the coefficient of $\polyVariableIt^{\varMonomial}$. Thus, one can simply define the polynomial $\functionf[\numberOfIterations]$ with the operators $\anOperatorBinary[0]$ and $\anOperatorBinary[1]$ and the Boolean function $\funcfForANF$. We use this notation to define the polynomials as shown in Table~\ref{table:ANFgolay}.

\section{Representation of a Recursion}\label{sec:encoding-with-an-iterative-process}

Let $\functionf[\indexIteration]$ and $\functiong[\indexIteration]$ for $\indexIteration=1,2,\dots,\numberOfIterations$ be the polynomials generated by
%Let $\functionf[0]$ and $\functiong[0]$ be an arbitrary function $\functionh\in\functionSpace$ where $\functionSpace$ is a function space. Assume that the following operations occur at the $\indexIteration$th step:
\begin{align}
\functionf[\indexIteration] = \operator[11][{\indexIteration}]\{\functionf[\indexIteration-1]\} + \operator[12][{\indexIteration}]\{\functiong[\indexIteration-1]\}\polyVariableIt^{2^\permutationShift[{\indexIteration}]}~, \nonumber\\
\functiong[\indexIteration] = \operator[21][{\indexIteration}]\{\functionf[\indexIteration-1]\} + \operator[22][{\indexIteration}]\{\functiong[\indexIteration-1]\}\polyVariableIt^{2^\permutationShift[{\indexIteration}]}~,
\label{eq:iterationBasic}
\end{align}
where  $\permutationShift[{\indexIteration}]$ is the $\indexIteration$th element of the sequence $\seqPermutationShift\triangleq(\permutationShift[{\indexIteration}])_{i=1}^{\numberOfIterations}$ defined by a permutation of $\{0,1,\dots,\numberOfIterations-1\}$,  $\operator[ij][{\indexIteration}]\in\{\operatorBinary[0][\indexIteration],\operatorBinary[1][\indexIteration]\}$  is a linear operator, and $\functionf[0]=\functiong[0]=\functionh$. Since $\seqPermutationShift$ includes all non-negative integers less than $\numberOfIterations$ and $\operatorBinary[0][\indexIteration]$ and $\operatorBinary[1][\indexIteration]$ are linear operators,  $\functionf[\numberOfIterations]$ and $\functiong[\numberOfIterations]$ can be obtained as 
\begin{align}
\functionf[\numberOfIterations] = \sum_{\varMonomial=0}^{2^\numberOfIterations-1} \overbrace{\operatorBinary[{\funcGfForANFdec[\numberOfIterations](\varMonomial)}][\numberOfIterations]\dots\operatorBinary[\funcGfForANFdec[2](\varMonomial)][2]\operatorBinary[\funcGfForANFdec[1](\varMonomial)][1]\{\functionh\}}^{\compositeOperatorF[\varMonomial][\functionh] }\polyVariableIt^{\varMonomial}~,
\label{eq:finalSequenceFBasic}
\end{align}
and
\begin{align}
\functiong[\numberOfIterations] = \sum_{\varMonomial=0}^{2^\numberOfIterations-1} \overbrace{ \operatorBinary[\funcGgForANFdec[\numberOfIterations](\varMonomial)][\numberOfIterations]\dots\operatorBinary[\funcGgForANFdec[2](\varMonomial)][2]\operatorBinary[\funcGgForANFdec[1](\varMonomial)][1]\{\functionh\}}^{\compositeOperatorG[\varMonomial][\functionh]  }\polyVariableIt^{\varMonomial}~,
\label{eq:finalSequenceGBasic}
\end{align}
where the Boolean functions  $\funcGfForANF[\indexIteration]:\integers^\numberOfIterations_2\rightarrow\integers_2$ and $\funcGgForANF[\indexIteration]:\integers^\numberOfIterations_2\rightarrow\integers_2$ indicate which of the two operators, i.e., $\operatorBinary[0][\indexIteration]$ and $\operatorBinary[1][\indexIteration]$, are involved in the constructions of  the composite operators  $\compositeOperatorF[\varMonomial][\functionh]$ and $\compositeOperatorG[\varMonomial][\functionh]$ by setting the subscripts of $\operatorBinary[{\funcGfForANFdec[\indexIteration](\varMonomial)}][\indexIteration]$ and $\operatorBinary[{\funcGgForANFdec[\indexIteration](\varMonomial)}][\indexIteration]$, respectively. 
%Hence, they are  sufficient to characterize $\functionf[\numberOfIterations]$ and $\functiong[\numberOfIterations]$.
The Boolean functions can be obtained in closed-form as follows:
\begin{lemma}
\label{th:framework}
Let \textup{$\vecArrangement[\indexIteration]$} be a configuration vector defined by
$ {\textup{$\vecArrangement[\indexIteration]^{\rm T}$}} \triangleq 
[
\binaryAsignment[11][{\indexIteration}]~\binaryAsignment[12][{\indexIteration}]~
\binaryAsignment[21][{\indexIteration}]~\binaryAsignment[22][{\indexIteration}] 
]\in \integers_2^4$, where $\binaryAsignment[ij][{\indexIteration}]=0$ if  $\operator[ij][{\indexIteration}] = \operatorBinary[0][{\indexIteration}]$ and  $\binaryAsignment[ij][{\indexIteration}]=1$ if  $\operator[ij][{\indexIteration}] = \operatorBinary[1][{\indexIteration}]$. Then,  $\funcGfForANF[\indexIteration](\seqx)$ and $\funcGgForANF[\indexIteration](\seqx)$ are the Boolean functions given by
\if\IEEEsubmission0
\begin{align}
\funcGfForANF[\indexIteration](\seqx) &=  
\begin{cases}
\displaystyle
\binaryAsignment[11][{\indexIteration}] (1 - \monomial[{\permutationMono[{\indexIteration}]}]) + 
\binaryAsignment[12][{\indexIteration}]\monomial[{\permutationMono[{\indexIteration}]}]~,
& \indexIteration=\numberOfIterations \\
\displaystyle
\binaryAsignment[11][{\indexIteration}](1- \monomial[{\permutationMono[{\indexIteration}]}])(1 - \monomial[{\permutationMono[{\indexIteration+1}]}]) \\~~~~+ 
\binaryAsignment[12][{\indexIteration}]{\monomial[{\permutationMono[{\indexIteration}]}]}(1 - \monomial[{\permutationMono[{\indexIteration+1}]}]) \\~~~~+
\binaryAsignment[21][{\indexIteration}](1 - \monomial[{\permutationMono[{\indexIteration}]}])\monomial[{\permutationMono[{\indexIteration+1}]}] \\~~~~+
\binaryAsignment[22][{\indexIteration}]{\monomial[{\permutationMono[{\indexIteration}]}]}\monomial[{\permutationMono[{\indexIteration+1}]}]~,
& \indexIteration<\numberOfIterations \\
\end{cases}~,	\label{eq:closedformANFsF}
\end{align}
\begin{align}
\funcGgForANF[\indexIteration](\seqx) &= 
\begin{cases}
\displaystyle
\binaryAsignment[21][{\indexIteration}] (1 - \monomial[{\permutationMono[{\indexIteration}]}]) + 
\binaryAsignment[22][{\indexIteration}]\monomial[{\permutationMono[{\indexIteration}]}]~,
& \indexIteration=\numberOfIterations \\
\displaystyle
\binaryAsignment[11][{\indexIteration}](1- \monomial[{\permutationMono[{\indexIteration}]}])(1 - \monomial[{\permutationMono[{\indexIteration+1}]}]) \\~~~~+ 
\binaryAsignment[12][{\indexIteration}]{\monomial[{\permutationMono[{\indexIteration}]}]}(1 - \monomial[{\permutationMono[{\indexIteration+1}]}]) \\~~~~+
\binaryAsignment[21][{\indexIteration}](1 - \monomial[{\permutationMono[{\indexIteration}]}])\monomial[{\permutationMono[{\indexIteration+1}]}] \\~~~~+
\binaryAsignment[22][{\indexIteration}]{\monomial[{\permutationMono[{\indexIteration}]}]}\monomial[{\permutationMono[{\indexIteration+1}]}]~,
& \indexIteration<\numberOfIterations \\
\end{cases}~,
\label{eq:closedformANFsG}
\end{align}
\else
\begin{align}
\funcGfForANF[\indexIteration](\seqx) &=  
\begin{cases}
\displaystyle
\binaryAsignment[11][{\indexIteration}] (1 - \monomial[{\permutationMono[{\indexIteration}]}]) + 
\binaryAsignment[12][{\indexIteration}]\monomial[{\permutationMono[{\indexIteration}]}]~,
& \indexIteration=\numberOfIterations \\
\displaystyle
\binaryAsignment[11][{\indexIteration}](1- \monomial[{\permutationMono[{\indexIteration}]}])(1 - \monomial[{\permutationMono[{\indexIteration+1}]}]) + 
\binaryAsignment[12][{\indexIteration}]{\monomial[{\permutationMono[{\indexIteration}]}]}(1 - \monomial[{\permutationMono[{\indexIteration+1}]}])  \\~~~~~+
\binaryAsignment[21][{\indexIteration}](1 - \monomial[{\permutationMono[{\indexIteration}]}])\monomial[{\permutationMono[{\indexIteration+1}]}] +
\binaryAsignment[22][{\indexIteration}]{\monomial[{\permutationMono[{\indexIteration}]}]}\monomial[{\permutationMono[{\indexIteration+1}]}]~,
& \indexIteration<\numberOfIterations \\
\end{cases}~,	\label{eq:closedformANFsF}
\end{align}
\begin{align}
\funcGgForANF[\indexIteration](\seqx) &= 
\begin{cases}
\displaystyle
\binaryAsignment[21][{\indexIteration}] (1 - \monomial[{\permutationMono[{\indexIteration}]}]) + 
\binaryAsignment[22][{\indexIteration}]\monomial[{\permutationMono[{\indexIteration}]}]~,
& \indexIteration=\numberOfIterations \\
\displaystyle
\binaryAsignment[11][{\indexIteration}](1- \monomial[{\permutationMono[{\indexIteration}]}])(1 - \monomial[{\permutationMono[{\indexIteration+1}]}]) +
\binaryAsignment[12][{\indexIteration}]{\monomial[{\permutationMono[{\indexIteration}]}]}(1 - \monomial[{\permutationMono[{\indexIteration+1}]}]) \\~~~~~+
\binaryAsignment[21][{\indexIteration}](1 - \monomial[{\permutationMono[{\indexIteration}]}])\monomial[{\permutationMono[{\indexIteration+1}]}] +
\binaryAsignment[22][{\indexIteration}]{\monomial[{\permutationMono[{\indexIteration}]}]}\monomial[{\permutationMono[{\indexIteration+1}]}]~,
& \indexIteration<\numberOfIterations \\
\end{cases}~,
\label{eq:closedformANFsG}
\end{align}
\fi
for $\indexIteration=1,2,\dots,\numberOfIterations$, where $\permutationMono[\indexIteration]=\numberOfIterations - \permutationShift[\indexIteration]$ is the $\indexIteration$th element of the sequence $\seqPermutationCompShift\triangleq(\permutationMono[\indexIteration])_{\indexIteration=1}^{\numberOfIterations}$. 
\end{lemma}

The proof for Lemma~\ref{th:framework} is given in Appendix~\ref{app:algConstruction}. 
 Since there exist $16$ possible options for the $\indexIteration$th configuration vector $\vecArrangement[\indexIteration]$, there also exist $16$ different  Boolean functions for each step in \eqref{eq:iterationBasic}.

\begin{example}
\rm Assume the operator configuration given by $\operator[11][{\indexIteration}]=\operator[12][{\indexIteration}]=\operator[21][{\indexIteration}]=\operatorBinary[0][{\indexIteration}]$ and $\operator[22][{\indexIteration}]=\operatorBinary[1][{\indexIteration}]$ for all iterations, $\numberOfIterations = 2$, and  $\seqPermutationShift=(0,1)$. 	
When $\indexIteration = 1$, $\functionf[1] =\operatorBinary[0][1]\{\functionh\}+ \operatorBinary[0][{1}]\{\functionh\}\polyVariableIt$ and $\functiong[1] = \operatorBinary[0][{1}]\{\functionh\}+ \operatorBinary[1][1]\{\functionh\}\polyVariableIt$. When $\indexIteration = 2$, $
\functionf[2] =
\operatorBinary[0][2]
\operatorBinary[0][1]
\{\functionh\}
+ 
\operatorBinary[0][2]
\operatorBinary[0][{1}]
\{\functionh\}
\polyVariableIt
+
\operatorBinary[0][2]
\operatorBinary[0][1]
\{\functionh\}
\polyVariableIt^2
+ 
\operatorBinary[0][2]
\operatorBinary[1][1]
\{\functionh\}
\polyVariableIt^3
$ and $
\functiong[2] = 
\operatorBinary[0][2]
\operatorBinary[0][1]
\{\functionh\}
+ 
\operatorBinary[0][2]
\operatorBinary[0][{1}]
({\functionh})
\polyVariableIt+
\operatorBinary[1][2]
\operatorBinary[0][1]
\{\functionh\}
\polyVariableIt^2
+ 
\operatorBinary[1][2]
\operatorBinary[1][1]
\{\functionh\}
\polyVariableIt^3
$.
%\fi
By investigating the distribution of the operators at the final polynomial,  the Boolean functions related to  the first and the second steps should produce the sequences
$\seqGf[1] = (0,0,0,1)$,
$\seqGf[2] = (0,0,0,0)$,
$\seqGg[1] = (0,0,0,1)$, and
$\seqGg[2] = (0,0,1,1)$.
\label{ex:signConstructionSequences}
\end{example}

\begin{example}
	\label{ex:sign}
	\rm Consider Example \ref{ex:signConstructionSequences}. Since the recursion is configured as   $\operator[11][{\indexIteration}]=\operator[12][{\indexIteration}]=\operator[21][{\indexIteration}]=\operatorBinary[0][{\indexIteration}]$ and $\operator[22][{\indexIteration}]=\operatorBinary[1][{\indexIteration}]$ for all iterations, the configuration vector is obtained as $ \vecArrangement[\indexIteration]^{\rm T} =[0~0~0~1]$ for $\indexIteration=\{1,2\}$. Hence, by using \eqref{eq:closedformANFsF} and  \eqref{eq:closedformANFsG}, $\funcGfForANF[\indexIteration](\seqx)$ and $\funcGgForANF[\indexIteration](\seqx)$ are obtained as
	\begin{align}
	\funcGfForANF[\indexIteration](\seqx) &= \begin{cases}
	0, 		& \text{if}~\indexIteration=\numberOfIterations\\
	\monomial[{\permutationMono[{\indexIteration}]}]\monomial[{\permutationMono[{\indexIteration+1}]}],     	& \text{if}~\indexIteration<\numberOfIterations
	\end{cases}~,\nonumber
	\end{align}
	and
	\begin{align}
	\funcGgForANF[\indexIteration](\seqx) = \begin{cases}
	\monomial[{\permutationMono[{\numberOfIterations}]}], 		& \text{if}~\indexIteration=\numberOfIterations\\
	\monomial[{\permutationMono[{\indexIteration}]}]\monomial[{\permutationMono[{\indexIteration+1}]}],     	& \text{if}~\indexIteration<\numberOfIterations
	\end{cases}~,\nonumber
	\end{align}
	respectively. Since $\permutationMono[\indexIteration] = \numberOfIterations - \permutationShift[\indexIteration]$ and $\seqPermutationShift=(0,1)$,  the Boolean functions that lead to the  sequences $\seqGf[1]$, $\seqGf[2]$, $\seqGg[1]$, and $\seqGf[2]$  are obtained as $\funcGfForANF[1](\seqx) = \monomial[{\permutationMono[{1}]}]\monomial[{\permutationMono[{2}]}] = \monomial[2]\monomial[{1}]$, $
	\funcGfForANF[2](\seqx) = 0$, $\funcGgForANF[1](\seqx)  = \monomial[{\permutationMono[{1}]}]\monomial[{\permutationMono[{2}]}] = \monomial[2]\monomial[{1}]$, and $
	\funcGfForANF[2](\seqx) = \monomial[{\permutationMono[{2}]}]  = \monomial[1]$, respectively,
	which result in the same sequences obtained in  Example \ref{ex:signConstructionSequences}. 
	Now assume that $\operatorBinary[0][{\indexIteration}]\{\functionh\}={\functionh}$ and $\operatorBinary[1][{\indexIteration}]\{\functionh\}={-\functionh}$. Since $\exponentialBase^{j\numberOfPointsForPSK/2}=-1$, we can then express the composite operator as $\compositeOperatorF[\varMonomial][\functionh] =\exponentialBase^{j\funcfForANFdec(\varMonomial)}\times\functionh$, where the underlying function determining the sign is $\funcfForANF(\seqx)=\frac{\numberOfPointsForPSK}{2}\sum_{\indexIteration=1}^{\numberOfIterations-1}\monomial[{\permutationMono[{\indexIteration}]}]\monomial[{\permutationMono[{\indexIteration+1}]}]$.
\end{example}

Lemma~\ref{th:framework} can be used for any recursion in the form given in \eqref{eq:iterationBasic} to identify underlying code. For example,  the underlying code for $\numberOfIterations$-bit Gray code  can be obtained as follows:
\begin{example}[Gray Code]
	\rm Assume the operator configuration given by $\operator[11][{\indexIteration}]=\operator[22][{\indexIteration}]=\operatorBinary[0][{\indexIteration}]$ and $\operator[12][{\indexIteration}]=\operator[21][{\indexIteration}]=\operatorBinary[1][{\indexIteration}]$ for all iterations and  $\seqPermutationShift=(0,1,\mydots,\numberOfIterations-1)$. 	
	In this case, the configuration vector is  $ \vecArrangement[\indexIteration]^{\rm T} =[0~1~1~0]$ for $\indexIteration=1,2,\mydots,\numberOfIterations$. Hence, by using \eqref{eq:closedformANFsF}, $\funcGfForANF[\indexIteration](\seqx)$ is obtained as
	\begin{align}
	\funcGfForANF[\indexIteration](\seqx) &= \begin{cases}
	\monomial[{\permutationMono[{\numberOfIterations}]}], 		& \text{if}~\indexIteration=\numberOfIterations\\
	(\monomial[{\permutationMono[{\indexIteration}]}]+\monomial[{\permutationMono[{\indexIteration+1}]}])_2,     	& \text{if}~\indexIteration<\numberOfIterations
	\end{cases}~,\label{eq:gray}
	\end{align}
where $\seqPermutationCompShift=(\numberOfIterations, \numberOfIterations-1,\mydots,1)$. As the recursion combines two polynomials after applying  $\operatorBinary[1][{\indexIteration}]$ to the second polynomial and the positions of  $(\indexIteration-1)$th operators  are reflected in $\functionf[\indexIteration-1]$ and $\functiong[\indexIteration-1]$, $\funcGfForANF[\indexIteration](\seqx)$  for $\indexIteration=1,2,\mydots,\numberOfIterations$  form the basis vectors of $\numberOfIterations$-bit Gray code, e.g., $\seqGf[1] = (0,1,1,0,0,1,1,0)$,
$\seqGf[2] = (0,0,1,1,1,1,0,0)$, and
$\seqGf[3] = (0,0,0,0,1,1,1,1)$ for $\numberOfIterations=3$.
\label{ex:gray}
\end{example}

\section{A Generic CS Construction}\label{sec:derivation-of-complementary-sequence-encoder}

In this section, we first utilize Lemma~\ref{th:framework} to develop a direct \ac{GCP} construction by re-expressing the recursion given in Theorem~\ref{th:golayIterative}. We then discuss how to generate distinct \ac{CS} without an auxiliary method and compare it with other constructions in the literature through examples.
\begin{theorem}[Main Theorem]
	\label{th:reduced}
	Let $(\seqGa,\seqGb)$ be a \ac{GCP} of length $\lengthGaGb\ge1$ and $\seqPermutationCompShift=(\permutationMono[\indexIteration])_{\indexIteration=1}^{\numberOfIterations}$ be a sequence defined by a permutation of $\{1,2,\dots,\numberOfIterations\}$. For any $\numberOfPointsForPSK\in\integersPositive$, $\scaleEexp[\indexIteration],\arbitraryScaleE\in\realNumbers$, and $\angleexpAll[\indexIteration],\arbitraryPhaseK, \arbitraryPhaseKPP \in[0,\numberOfPointsForPSK)$ for $\indexIteration=1,2,\mydots,\numberOfIterations$ and $\exponentialBase=\constante^{\frac{2\pi}{\numberOfPointsForPSK}}$, let
	\begin{align}
	&\funcfForFinalAmplitude(\seqx)
	=\scaleEexp[\numberOfIterations]\monomial[{\permutationMono[{\numberOfIterations}]}]+{ \sum_{\indexIteration=1}^{\numberOfIterations-1}\scaleEexp[\indexIteration](\monomial[{\permutationMono[{\indexIteration}]}] +\monomial[{\permutationMono[{\indexIteration+1}]}])_2 }+\arbitraryScaleE\label{eq:realPartReduced}~,
	\\
	&\funcgForFinalAmplitude(\seqx)
	=\scaleEexp[\numberOfIterations](1+\monomial[{\permutationMono[{\numberOfIterations}]}])_2+{ \sum_{\indexIteration=1}^{\numberOfIterations-1}\scaleEexp[\indexIteration](\monomial[{\permutationMono[{\indexIteration}]}] +\monomial[{\permutationMono[{\indexIteration+1}]}])_2 }+\arbitraryScaleE\label{eq:realPartReducedG}~,
	\\	
	&\funcfForFinalPhase(\seqx)
	= {\frac{\numberOfPointsForPSK}{2}\sum_{\indexIteration=1}^{\numberOfIterations-1}\monomial[{\permutationMono[{\indexIteration}]}]\monomial[{\permutationMono[{\indexIteration+1}]}]}+\sum_{\indexIteration=1}^\numberOfIterations \angleexpAll[\indexIteration]\monomial[{\permutationMono[{\indexIteration}]}]+  \arbitraryPhaseK\label{eq:imagPartReduced}~,
	\\	
	&\funcgForFinalPhase(\seqx)
	= {\frac{\numberOfPointsForPSK}{2}\left(\monomial[{\permutationMono[{\numberOfIterations}]}]+\sum_{\indexIteration=1}^{\numberOfIterations-1}\monomial[{\permutationMono[{\indexIteration}]}]\monomial[{\permutationMono[{\indexIteration+1}]}]\right)}_2+\sum_{\indexIteration=1}^\numberOfIterations \angleexpAll[\indexIteration]\monomial[{\permutationMono[{\indexIteration}]}]+  \arbitraryPhaseKPP\label{eq:imagPartReducedG}~,	
	\\
	&\funcfForCommonShift(\seqx) =\sum_{\indexIteration=1}^\numberOfIterations\separationGolay[\indexIteration]\monomial[{\permutationMono[{\indexIteration}]}]~,
	\label{eq:shift}
	\\
	&\funcfForCommonOrder(\seqx,\polyVariable)=\polySeq[{\seqGaP}][\polyVariable](1+\monomial[{\permutationMono[{1}]}])_2+\polySeq[{\seqGbP}][\polyVariable]\monomial[{\permutationMono[{1}]}]~.\label{eq:order}
	\end{align}
	Then, the sequences $\seqGc$ and $\seqGd$ associated with the polynomials given by
	\begin{align}
	\polySeq[{\seqGcP}][\polyVariable] &= 
	\sum_{\varMonomial=0}^{2^\numberOfIterations-1}
	\funcfForCommonOrderDec(\varMonomial,\polyVariable)
	\exponentialBase^{\funcfForFinalAmplitudeDec(\varMonomial)   + \constantj \funcfForFinalAmplitudeDec(\varMonomial)}
	\polyVariable^{\funcfForCommonShiftDec(\varMonomial)+\varUpsample\varMonomial}~,\label{eq:encodedFOFDMonly}\\
	\polySeq[{\seqGdP}][\polyVariable] &= 
	\sum_{\varMonomial=0}^{2^\numberOfIterations-1}
	\funcfForCommonOrderDec(\varMonomial,\polyVariable)
	\exponentialBase^{\funcgForFinalAmplitudeDec(\varMonomial)   + \constantj \funcgForFinalPhaseDec(\varMonomial)}
	\polyVariable^{\funcfForCommonShiftDec(\varMonomial)+\varUpsample\varMonomial}~,\label{eq:encodedGOFDMonly}	
	\end{align}
    construct a \ac{GCP}.
\end{theorem}

The proof is given in Appendix~\ref{app:red} in detail.

Theorem~\ref{th:reduced} shows how each element of the generated \acp{CS} is formed explicitly as compared to Theorem~\ref{th:golayIterative}. As described in Section~\ref{sec:pre}, while $\funcfForFinalAmplitude(\seqx)$ alters the amplitude of $\exponentialBase^{\funcfForFinalAmplitude(\seqx)   + \constantj \funcfForFinalPhase(\seqx)}$, $\funcfForFinalPhase(\seqx)$ changes its phase.
The function $\funcfForCommonShift(\seqx)$  determines the position of the seed sequences in the constructed \ac{CS} by offsetting the order of the polynomial $\funcfForCommonOrder(\seqx,\polyVariable)$. The seed sequences are also scaled by different complex values, i.e., $\exponentialBase^{\funcfForFinalAmplitude(\seqx)   + \constantj \funcfForFinalPhase(\seqx)}$ and $\exponentialBase^{\funcgForFinalAmplitude(\seqx)   + \constantj \funcgForFinalPhase(\seqx)}$.

The proof of Theorem~\ref{th:reduced} results in the following conclusions:
\begin{corollary}
\label{co:RecAlg}
 Let $\scaleA[\indexIteration]=\exponentialBase^{\scaleAexp[\indexIteration]}$, $\scaleB[\indexIteration]=\exponentialBase^{\scaleBexp[\indexIteration]}$, 
$\angleScaleA[\indexIteration][]=\exponentialBase^{\constantj\angleScaleAexp[\indexIteration]}$,
$\angleScaleB[\indexIteration][]=\exponentialBase^{\constantj\angleScaleBexp[\indexIteration]}$,
and $\angleGolay[\indexIteration]=\exponentialBase^{\constantj\angleexp[\indexIteration]}$. 
	Theorem~\ref{th:golayIterative} and Theorem~\ref{th:reduced} lead to  identical \ac{GCP} if $\permutationMono[\indexIteration]=\numberOfIterations - \permutationShift[\indexIteration]$,
	$\separationGolay[\indexIteration|\indexIteration]=\separationIterative[\indexIteration]$, 
	 $\scaleEexp[\indexIteration|\indexIteration]=\scaleBexp[\indexIteration]-\scaleAexp[\indexIteration]$ for $\indexIteration=1,2,\mydots,\numberOfIterations$, $\arbitraryScaleE= \sum_{\indexIteration=1}^{\numberOfIterations}\scaleAexp[\indexIteration]$,  
	$\angleexpAll[1] =\angleexp[1]+\angleScaleBexp[1]-\angleScaleAexp[1]$, $ 
	\angleexpAll[\indexIteration|\indexIteration=2,3,\mydots,\numberOfIterations]=\angleexp[\indexIteration]+\angleScaleBexp[\indexIteration]-\angleScaleAexp[\indexIteration]-\angleScaleBexp[\indexIteration-1]-\angleScaleAexp[\indexIteration-1]$, $\arbitraryPhaseK =  \sum_{\indexIteration=1}^{\numberOfIterations}\angleScaleAexp[\indexIteration]$, and $\arbitraryPhaseKPP = -\angleScaleBexp[{\numberOfIterations}] + \sum_{\indexIteration=1}^{\numberOfIterations-1}\angleScaleAexp[{\indexIteration}]$.
\end{corollary}

\begin{corollary}
The length of the constructed \acp{CS} in \eqref{eq:encodedFOFDMonly} and \eqref{eq:encodedGOFDMonly} are
  $\varUpsample(2^\numberOfIterations-1)+\lengthGaGb+\sum_{\indexIteration=1}^\numberOfIterations\separationGolay[\indexIteration]$.
\end{corollary}
\begin{proof}
The length is equal to
$\max_{\varMonomial} {\varUpsample\varMonomial+ \lengthGaGb+\funcfForCommonShiftDec(\varMonomial)}$, which gives $\varUpsample(2^\numberOfIterations-1)+\lengthGaGb+\sum_{\indexIteration=1}^\numberOfIterations\separationGolay[\indexIteration]$
since the function $\funcfForCommonShiftDec(\varMonomial)$ gets its maximum and minimum values when $\varMonomial$ is maximum, i.e., $2^\numberOfIterations-1$, and minimum, i.e., $0$, respectively.
\end{proof}

In the rest of the study, we provide our discussions based on $\polySeq[{\seqGcP}][\polyVariable]$ as  $\polySeq[{\seqGcP}][\polyVariable]$  and $\polySeq[{\seqGdP}][\polyVariable]$ share the same properties.

\def\setMonomials[#1]{{\mathcal{M}_{#1}}}
\def\varMonomialAnother{{y}}
\subsection{Distinct CSs with a Fixed Support}

\subsubsection{Avoiding Overlapping} 
For a given $\varMonomial$, the order of the  polynomial $\funcfForCommonOrderDec(\varMonomial,\polyVariable)\polyVariable^{\varUpsample\varMonomial+\funcfForCommonShiftDec(\varMonomial)}$ is $\varUpsample\varMonomial+\funcfForCommonShiftDec(\varMonomial)+\lengthGaGb-1$. Hence, the seed sequences are placed side-by-side for $\varUpsample=\lengthGaGb$ and $\funcfForCommonShiftDec(\varMonomial)=0$ for all $\varMonomial$. On the other hand,  for different values of $\varMonomial$, the resulting polynomials can consist of the same monomials for $\funcfForCommonShiftDec(\varMonomial)\neq0$ or $\varUpsample\neq\lengthGaGb$. Thus, $\funcfForCommonShift(\seqx)$ can change the phase and amplitude of the elements through the superposition of the seed sequences. This issue can  be avoided if
\begin{align}
\separationGolay[k|{\permutationMono[{k}]=a}] \ge \sum_{\ell\in \{l| {\permutationMono[{l}]>a}\}} \separationGolay[\ell],~  \varUpsample\ge\lengthGaGb,
\label{eq:conditionNoOverlapping}
\end{align}
for  $1\le a\le \numberOfIterations-1$. For example, for $\seqPermutationCompShift=(\numberOfIterations, \numberOfIterations-1,\mydots,1)$, several conditions can be listed as $ \separationGolay[m] \ge  \separationGolay[m-1]+\separationGolay[m-2]+\cdots+\separationGolay[1]$ for $a=1$,  $ \separationGolay[m-1] \ge  \separationGolay[m-2]+\separationGolay[m-3]+\cdots+1$ for $a=2$, and $ \separationGolay[2] \ge  \separationGolay[1]$ for $a=\numberOfIterations-1$. The conditions in \eqref{eq:conditionNoOverlapping} can be derived by evaluating  the Boolean functions in \eqref{eq:shift}. For instance, $\monomial[1]$ in \eqref{eq:shift} shifts the last $2^{\numberOfIterations-1}$ seed sequences  by $\separationGolay[k|{\permutationMono[{k}]=1}]$ as the last $2^{\numberOfIterations-1}$ elements of the corresponding sequence for $\monomial[1]$ is $1$. Therefore, it gives a room of size $\separationGolay[k|{\permutationMono[{k}]=1}]$ for other shift parameters for  $\varUpsample=\lengthGaGb$ as expressed in \eqref{eq:conditionNoOverlapping} for $a=1$. 

\subsubsection{Fixed Support}
Let $\funcfForCommonShift(\seqx)$  be a fixed function for all $\seqPermutationCompShift$ by using a fixed set of $\separationGolay[{\permutationMono[{\indexIteration}]}]$ for $\indexIteration=1,2,\mydots,\numberOfIterations$. In this case, the supports of the seed sequences in the \ac{CS} do not change for different $\seqPermutationCompShift$ values, which is instrumental to generate \acp{CS} with a certain support.

\subsubsection{Distinct Sequences}
\label{subsec:piReverse}
Let $\funcfForCommonShift(\seqx)$ be a fixed function where its  coefficients satisfy \eqref{eq:conditionNoOverlapping}. Then, the superpositions of the seed sequences are avoided and the supports of the seed sequences are fixed. Therefore, the functionality of $\funcfForCommonShift(\seqx)$   in \eqref{eq:encodedFOFDMonly} is completely separated the ones for $\funcfForFinalPhase(\seqx)$, $\funcfForFinalAmplitude(\seqx)$, and $\funcfForCommonOrder(\seqx,\polyVariable)$. Hence, for distinct  $\funcfForFinalAmplitude(\seqx)$, $\funcfForFinalPhase(\seqx)$, or $\funcfForCommonOrder(\seqx,\polyVariable)$, the \ac{CS} $\seqGc$ is distinct.

The functions  $\funcfForFinalAmplitude(\seqx)$, $\funcfForFinalPhase(\seqx)$, or $\funcfForCommonOrder(\seqx,\polyVariable)$ lead to sets of distinct sequences  for a given $\seqPermutationCompShift$. On the other hand, for different $\seqPermutationCompShift$, their combined impact result in distinct \acp{CS} under certain conditions: 
\begin{itemize}
	\item For  $\lengthGaGb=1$, when the elements of  $\seqPermutationCompShift$ are reversed, $\sum_{\indexIteration=1}^{\numberOfIterations-1}\monomial[{\permutationMono[{\indexIteration}]}]\monomial[{\permutationMono[{\indexIteration+1}]}]$ does not change. Therefore, $\seqPermutationCompShift$ and the reversed $\seqPermutationCompShift$ do not change the set of sequences generated by $\funcfForFinalPhase(\seqx)$ if  $\angleexpAll[\indexIteration],\arbitraryPhaseK \in\integers_{\numberOfPointsForPSK}$. However, if one of $\{\angleexpAll[\indexIteration]\}$ is offset by a non-integer value for even $\numberOfIterations$, $\seqPermutationCompShift$ and the reversed $\seqPermutationCompShift$  lead to different set of sequences. For odd $\numberOfIterations$, $\seqPermutationCompShift$ and the reversed $\seqPermutationCompShift$ can be used if one of $\{\angleexpAll[\indexIteration|\indexIteration\neq\ceil{\numberOfIterations/2}]\}$ is offset by an non-integer value. Also, all permutations generate a set of distinct \acp{CS} when $\scaleEexp[\numberOfIterations]\neq \scaleEexp[\indexIteration\neq\numberOfIterations]$ since $\funcfForFinalAmplitude(\seqx)$ yields different functions for $\seqPermutationCompShift$ and reversed $\seqPermutationCompShift$ even $\funcfForCommonPhaseA(\seqx)$ remains unchanged. 
	
	\item For $\lengthGaGb>1$ ($\seqGa\neq k\seqGb$ for $k\in\complexNumbers$), each permutation lead to a distinct set of \acp{CS}. This is because the polynomial $\funcfForCommonOrder(\seqx,\polyVariable)$ is a function of $\monomial[{\permutationMono[{1}]}]$. When $\seqPermutationCompShift$ is reversed, the placements of the seed sequences in the \ac{CS} $\seqGc$ changes. 
\end{itemize}

\subsection{Comparisons with Other Constructions and Examples}
\label{subsec:comparison}

Theorem~\ref{th:reduced} under the condition \eqref{eq:conditionNoOverlapping} provides further insight into \acp{CS} as follows: 

\subsubsection{Independent functions for amplitude and phase} 
Theorem~\ref{th:reduced} extends the Davis and Jedwab's  construction \cite{davis_1999} by providing independent functions that manipulate the amplitude and phase of the elements of the generated \acp{CS}, where the one for phase coincides with the result in \cite{davis_1999}. The parameters $\scaleEexp[\indexIteration]$ for $\indexIteration=1,2,\mydots,\numberOfIterations$ and $\arbitraryScaleE$ change the real part of the exponent of $\exponentialBase$, which is beneficial to synthesize \acp{CS} with various amplitude levels without affecting the phase of the elements. The offset method, e.g.,   \cite{robing_2001, Chong_2003, Zeng_2014, Chong_2002, Li_2008, Lee_2006, Chang_2010, Jiang_2016,Tarokh_2003, Liu_2014, Li_2010}, and para-unitary matrix-based construction, e.g., \cite{budisin_2018},  do not yield concise expressions related to the phase, amplitude, or seed sequences for synthesizing \acp{CS}. As we shown in  Section~\ref{sec:constellation} and Section~\ref{sec:encDec}, the independent pseudo-Boolean functions for amplitude and phase can be exploited to synthesize \acp{CS} with various constellation and develop algorithms for receiver.

\begin{example}
	\rm
	Let $\numberOfIterations=3$, $\varUpsample=1$, $\seqPermutationCompShift=(3,2,1)$, 
	$\scaleEexp[1]=\frac{\numberOfPointsForPSK}{2\pi}\ln(3)$, 
	$\scaleEexp[\indexIteration|\indexIteration=2,3]=0$,
	$\arbitraryScaleE=0$,
	$\numberOfPointsForPSK=4$,
	$\angleexpAll[\indexIteration|\indexIteration=1,2,3]=0$, $\arbitraryPhaseK=0$,
	$\separationGolay[\indexIteration|\indexIteration=1,2,3]=0$, $\seqGa =\seqGb = (1)$. Based on Theorem~\ref{th:reduced}, the basis vectors for $\funcfForFinalAmplitude(\seqx)$ can be listed as
	$	\monomial[{\permutationMono[{3}]}] = (0,0,0,0 , 1 , 1 , 1 , 1)$, $\monomial[{\permutationMono[{2}]}] +\monomial[{\permutationMono[{3}]}]  =  (0, 0, 1, 1, 1, 1, 0, 0)$, $\monomial[{\permutationMono[{1}]}] +\monomial[{\permutationMono[{2}]}] =  (0, 1, 1, 0, 0, 1, 1, 0)$, ${\bf 1} = ( 1, 1, 1, 1, 1, 1, 1, 1)$. The function $\funcfForFinalAmplitude(\seqx)$ leads to the sequence $(0,\scaleEexp[1],\scaleEexp[1],0,0,\scaleEexp[1],\scaleEexp[1],0)$ while the function $\funcfForFinalPhase(\seqx)$ gives the sequence
	$(0,0,0,2,0,0,2,0)$ independently. The function $\funcfForCommonOrder(\seqx,\polyVariable) =\polySeq[{\seqGaP}][\polyVariable](1-\monomial[{\permutationMono[{1}]}])_2+\polySeq[{\seqGbP}][\polyVariable]\monomial[{\permutationMono[{1}]}] = \polySeq[{\seqGaP}][\polyVariable](1-\monomial[3])_2+\polySeq[{\seqGbP}][\polyVariable]\monomial[3] = 1$ results in $(1,1,1,1,1,1,1,1)$. As $\separationGolay[\indexIteration|\indexIteration=1,2,3]=0$, $\funcfForCommonShift(\seqx)=0$. Finally, combining all functions with \eqref{eq:encodedFOFDMonly}, the sequence $\seqGc$ is obtained as $(1,3,3,-1,1,3,-3,1)$. 
	\label{ex:synthesis}
\end{example} 

Note that for $\seqPermutationCompShift=(\numberOfIterations, \numberOfIterations-1,\mydots,1)$, the amplitude of the elements of the \ac{CS} are determined by the linear combinations of the columns of the $\numberOfIterations$-bit Gray code (see \eqref{eq:gray} and \eqref{eq:realPartReduced}) over $\realNumbers$ since the variables that change the amplitudes  in Theorem~\ref{th:golayIterative} alternate as in Example~\ref{ex:gray}.

\subsubsection{Enumeration under the presence of a seed GCP} 
Theorem~\ref{th:reduced}  gives contiguous \acp{CS} of length $\lengthGaGb\cdot 2^\numberOfIterations$ when a seed \ac{GCP} $(\seqGa,\seqGb)$ of length $\lengthGaGb>1$ is utilized for $\varUpsample=\lengthGaGb$ and $\funcfForCommonShift(\seqx)=0$. 
It reduces to the  method in \cite{davis_1999} for $\funcfForFinalAmplitude(\seqx)=0$, $\angleexpAll[\indexIteration],\arbitraryPhaseK,\arbitraryPhaseKPP\in\integers_\numberOfPointsForPSK$, and $\seqGa=\seqGb=1$.  However, under the same configuration for  $\lengthGaGb>1$, 
it determines $\numberOfIterations!\cdot\numberOfPointsForPSK^{\numberOfIterations+1}$ \acp{CS}, i.e., doubles the number in \cite{davis_1999}, as all permutations are valid for generating distinct \acp{CS} in the presence of the seed sequences.

\begin{example}
	\rm
	Consider the parameters in Example~\ref{ex:synthesis} and replace $\seqGa$ and $\seqGb$ as $\seqGa = (1,\constantj,1,1,1,-1)$, and $\seqGb = (1,\constantj,1,-1,-1,1)$ \cite{holzmann_1991}. For $\seqPermutationCompShift=(3,2,1)$ and $\varUpsample=6$, the function $\funcfForCommonOrder(\seqx,\polyVariable) = \polySeq[{\seqGaP}][\polyVariable](1-\monomial[{\permutationMono[{1}]}])_2+\polySeq[{\seqGbP}][\polyVariable]\monomial[{\permutationMono[{1}]}] = \polySeq[{\seqGaP}][\polyVariable](1-\monomial[3])_2+\polySeq[{\seqGbP}][\polyVariable]\monomial[3] $ results in $(\polySeq[{\seqGaP}][\polyVariable],\polySeq[{\seqGbP}][\polyVariable],\polySeq[{\seqGaP}][\polyVariable],\polySeq[{\seqGbP}][\polyVariable],\polySeq[{\seqGaP}][\polyVariable],\polySeq[{\seqGbP}][\polyVariable],\polySeq[{\seqGaP}][\polyVariable],$ $\polySeq[{\seqGbP}][\polyVariable])$. Therefore, the resulting sequence $\seqGc$ is length of $48$ and can be expressed as
	$(\seqGa,3\cdot\seqGb,3\cdot\seqGa,-\seqGb,\seqGa,3\cdot\seqGb,-3\cdot\seqGa,\seqGb)$
	from \eqref{eq:encodedFOFDMonly}. Now assume the order of the elements in $\pi$ are reversed, $\scaleEexp[\indexIteration|\indexIteration=1,3]=0$, and $\scaleEexp[2]=\frac{2}{\pi}\ln(3)$. The function $\funcfForCommonOrder(\seqx,\polyVariable) = \polySeq[{\seqGaP}][\polyVariable](1-\monomial[1])_2+\polySeq[{\seqGbP}][\polyVariable]\monomial[1]$ then gives
	$(\polySeq[{\seqGaP}][\polyVariable],\polySeq[{\seqGaP}][\polyVariable],\polySeq[{\seqGaP}][\polyVariable],$ $\polySeq[{\seqGaP}][\polyVariable],\polySeq[{\seqGbP}][\polyVariable],\polySeq[{\seqGbP}][\polyVariable],\polySeq[{\seqGbP}][\polyVariable],\polySeq[{\seqGbP}][\polyVariable])$,
	while $\funcfForFinalAmplitude(\seqx)$ and $\funcfForFinalPhase(\seqx)$ remain the same. The final sequence is obtained as  $(\seqGa,3\cdot\seqGa,3\cdot\seqGa,-\seqGa,\seqGb,3\cdot\seqGb,-3\cdot\seqGb,\seqGb)$, which is different from the original one as the positions of the seed sequences changes.
	\label{ex:init}
\end{example} 

In \cite{fiedler2008_multi}, the multi-dimensional  arrays were exploited for generating \acp{CS} by projecting the arrays to a lower dimension, where the projection operation  re-orders the elements of the multi-dimensional arrays. As compared to \cite{fiedler2008_multi}, we assume that the seed \acp{CS} are given in this study. Our construction can be inferred as a projection from two-dimensional array to a single dimension in \cite{fiedler2008_multi} for $\varUpsample=\lengthGaGb$. However, for $\varUpsample\neq\lengthGaGb$, our construction extends the definition of projection by allowing zero-valued elements or summation of the elements of the seed \acp{CS}. We also identify the seed sequence positions for a given $\pi$, which is useful for multiplexing as done in Section~\ref{sec:numerical}.

\subsubsection{Flexible support} In the literature, the constructions mainly focus on contiguous \acp{CS}, i.e., \acp{CS} with no zero-valued elements \cite{robing_2001, Chong_2003, Zeng_2014, Chong_2002, Li_2008, Lee_2006, Chang_2010, Jiang_2016,Tarokh_2003, Liu_2014, Li_2010, budisin_2018}. 
In contrast,
Theorem~\ref{th:reduced}  explains how to generate distinct non-contiguous \acp{CS} under the condition given in \eqref{eq:conditionNoOverlapping}. In \cite{Zhou_2020}, 
various resource allocations
for complementary sets were considered for preamble design. However, the maximum PMEPR of the resulting waveform reaches 4 dB. In \cite{Liu_2018}, several correlation bounds under non-contiguous resource allocation and several constructions were discussed.   However, systematic construction of the zero-valued elements for \acp{CS} with \eqref{eq:shift} was not investigated in these studies.
\begin{example}
	\rm\label{ex:noncontiguous}
	Consider the parameters given in Example~\ref{ex:synthesis}. Let $\seqGa = (1,\constantj,1)$, $\seqGb = (1,1,-1)$ \cite{holzmann_1991} and $\varUpsample=3$. The final sequence $\seqGc$ is $(\seqGa,3\cdot\seqGb,3\cdot\seqGa,-\seqGb,\seqGa,3\cdot\seqGb,-3\cdot\seqGa,\seqGb)$. Now assume that $\separationGolay[3]=60$ and $\separationGolay[\indexIteration|\indexIteration=1,2]=0$,  
	The function $\funcfForCommonShift(\seqx) $ leads to $(0,0,0,0,60,60,60,60)$. Hence, from \eqref{eq:encodedFOFDMonly}, the support of the last four parts of $\seqGc$ are shifted by $60$. The resulting sequence can be expressed as
	\begin{align}
		\seqGc = (\seqGa,3\cdot\seqGb,3\cdot\seqGa,-\seqGb,\overbrace{0,\mydots,0}^{\separationGolay[1]=60},\seqGa,3\cdot\seqGb,-3\cdot\seqGb,\seqGa)~.\nonumber
	\end{align}
	In other words, a non-contiguous \ac{CS} with two clusters of length of $12$ separated by $\separationGolay[2]=60$ zero-valued elements is obtained where the alphabet on the clusters remains the same. Under the same condition, assume that $\seqPermutationCompShift=(1,2,3)$. Let $\separationGolay[1]=60$ and $\separationGolay[\indexIteration|\indexIteration=2,3]=0$. Since  $\funcfForCommonShift(\seqx)$ still gives $(0,0,0,0,60,60,60,60)$ the resulting sequence keeps the original support as
	\begin{align}
		\seqGc = (\seqGa,3\cdot\seqGa,3\cdot\seqGa,-\seqGa,\overbrace{0,\mydots,0}^{\separationGolay[3]=60},\seqGb,3\cdot\seqGb,-3\cdot\seqGb,\seqGb)~\nonumber
	\end{align}
	while the position of the seed sequence change.
\end{example}

\subsubsection{Continuous parameters} Theorem~\ref{th:reduced} provides a framework for synthesizing \acp{CS} with various alphabets through the continuous parameters   $\scaleEexp[\indexIteration]$ and $\arbitraryScaleE$, and $\angleexpAll[\indexIteration]$ and $\arbitraryPhaseK$, as opposed to existing work on \acp{CS} \cite{paterson_2000, Li_2010,budisin_2018}. For example, since $\angleexpAll[\indexIteration]$ and $\arbitraryPhaseK$ are in $[0,\numberOfPointsForPSK)$, i.e., a range rather than $\integers_\numberOfPointsForPSK$, it offers a  flexibility in phase. Hence, it is possible to generate \acp{CS} with various alphabets as  discussed in Section~\ref{sec:constellation} in detail.

\section{Complementary Sequences with Equiangular Sub-symmetric Constellations}\label{sec:constellation}
%We enumerate \acp{CS} for a given \ac{ESC} by determining the sets for $\arbitraryPhaseK$ and $\angleexpAll[\indexIteration]$, and the values of $\arbitraryScaleE$, $\scaleEexp[\indexIteration]$,  and $\numberOfPointsForPSK$ in Theorem~\ref{th:reduced} by exploiting the equiangularity and sub-symmetricity of $\constellationPoints$. 
Consider two points $\pointQAM_\indexU$ and $\pointQAM_\indexV$ in $\Squad$ indexed by $\indexU,\indexV\in\{1,2,\mydots,\numberOfPointsOnConstellation\}$, where $\numberOfPointsForPSK$ is an even number\footnote{As the coset term in \eqref{eq:imagPartReduced} can flip the sign of the elements of \acp{CS}, we ensure that  $\pointQAM_\indexU$ and $-\pointQAM_\indexU$ are in $\constellationPoints$ with an even $\numberOfPointsForPSK$ as in \cite{paterson_2000}.}. 
By choosing $\arbitraryScaleE=\frac{\numberOfPointsForPSK}{2\pi}\ln\ratioBetweenDistanceAndInner[\indexU]$  and $\scaleEexp[\indexChoosen]=\frac{\numberOfPointsForPSK}{2\pi}\ln\frac{\ratioBetweenDistanceAndInner[\indexV]}{\ratioBetweenDistanceAndInner[\indexU]}$ while $\scaleEexp[\indexIteration|\indexIteration\neq\indexChoosen]=  0$ for $\indexIteration\in\{1,2,\mydots,\numberOfIterations\}$ in \eqref{eq:realPartReduced} for $\indexChoosen\in\{1,\mydots,\numberOfIterations\}$, we equate the amplitudes of the elements of the \acp{CS}  to either $\ratioBetweenDistanceAndInner[\indexU]$ or $\ratioBetweenDistanceAndInner[\indexV]$.  Since our construction also admits real values for \eqref{eq:imagPartReduced}, 
we set $\arbitraryPhaseK=\arbitraryPhaseKBit+\frac{\numberOfPointsForPSK}{2\pi}\phaseOffset$ and $\angleexpAll[\indexIteration]= \angleexpAllBit[\indexIteration]+\frac{\numberOfPointsForPSK}{2\pi}\phaseOffsetl[\indexIteration]$ for $\phaseOffset,\phaseOffsetl[\indexIteration]\in\realsNonnegative$ and $\arbitraryPhaseKBit,\angleexpAll[\indexIteration]\in\integers_{\numberOfPointsForPSK}$ for $\indexIteration\in\{1,2,\mydots,\numberOfIterations\}$. While $\indexChoosen$ determines the parameters to be offset, $\indexU$ and $\indexV$ identify where to offset.

Consider the case $\indexU=\indexV$. Due to the equiangularity property, an \ac{ESC} consists of the points in $\Squad$ rotating with an integer multiple of $2\pi/\numberOfPointsForPSK$ radian. Hence,  by offsetting $\arbitraryPhaseKBit$ by $\phaseOffset=\angleBetweenPointAndXYDiagonal[\indexU]$, the rotated and scaled $\numberOfPointsForPSK$-PSK \ac{GDJ} sequences on an \ac{ESC} alphabet can be obtained. 
Since there are $\numberOfPointsOnConstellation$ elements in $\Squad$, only $\numberOfPointsOnConstellation$ valid permutations  exist for $(\indexU ,\indexV)$, where each permutation generates  $\unitSequences\triangleq\frac{\numberOfIterations!}{2} \numberOfPointsForPSK^{\numberOfIterations+1}$ sequences for  $\lengthGaGb=1$ and $  2\unitSequences$ sequences for $\lengthGaGb>1$, based on Section~\ref{subsec:piReverse}.

Now, consider the case $\indexU\neq\indexV$. Assume that $\angleexpAll[{1}]=\angleexpAllN[{1}]$,  $\angleexpAll[{\indexIteration}]=\angleexpAllN[{\indexIteration}]\pm\angleexpAllN[{\indexIteration-1}]$ for $\indexIteration=2,\mydots,\numberOfIterations$, and $\arbitraryPhaseK=\arbitraryPhaseKN$. We then write
$
	\sum_{\indexIteration=1}^\numberOfIterations \angleexpAll[\indexIteration]\monomial[{\permutationMono[{\indexIteration}]}]+  \arbitraryPhaseK=\angleexpAllN[{\numberOfIterations}]\monomial[{\permutationMono[{\numberOfIterations}]}]+\sum_{\indexIteration=1}^{\numberOfIterations-1}\angleexpAllN[{\indexIteration}]\left( (\monomial[{\permutationMono[{\indexIteration}]}])_2
	\pm(\monomial[{\permutationMono[{\indexIteration+1}]}])_2\right)+  \arbitraryPhaseKN$, which allows us to express the phase and amplitude functions in similar forms. For example, if $\angleexpAll[{\indexIteration}]=\angleexpAllN[{\indexIteration}]-\angleexpAllN[{\indexIteration-1}]$,
\begin{align}
(\monomial[{\permutationMono[{\indexIteration}]}])_2
-(\monomial[{\permutationMono[{\indexIteration+1}]}])_2\in
\begin{cases}
	\{0\},	& \text{if}~(\monomial[{\permutationMono[{\indexIteration}]}]
	+\monomial[{\permutationMono[{\indexIteration+1}]}])_2 =0  	\\
	\{-1,1\},&	\text{if}~(\monomial[{\permutationMono[{\indexIteration}]}]
	+\monomial[{\permutationMono[{\indexIteration+1}]}])_2 =1 \\
\end{cases}.\label{eq:vrot}
\end{align}
Therefore,  $\angleexpAllN[{\indexChoosen}]$ and  $\scaleEexp[\indexChoosen]$ alter the same elements of the sequence. Since the non-zero values of $(\monomial[{\permutationMono[{\indexChoosen}]}])_2
-(\monomial[{\permutationMono[{\indexChoosen+1}]}])_2$ are either $-1$ or $1$, the phase function rotates the  elements scaled with $\scaleEexp[\indexChoosen]$  in the counterclockwise and the clockwise directions equally on the unit circle. For $\angleexpAllN[{\indexChoosen}]=\angleBetweenPointAndXYDiagonal[\indexV ]$, the amount of the phase rotation is $\angleBetweenPointAndXYDiagonal[\indexV ]$. If $\angleexpAll[{\indexIteration}]=\angleexpAllN[{\indexIteration}]+\angleexpAllN[{\indexIteration-1}]$,
\begin{align}
	1-(\monomial[{\permutationMono[{\indexIteration}]}])_2
	-(\monomial[{\permutationMono[{\indexIteration+1}]}])_2\in
	\begin{cases}
		\{0\},	& \text{if}~(\monomial[{\permutationMono[{\indexIteration}]}]
		+\monomial[{\permutationMono[{\indexIteration+1}]}])_2 =1  	\\
		\{-1,1\},&	\text{if}~(\monomial[{\permutationMono[{\indexIteration}]}]
		+\monomial[{\permutationMono[{\indexIteration+1}]}])_2 =0 \\
	\end{cases},\label{eq:urot}
\end{align}
which implies that the amount of phase rotation in the counterclockwise and the clockwise directions  for the elements that are {\em not} scaled by $\scaleEexp[\indexChoosen]$ is $\angleBetweenPointAndXYDiagonal[\indexU]$ for $\arbitraryPhaseKN=\angleBetweenPointAndXYDiagonal[\indexU ]$ and $\angleexpAllN[{\indexChoosen}]=-\angleBetweenPointAndXYDiagonal[\indexU ]$. 

The rotation in the counterclockwise and the clockwise directions equally is exactly aligned with  the sub-symmetricity property of an \ac{ESC}. Equations \eqref{eq:vrot} and \eqref{eq:urot} allow us to alter the different elements of the sequence,  for given $\indexChoosen$, $\indexU$, and $\indexV$. Their combined impacts on the phase offset parameters can be calculated as $\phaseOffset=\angleBetweenPointAndXYDiagonal[\indexU]$,
$\phaseOffsetl[\indexChoosen] = \angleBetweenPointAndXYDiagonal[\indexV ]-\angleBetweenPointAndXYDiagonal[\indexU]$, $\phaseOffsetll[\indexChoosen+1]=-\angleBetweenPointAndXYDiagonal[\indexV ]-\angleBetweenPointAndXYDiagonal[\indexU ]$ if $\indexChoosen<\numberOfIterations$, and $\phaseOffsetll[\indexIteration|\indexIteration\neq\indexChoosen,\indexChoosen+1]=0$ for $\indexChoosen\in\{1,2,\mydots,\numberOfIterations\}$.

For $\lengthGaGb>1$, the total number of $(\indexU ,\indexV)$ permutations where $\indexU\neq\indexV$ is $\numberOfPointsOnConstellation^2-\numberOfPointsOnConstellation$ and  $\indexChoosen\in\{1,2,\mydots,\numberOfIterations\}$. This implies that the total number \acp{CS} with an \ac{ESC} generated through the sub-symmetricity is $2((\numberOfPointsOnConstellation^2-\numberOfPointsOnConstellation)\numberOfIterations)\cdot\unitSequences$. For $\lengthGaGb=1$, $\seqPermutationCompShift$ and the reversed $\seqPermutationCompShift$ give different \acp{CS} when $\ratioBetweenDistanceAndInner[\indexU]\neq\ratioBetweenDistanceAndInner[\indexV]$ for $\indexChoosen=\numberOfIterations$ based on the discussions in Section~\ref{subsec:piReverse}. In addition, %when $|\angleBetweenPointAndXYDiagonal[\indexU]|\neq|\angleBetweenPointAndXYDiagonal[\indexV]|$ for $\indexChoosen=\numberOfIterations$, all permutations generate distinct sequences. %This is due to the fact that  $\seqPermutationCompShift$ and $\indexChoosen=\ell'$ and the reversed $\seqPermutationCompShift$ with  $\indexChoosen=\numberOfIterations-\ell'$ lead to the same functions for $\ell'<\numberOfIterations$. 
 $\seqPermutationCompShift$ and the reversed $\seqPermutationCompShift$ yield different \acp{CS} for $\indexChoosen=\numberOfIterations$ if $\phaseOffsetl[\numberOfIterations] = \angleBetweenPointAndXYDiagonal[\indexV ]-\angleBetweenPointAndXYDiagonal[\indexU]\neq 0$  and $\angleBetweenPointAndXYDiagonal[\indexU]+\angleBetweenPointAndXYDiagonal[\indexV]\neq0$, which implies that $|\angleBetweenPointAndXYDiagonal[\indexU]|\neq|\angleBetweenPointAndXYDiagonal[\indexV]|$ should hold. Since these special cases generate $2\unitSequences$  different \acp{CS} for $\indexChoosen=\numberOfIterations$ while others lead to $\unitSequences$ different \acp{CS}, the total number of \acp{CS} can be calculated as
$((\numberOfPointsOnConstellation^2-\numberOfPointsOnConstellation)\numberOfIterations+\numberOfPointsNotEquidistance+\numberOfPointsEquidistanceButDifferentAbsAngle)\cdot\unitSequences$, where  $\numberOfPointsNotEquidistance$ and $\numberOfPointsEquidistanceButDifferentAbsAngle$ are the number of $(\indexU,\indexV)$ permutations for $\ratioBetweenDistanceAndInner[\indexU ]\neq\ratioBetweenDistanceAndInner[\indexV ]$ and
the number of $(\indexU,\indexV)$ permutations for $\ratioBetweenDistanceAndInner[\indexU ]=\ratioBetweenDistanceAndInner[\indexV ]$
and $|\angleBetweenPointAndXYDiagonal[\indexU]|\neq|\angleBetweenPointAndXYDiagonal[\indexV]|$, respectively.

We summarize how the phase and amplitude parameters are modified for given $\indexU$, $\indexV$, and $\indexChoosen$ for an arbitrary \ac{ESC} in Algorithm~\ref{alg:ESC}. The number of distinct \acp{CS} are calculated as follows:
\begin{corollary}
	\label{co:numberOfSequences}
	The numbers of distinct \acp{CS} with an \ac{ESC} of $\numberOfPointsForPSK\cdot\numberOfPointsOnConstellation$ points are at least  $((\numberOfPointsOnConstellation^2-\numberOfPointsOnConstellation)\numberOfIterations+\numberOfPointsNotEquidistance+\numberOfPointsEquidistanceButDifferentAbsAngle+\numberOfPointsOnConstellation)\cdot\unitSequences$ and $2((\numberOfPointsOnConstellation^2-\numberOfPointsOnConstellation)\numberOfIterations+\numberOfPointsOnConstellation)\cdot\unitSequences$ for $\lengthGaGb=1$ and $\lengthGaGb>1$, respectively.
	\label{co:enum}
\end{corollary}
Note that non-contiguous \acp{CS} with a certain support can be generated by fixing $\funcfForCommonShift(\seqx)$ as in Example~\ref{ex:noncontiguous}. Hence, for a given support, the number of distinct non-contiguous \acp{CS} is equal to the number of distinct contiguous \acp{CS}.

\def\minimumRadius{r_{\rm min}}
\def\minimumDiff{\delta_{\rm min}}
\def\minDist{d_{\rm min}}

	\subsection{Spectral Efficiency and Minimum Distance}
	Based on Corollary~\ref{co:numberOfSequences}, the \ac{SE} $\spectralEfficient$ with  \acp{CS} with an arbitrary \ac{ESC} can be as high as $ \floor{\log_2((\numberOfPointsOnConstellation^2-\numberOfPointsOnConstellation)\numberOfIterations+\numberOfPointsNotEquidistance+\numberOfPointsEquidistanceButDifferentAbsAngle+\numberOfPointsOnConstellation)}\cdot\unitSequences/2^\numberOfIterations$ bit/s/Hz and $\floor{\log_2(2((\numberOfPointsOnConstellation^2-\numberOfPointsOnConstellation)\numberOfIterations+\numberOfPointsOnConstellation)\cdot\unitSequences)}/(2^\numberOfIterations\cdot\lengthGaGb)$ bit/s/Hz for $\lengthGaGb=1$ and $\lengthGaGb>1$, respectively. 	To infer the  minimum Euclidean distance, denoted by $\minDist$, let $\minimumRadius\triangleq\min_{\pointQAM_\indexU\in\Squad}|\pointQAM_\indexU|$ and $\minimumDiff\triangleq\min_{\pointQAM_\indexU,\pointQAM_\indexV\in\Squad,\indexU\neq\indexV}|\pointQAM_\indexU-\pointQAM_\indexV|$. 	In \cite[Theorem~10]{davis_1999}, it was shown that the minimum Lee distances for the phase function $\funcfForFinalPhase(\seqx)$ are $2^{\numberOfIterations-1}$ and $2^{\numberOfIterations-2}$ for  $\numberOfPointsForPSK>2$ and $\numberOfPointsForPSK=2$, respectively. Therefore, it can be shown that the minimum Euclidean distance for all sequences generated with $\minimumRadius\exponentialBase^{ \constantj \funcfForFinalPhase(\seqx)}$ is $2^{{(\numberOfIterations+1)}/2}\minimumRadius\sin{(\pi/\numberOfPointsForPSK)}$ for $\numberOfPointsForPSK>2$ and $2^{{\numberOfIterations}/2}\minimumRadius\sin{(\pi/\numberOfPointsForPSK)}$ for $\numberOfPointsForPSK=2$. Also, the amplitude of  $2^{\numberOfIterations-1}$ of the elements of \acp{CS} are set to either $\ratioBetweenDistanceAndInner[\indexU]$ and  $\ratioBetweenDistanceAndInner[\indexV]$ in this study. Therefore, for a given $\funcfForFinalPhase(\seqx)$, the minimum distance is less than or equal to $2^{(\numberOfIterations-1)/2}\minimumDiff$. By combining these results, for $\lengthGaGb=1$ and $\seqGa=\seqGb=(1)$,
	\begin{align}
		\minDist\le
		\begin{cases}
	\min\{2^\frac{\numberOfIterations+1}{2}\sin\left(\frac{\pi}{\numberOfPointsForPSK}\right)\minimumRadius,2^\frac{\numberOfIterations-1}{2}\minimumDiff\},
		& \numberOfPointsForPSK>2	\\
		\min\{2^\frac{\numberOfIterations}{2}\sin\left(\frac{\pi}{\numberOfPointsForPSK}\right)\minimumRadius,2^\frac{\numberOfIterations-1}{2}\minimumDiff\},
& \numberOfPointsForPSK=2
		\end{cases}.
		\label{eq:dminBound}
	\end{align}
	When the seed sequences exist, $\minDist$ is a function of the seed sequences in general.%In this study, we use seed sequences for orthogonal user multiplexing. In this case, \eqref{eq:dminBound} is scaled by the norm of seed sequences for each user.

Our numerical analysis on $\minDist$ shows that \eqref{eq:dminBound} is exact in certain scenarios, e.g., $4\sqrtofNumPointsInQuad^2$-QAM for $(\sqrtofNumPointsInQuad,\numberOfIterations)=\{(2,2),(2,3),(3,2),(3,3)\}$, $\numberOfPointsForPSK\sqrtofNumPointsInQuad$-STAR for $(\sqrtofNumPointsInQuad,\numberOfIterations)=\{(2,2),(2,3),(2,4), (3,2),(3,3),(3,4)\}$ for $\numberOfPointsForPSK=4$). However, we do not have a proof that shows the tightness of \eqref{eq:dminBound} in this study.

\def\isReversedPiValid{{isReversedPiValid}}

\if\IEEEsubmission0
\else
\renewcommand{\baselinestretch}{1}
\fi
\begin{algorithm}[t]{
	\scriptsize
	\caption{\small Offsetting amplitude and phase parameters   for an arbitrary ESC}\label{alg:ESC}
	\SetKwInput{KwInput}{Input}                % Set the Input
	\SetKwInput{KwOutput}{Output}              % set the Output
	\DontPrintSemicolon
	
	\KwInput{$\Squad,\numberOfPointsForPSK,\lengthGaGb,\numberOfIterations,\indexU,\indexV,\indexChoosen$, and $\arbitraryPhaseKBit,\angleexpAllBit[\indexIteration]\in\integers_{\numberOfPointsForPSK}$ for $\indexIteration=1,2,\mydots,\numberOfIterations$}
	\KwOutput{$\arbitraryScaleE,\scaleEexp[\indexIteration],\arbitraryPhaseK,\angleexpAll[\indexIteration]$  for $\indexIteration=1,2,\mydots,\numberOfIterations$, \isReversedPiValid}
	
	% Set Function Names
	\SetKwFunction{FMain}{calculateParametersWithOffsets}
	
	\SetKwProg{Fn}{Function}{}{}
	\Fn{\FMain}{
		$\arbitraryScaleE\leftarrow\frac{\numberOfPointsForPSK}{2\pi}\ln\ratioBetweenDistanceAndInner[\indexU]$, $\scaleEexp[\indexChoosen]\leftarrow\frac{\numberOfPointsForPSK}{2\pi}\ln\frac{\ratioBetweenDistanceAndInner[\indexV]}{\ratioBetweenDistanceAndInner[\indexU]}$\\ $\scaleEexp[\indexIteration|\indexIteration\neq\indexChoosen]\leftarrow0$ for $\indexIteration=1,2,\mydots,\numberOfIterations$\\
		\eIf{$\indexU=\indexV$}{
			$\arbitraryPhaseK\leftarrow\arbitraryPhaseKBit+\frac{\numberOfPointsForPSK}{2\pi}\angleBetweenPointAndXYDiagonal[\indexU]$
		}{
			
     		$\arbitraryPhaseK\leftarrow\arbitraryPhaseKBit+\frac{\numberOfPointsForPSK}{2\pi}\angleBetweenPointAndXYDiagonal[\indexU]$\\
			$\angleexpAll[\indexChoosen]\leftarrow\angleexpAllBit[\indexChoosen]+\frac{\numberOfPointsForPSK}{2\pi}\angleBetweenPointAndXYDiagonal[\indexV ]-\frac{\numberOfPointsForPSK}{2\pi}\angleBetweenPointAndXYDiagonal[\indexU]$\\
			$\angleexpAll[\indexChoosen+1]\leftarrow\angleexpAllBit[\indexChoosen+1]-\frac{\numberOfPointsForPSK}{2\pi}\angleBetweenPointAndXYDiagonal[\indexV ]-\frac{\numberOfPointsForPSK}{2\pi}\angleBetweenPointAndXYDiagonal[\indexU]$ (if $\indexChoosen+1\le\numberOfIterations$)
		}
		\eIf{$\lengthGaGb>1$}{\isReversedPiValid $\leftarrow1$}{\eIf{$\indexChoosen=\numberOfIterations$ and ($\ratioBetweenDistanceAndInner[\indexU]\neq \ratioBetweenDistanceAndInner[\indexV]$ or $|\angleBetweenPointAndXYDiagonal[\indexU]|\neq |\angleBetweenPointAndXYDiagonal[\indexV]|$) }{\isReversedPiValid $\leftarrow1$}{\isReversedPiValid $\leftarrow0$}}
	%	\KwRet $\arbitraryScaleE,\scaleEexp[\indexIteration],\arbitraryPhaseK,\angleexpAll[\indexIteration]$  for $\indexIteration=1,2,\mydots,\numberOfIterations$, isReversedPiValid\;
	}\;}
\vspace{-0mm}
\end{algorithm}
\renewcommand{\baselinestretch}{\baselineSize}

\subsection{Examples}

\subsubsection{$4\sqrtofNumPointsInQuad^2$-QAM}
 The number of $(\indexU ,\indexV)$ permutations for $\ratioBetweenDistanceAndInner[\indexU ]=\ratioBetweenDistanceAndInner[\indexV ]$ and $|\angleBetweenPointAndXYDiagonal[\indexU]|=|\angleBetweenPointAndXYDiagonal[\indexV]|$ can be calculated  $\sqrtofNumPointsInQuad+4\sqrtofNumPointsInQuad(\sqrtofNumPointsInQuad-1)/2=2\sqrtofNumPointsInQuad^2-\sqrtofNumPointsInQuad$. Hence, $\numberOfPointsNotEquidistance+\numberOfPointsEquidistanceButDifferentAbsAngle$ must be equal to $\sqrtofNumPointsInQuad^4-(2\sqrtofNumPointsInQuad^2-\sqrtofNumPointsInQuad)=\sqrtofNumPointsInQuad^4-2\sqrtofNumPointsInQuad^2+\sqrtofNumPointsInQuad$.
Thus, the total number $4\sqrtofNumPointsInQuad^2$-\ac{QAM} \acp{CS} can be calculated as $((\sqrtofNumPointsInQuad^4-\sqrtofNumPointsInQuad^2)(\numberOfIterations+1)+\sqrtofNumPointsInQuad)\cdot\unitSequences$ and $2((\sqrtofNumPointsInQuad^4-\sqrtofNumPointsInQuad^2)\numberOfIterations+\sqrtofNumPointsInQuad^2)\cdot\unitSequences$  for $\lengthGaGb=1$ and $\lengthGaGb>1$, respectively. For $4\sqrtofNumPointsInQuad^2$-\ac{QAM}, $\minimumRadius=\sqrt{3/(4\sqrtofNumPointsInQuad^2-1)}$ and $\minimumDiff=\sqrt{6/(4\sqrtofNumPointsInQuad^2-1)}$. Therefore, $\minDist\le2^{\numberOfIterations/2}\sqrt{3/(4\sqrtofNumPointsInQuad^2-1)}$ for $\lengthGaGb=1$.

Our result extends the enumerations in \cite{Li_2010} via offset method and Case I-II results in \cite{budisin_2018} via para-unitary matrices. In \cite{Li_2010}, for $\lengthGaGb=1$, Li showed that there exist at least $((\numberOfIterations+1)4^{2(q-1)}-(m+1)4^{(q-1)}+2^{q-1})\cdot\unitSequences$ \acp{CS} for $4^q$-QAM alphabet. For $\sqrtofNumPointsInQuad =  2^{q-1}$, we obtain the same result when it is plugged into $(\sqrtofNumPointsInQuad^4-\sqrtofNumPointsInQuad^2)(\numberOfIterations+1)+\sqrtofNumPointsInQuad$ without using an auxiliary method such as Gaussian integers. As a new result, for $\lengthGaGb>1$, we show that the total number of sequences is 
%$2((\sqrtofNumPointsInQuad^4-\sqrtofNumPointsInQuad^2)\numberOfIterations+\sqrtofNumPointsInQuad^2)\cdot \unitSequences=$
$2(\numberOfIterations4^{2(q-1)}-m4^{(q-1)}+2^{2q-2})\cdot\unitSequences$ for $4^q$-QAM alphabet. Note that our framework also enumerates the \acp{CS} for non-square constellations, e.g., $4\sqrtofNumPointsInQuad^2=36$ points in $\constellationPoints$ for $\sqrtofNumPointsInQuad=3$.  

In this study, $\funcfForFinalAmplitude(\seqx)$ is utilized to generate \acp{CS} with two distinct amplitude levels.
%A natural question  would be then if there exists any \ac{CS} with more than two amplitude levels on the \ac{QAM} constellation. 
By exploiting the grid nature of \ac{QAM} alphabet, it is possible to obtain the phase and amplitude parameters with a computer search for \acp{CS} with more than two amplitude levels, as done in \cite{budisin_2018} with para-unitary matrix construction. However, we do not have a systematic rule to enumerate \ac{QAM}-\acp{CS} with more than two amplitude levels. Since these sequences also require a large $q$ \cite{wang2020newQAM,budisin_2018}, it can decrease $\minDist$ drastically without a particular \ac{CS} selection.

\subsubsection{$\numberOfPointsForPSK\numberOfPointsOnConstellation$-Star}  For this case, $(\numberOfPointsNotEquidistance,\numberOfPointsEquidistanceButDifferentAbsAngle)=(\numberOfPointsOnConstellation(\numberOfPointsOnConstellation-1),0)$. Thus, there are at least  $((\numberOfPointsOnConstellation^2-\numberOfPointsOnConstellation)(\numberOfIterations+1)+\numberOfPointsOnConstellation)\cdot\unitSequences$ and $2((\numberOfPointsOnConstellation^2-\numberOfPointsOnConstellation)\numberOfIterations+\numberOfPointsOnConstellation)\cdot\unitSequences$  \acp{CS} with an $\numberOfPointsForPSK\numberOfPointsOnConstellation$-Star alphabet for $\lengthGaGb=1$ and $\lengthGaGb>1$, respectively.

\subsubsection{DVB-16/32APSK}  For DVB-16APSK and DVB-32APSK, $(\numberOfPointsNotEquidistance,\numberOfPointsEquidistanceButDifferentAbsAngle)=(6,4)$ and $(\numberOfPointsNotEquidistance,\numberOfPointsEquidistanceButDifferentAbsAngle)=(38,14)$, respectively. Thus, the  numbers of  \acp{CS} with DVB-16APSK constellation are at least  $(240\numberOfIterations+26)\cdot\unitSequences$ and $(480\numberOfIterations+32)\cdot\unitSequences$ for $\lengthGaGb=1$ and $\lengthGaGb>1$. For DVB-32APSK constellation, there are at least  $(992\numberOfIterations+84)\cdot\unitSequences$ and $(1984\numberOfIterations+64)\cdot\unitSequences$ \acp{CS} for $\lengthGaGb=1$ and $\lengthGaGb>1$, respectively. 

\subsubsection{Modified IEEE 802.11ay 64-NUC} For modified IEEE 802.11ay 64-NUC constellation, $(\numberOfPointsNotEquidistance,\numberOfPointsEquidistanceButDifferentAbsAngle)=(228,0)$ and the  numbers of \acp{CS} are at least $(4032\numberOfIterations+292)\cdot\unitSequences$ and $(8064\numberOfIterations+128)\cdot\unitSequences$ for $\lengthGaGb=1$ and $\lengthGaGb>1$, respectively. 

%To the best of our knowledge, enumerations of \acp{CS} with $\numberOfPointsForPSK\numberOfPointsOnConstellation$-Star, a modified IEEE 802.11ay 64-NUC, DVB-16/32APSK, or, a family of constellations different from uniform constellations have not been reported in the literature.

\subsection{Constant-Modulus CSs with an ESC}
Using a constellation different from \ac{PSK} for \acp{CS} can increase the mean power of the \ac{OFDM} symbol. %For example, if the elements of a \ac{CS} utilize only the outer part of the constellation, the mean power of the \ac{OFDM} symbol is scaled although the fluctuations within the \ac{OFDM} symbol is low.
This issue demotes the main motivation of using \acp{CS} with a larger constellation for a practical system since the \ac{PMEPR} can be higher than $3$~dB \cite{Li_2010}. On the other hand, our framework provides a way to control the mean power of \ac{OFDM} symbol through $\{(\indexU,\indexV),\indexChoosen\}$ for a constellation different from \ac{PSK}. The choice of  $\arbitraryScaleE=\frac{\numberOfPointsForPSK}{2\pi}\ln\ratioBetweenDistanceAndInner[\indexU]$, $\scaleEexp[\indexChoosen]=\frac{\numberOfPointsForPSK}{2\pi}\ln\frac{\ratioBetweenDistanceAndInner[\indexV]}{\ratioBetweenDistanceAndInner[\indexU]}$, and $\scaleEexp[\indexIteration|\indexIteration\neq\indexChoosen]=  0$  in Theorem~\ref{th:reduced} scales halves of the elements to $\ratioBetweenDistanceAndInner[\indexU]$ and $\ratioBetweenDistanceAndInner[\indexV]$. Hence, the  \ac{OFDM} symbol power can be fixed   through constant-modulus \acp{CS} if $(\ratioBetweenDistanceAndInner[\indexU]^2+\ratioBetweenDistanceAndInner[\indexV]^2)/2=1$ holds. To retain the \ac{PMEPR} benefit of \acp{CS}, we consider only $\{(\indexU,\indexV),\indexChoosen\}$ combinations that satisfies $(\ratioBetweenDistanceAndInner[\indexU]^2+\ratioBetweenDistanceAndInner[\indexV]^2)/2=1$. 
Under this constraint, the number of distinct \acp{CS} reduces, but it can still  be  high if the constellation is designed well. For example, if $\numberOfPointsOnConstellation=4$ for $\numberOfPointsForPSK\numberOfPointsOnConstellation$-Star constellation, $(\ratioBetweenDistanceAndInner[1]^2,\ratioBetweenDistanceAndInner[2]^2,\ratioBetweenDistanceAndInner[3]^2,\ratioBetweenDistanceAndInner[4]^2)$ can be chosen as $(2-b^2,2-a^2,a^2,b^2)$ for any $1<a<b<\sqrt{2}$. For \ac{QAM}, there exist $(\indexU,\indexV)$ permutations that satisfy  $(\ratioBetweenDistanceAndInner[\indexU]^2+\ratioBetweenDistanceAndInner[\indexV]^2)/2=1$  and the valid permutations can be easily identified through a computer search. It is also worth noting that 
%although our construction results in constant-modulus \acp{CS} with a constellation different from \ac{PSK}, 
ensuring a larger $\minDist$ with a constellation optimization or a constant-modulus \ac{CS} selection algorithm is not considered in this work.

\section{Encoder and Decoder}\label{sec:encDec}
%In this section, we develop an encoder utilizing \acp{CS} with an \ac{ESC} and a low-complexity \ac{ML} decoder. We aim at a low-\ac{PMEPR} joint coding-and-modulation scheme for non-contiguous resource allocations shared by multiple users.

\subsection{Encoder}
\def\Nuvl{N_{\indexU,\indexV,\indexChoosen}}
\def\bitsuvl{K_{\indexU,\indexV,\indexChoosen}}
\def\Indexuvlpi{i_{\indexU,\indexV,\indexChoosen,\seqPermutationCompShift}}
\def\IndexuvlpiD{\hat{i}_{\indexU,\indexV,\indexChoosen,\seqPermutationCompShift}}
\def\Indexuvl{i_{\indexU,\indexV,\indexChoosen}}
\def\IndexuvlD{\hat{i}_{\indexU,\indexV,\indexChoosen}}
\def\Indexpi{i_{\seqPermutationCompShift}}
\def\IndexpiD{\hat{i}_{\seqPermutationCompShift}}
The proposed encoder encodes the information bits with $\{(\indexU,\indexV),\indexChoosen\}$ combinations, $\arbitraryPhaseKBit,\angleexpAllBit[\indexIteration]\in\integers_{\numberOfPointsForPSK}$ for $\indexIteration\in\{1,2,\mydots,\numberOfIterations\}$, and the permutations of $\seqPermutationCompShift$, where 
$\arbitraryPhaseK=\arbitraryPhaseKBit+\frac{\numberOfPointsForPSK}{2\pi}\phaseOffset$, $\angleexpAll[\indexIteration]=\angleexpAllBit[\indexIteration]+\frac{\numberOfPointsForPSK}{2\pi}\phaseOffsetl[\indexIteration]$   for $\indexIteration\in\{1,2,\mydots,\numberOfIterations\}$ and $\phaseOffset$, $\phaseOffsetl[\indexIteration]$, $\arbitraryScaleE$, and  $\scaleEexp[\indexIteration]$ are functions of $\{(\indexU,\indexV),\indexChoosen\}$ based on Algorithm~\ref{alg:ESC}, and $\numberOfPointsForPSK$ is assumed to be power-of-two. We utilize orthogonal seed \acp{GCP} to support orthogonal multiple access in the uplink by using the properties of modulated unimodular sequences \cite{Benedetto_2009} and use a fixed  $\permutationMono[1]$ to keep the locations of the seed sequences at a cost of maximum $\ceil{\log_2{\numberOfIterations}}$ bits. Therefore, the same subcarriers can be shared by $\lengthGaGb$ users for a seed \ac{GCP} of length $\lengthGaGb$.

Based on Corollary~\ref{co:enum}, we transmit $\floor{\log_2{(\Nuvl(\numberOfIterations-1)!})}+{(\numberOfIterations+1)\log_2{\numberOfPointsForPSK}}$ information bits over $\lengthGaGb2^\numberOfIterations$ subcarriers, where $\Nuvl$ is the number of $\{(\indexU,\indexV),\indexChoosen\}$ combinations and equal to $ (\numberOfPointsOnConstellation^2-\numberOfPointsOnConstellation)\numberOfIterations+\numberOfPointsOnConstellation$. To utilizes all available constellation points, $\bitsuvl=\floor{\log_2{(\Nuvl(\numberOfIterations-1)!})}$ information bits is first converted to an integer $\Indexuvlpi\in\{0,1,\mydots,2^{\bitsuvl}-1\}$. It is then decomposed as $\Indexuvlpi=\Indexpi\Nuvl+\Indexuvl$, where $\Indexuvl\in\{0,1,\mydots,\Nuvl-1\}$. The indices $\Indexuvl$ and  $\Indexpi$ are utilized to identify a combination of $\{(\indexU,\indexV),\indexChoosen\}$ and a permutation of $\seqPermutationCompShift$, respectively. We use factoradic via Lehmer code to construct a bijective mapping from integers to permutations.

%The \ac{SE} considering multiple users can be calculated as $\spectralEfficient=\floor{2\log_2{\numberOfPointsOnConstellation}+(\numberOfIterations+1)\log_2{\numberOfPointsForPSK}+\log_2{(\numberOfIterations-1)!}}/2^\numberOfIterations$ since the length of \acp{CS} is $\lengthGaGb$. 
%For $\lengthGaGb=1$, assuming that only half of permutations can be utilized, the \ac{SE} can be calculated as $\spectralEfficient=\floor{2\log_2{\numberOfPointsOnConstellation}+(\numberOfIterations+1)\log_2{\numberOfPointsForPSK}+\log_2{(\numberOfIterations!/2)}}/2^\numberOfIterations$.

%On the other hand, if $\numberOfPointsForPSK$ and $\numberOfPointsOnConstellation$ are power-of-two, the bit mappings for $(\indexU ,\indexV )$, $\arbitraryPhaseK,\angleexpAll[\indexIteration]$, and $\seqPermutationCompShift$ can be separated. In this study, we utilize factoradics to construct a mapping from bits to the permutations.
\subsection{Decoder}

\def\order{r}
\def\referenceDerivative{\ell}
\def\referenceDerivativeNext{\ell'}
\def\receivedElement[#1]{r_{#1}}
\def\channel[#1]{c_{#1}}
\def\noise[#1]{n_{#1}}
\def\parametersReal{\theta_{\rm r}}
\def\parametersImag{\theta_{\rm i}}
\def\parametersRealg{{{\psi}}_{\rm r}}
\def\parametersImagg{{{\psi}}_{\rm i}}
\def\parametersRealf{{{\phi}}_{\rm r}}
\def\parametersImagf{{{\phi}}_{\rm i}}

\def\parametersRealEstimate{{\tilde{\theta}}_{\rm r}}
\def\parametersImagEstimate{{\tilde{\theta}}_{\rm i}}
\def\parametersRealEstimateRed{{\tilde{\psi}}_{\rm r}}
\def\parametersImagEstimateRed{{\tilde{\psi}}_{\rm i}}

\def\weightedElement[#1]{s_{#1}}
\def\channelPowerElement[#1]{h_{#1}}
\def\weightedElementAfterSum[#1][#2]{s_{#1}^{#2}}
\def\channelPowerElementAfterSum[#1][#2]{h_{#1}^{#2}}

\def\funcfForFinalAmplitudeDerivate[#1][#2]{f_{\rm r}^{#1}(#2)}
\def\funcfForFinalPhaseDerivate[#1][#2]{f_{\rm i}^{#1}(#2)}
\def\funcgForFinalAmplitudeDerivate[#1][#2]{g_{\rm r}^{#1}(#2)}
\def\funcgForFinalPhaseDerivate[#1][#2]{g_{\rm i}^{#1}(#2)}
\def\mappingFunction[#1]{{I_{\referenceDerivative}(#1)}}
\def\algorithmDecoder{A}

\def\funcfForFinalAmplitudeDerivateDec[#1][#2]{\check{f}_{\rm r}^{#1}(#2)}
\def\funcfForFinalPhaseDerivateDec[#1][#2]{\check{f}_{\rm i}^{#1}(#2)}
\def\funcgForFinalAmplitudeDerivateDec[#1][#2]{\check{g}_{\rm r}^{#1}(#2)}
\def\funcgForFinalPhaseDerivateDec[#1][#2]{\check{g}_{\rm i}^{#1}(#2)}

\def\weightedSequence[#1]{\textit{{\textbf{s}}}_{#1}}
\def\weightedChannel[#1]{\textit{{\textbf{h}}}_{#1}}
\def\numberOfSequences{L}
\def\setOfweightedSequence{\textit{\textbf{S}}_{\numberOfIterations}}
\def\setOfweightedChannel{\textit{\textbf{H}}_{\numberOfIterations}}
\def\setOfpermutations{\textit{\textbf{L}}_{\numberOfIterations}}
\def\setOfOffsets{\textit{\textbf{O}}_{\numberOfIterations}}

\def\setOfweightedSequenceSub{\textit{\textbf{S}}_{\numberOfIterations-1}}
\def\setOfweightedChannelSub{\textit{\textbf{H}}_{\numberOfIterations-1}}
\def\setOfpermutationsSub{\textit{\textbf{L}}_{\numberOfIterations-1}}
\def\setOfOffsetSub{\textit{\textbf{O}}_{\numberOfIterations-1}}

\def\setPhaseOffset{{\textit{\textbf{D}}}_{\numberOfIterations}}
\def\setPhaseOffsetSub{{\textit{\textbf{D}}}_{\numberOfIterations-1}}
\def\setScaleOffset{{\textit{\textbf{E}}}_{\numberOfIterations}}
\def\setScaleOffsetSub{{\textit{\textbf{E}}}_{\numberOfIterations-1}}

\def\setParameters{{{\bf \Phi}}_{\numberOfIterations}}
\def\setPhase{{{\bf \Theta}}_{\numberOfIterations}}
\def\setPi{{\bf \Pi}_{\numberOfIterations}}
\def\setScores{{\bf \Sigma}_{\numberOfIterations}}

\def\setParametersSub{{{\bf \Phi}}_{\numberOfIterations-1}}
\def\setPhaseSub{{{\bf \Theta}}_{\numberOfIterations-1}}
\def\setPiSub{{\bf \Pi}_{\numberOfIterations-1}}
\def\setScoresSub{{\bf \Sigma}_{\numberOfIterations-1}}

\def\indexOptimum{\tilde{n}}
\def\indexSequence{i}
\def\indexEnumaration{n}
\def\indexOffset{j}
\def\score{\Sigma}

\def\numberOfGoodSeqences{{N_{\rm best}}}
\def\setOfgoodSequenceIndex{\textit{\textbf{N}}_{\numberOfIterations}}
\def\EbNO{E_{\rm b}/N_{0}}
\def\SNR{SNR}
\def\indexRecursion{i}

At the receiver side, we first separate the users by using the orthogonality of the seed sequences and exploiting the fixed  $\funcfForCommonOrder(\seqx,\polyVariable)$. We then concatenate the matched filter outputs, i.e., the sequence length reduces to $2^\numberOfIterations$ for each user. The $\varMonomial$th element of the sequence of a user can be expressed as
$
\receivedElement[\varMonomial] = \channel[\varMonomial]\exponentialBase^{\funcfForFinalAmplitudeDec(\varMonomial)   + \constantj \funcfForFinalPhaseDec(\varMonomial)} +\noise[\varMonomial]
$,
where $ \channel[\varMonomial]$ and $\noise[\varMonomial]$ are the complex channel and the noise coefficients, respectively.  Assuming that $ \channel[\varMonomial]$ is available at the receiver, an \ac{ML} decoder is equivalent to a minimum distance decoder for \ac{AWGN}, i.e.,
\if \IEEEsubmission 0
\begin{align}
	\{\parametersRealEstimate, \parametersImagEstimate\}& = \arg\min_{\parametersReal,\parametersImag} \sum_{\varMonomial=0}^{2^\numberOfIterations-1}   |\channel[\varMonomial]\exponentialBase^{\funcfForFinalAmplitudeDec(\varMonomial;\parametersReal)   + \constantj \funcfForFinalPhaseDec(\varMonomial; \parametersImag)} -  \receivedElement[\varMonomial] |^2
	\nonumber \\ &
	\hspace{-5mm}=\arg\max_{\parametersReal,\parametersImag} \Re\left\{ \sum_{\varMonomial=0}^{2^\numberOfIterations-1}  \exponentialBase^{\funcfForFinalAmplitudeDec(\varMonomial;\parametersReal)-\constantj \funcfForFinalPhaseDec(\varMonomial; \parametersImag)} \weightedElement[\varMonomial]  \shortMinus  \exponentialBase^{2\funcfForFinalAmplitudeDec(\varMonomial;\parametersReal)}\channelPowerElement[\varMonomial]\right\}~,
	\label{eq:mlNew}
\end{align}
\else
\begin{align}
	\{\parametersRealEstimate, \parametersImagEstimate\}& = \arg\min_{\parametersReal,\parametersImag} \sum_{\varMonomial=1}^{2^\numberOfIterations-1}   |\channel[\varMonomial]\exponentialBase^{\funcfForFinalAmplitude(\seqx;\parametersReal)   + \constantj \funcfForFinalPhase(\seqx; \parametersImag)} -  \receivedElement[\varMonomial] |^2
	=\arg\max_{\parametersReal,\parametersImag} \Re\left\{ \sum_{\varMonomial=1}^{2^\numberOfIterations-1}  \exponentialBase^{\funcfForFinalAmplitude(\seqx;\parametersReal)-\constantj \funcfForFinalPhase(\seqx; \parametersImag)} \weightedElement[\varMonomial]  \shortMinus  \exponentialBase^{2\funcfForFinalAmplitude(\seqx;\parametersReal)}\channelPowerElement[\varMonomial]\right\}~,
	\label{eq:mlNew}
\end{align}
\fi
where $\parametersReal \cup\parametersImag = \{(\indexU, \indexV),\indexChoosen,\seqPermutationCompShift,\arbitraryPhaseK,\angleexpAll[1],\mydots,\angleexpAll[\numberOfIterations] \}$, $\weightedElement[\varMonomial] = \channel[\varMonomial]^* \receivedElement[\varMonomial]$, and
$\channelPowerElement[\varMonomial]={|\channel[\varMonomial]|^2}/{2}$.
However, solving \eqref{eq:mlNew} is not easy task and the complexity with a brute-force search can be very high.

\subsubsection{Principle}
To achieve a low-complexity \ac{ML} detector, we decompose  $\funcfForFinalAmplitude(\seqx;\parametersReal)$ and $\funcfForFinalPhase(\seqx;\parametersImag)$ as $\funcfForFinalAmplitude(\seqx;\parametersReal)=\funcgForFinalAmplitudeDerivate[\referenceDerivative][\seqx^\referenceDerivative;\parametersRealg]+\monomial[\referenceDerivative]\funcfForFinalAmplitudeDerivate[\referenceDerivative][\seqx^\referenceDerivative;\parametersRealf]$ and $\funcfForFinalPhase(\seqx;\parametersImag)=\funcgForFinalPhaseDerivate[\referenceDerivative][\seqx^\referenceDerivative;\parametersImagg]+\monomial[\referenceDerivative]\funcfForFinalPhaseDerivate[\referenceDerivative][\seqx^\referenceDerivative;\parametersImagf]$ for $\referenceDerivative=\{1,2,\mydots,\numberOfIterations\}$, respectively, where $\seqx^\referenceDerivative\triangleq(\monomial[1],\monomial[2], \mydots, \monomial[\referenceDerivative-1],\monomial[\referenceDerivative+1], \mydots, \monomial[\numberOfIterations])$. While the order of $\funcgForANF^\referenceDerivative(\seqx^\referenceDerivative)$ is  $\order$,  $\funcfForANF^\referenceDerivative(\seqx^\referenceDerivative)$ is a polynomial with the order of $\order-1$ and can be calculated as $\funcfForANF^\referenceDerivative(\seqx^\referenceDerivative)={\partial \funcfForANF(\seqx)}/{\partial \monomial[{\referenceDerivative}]}$. Both  $\funcfForANF^\referenceDerivative(\seqx^\referenceDerivative)$ and $\funcgForANF^\referenceDerivative(\seqx^\referenceDerivative)$ lead to the sequences of length $2^{\numberOfIterations-1}$. By using these decompositions,
%of  $\funcfForFinalAmplitude(\seqx;\parametersReal)$ and $\funcfForFinalPhase(\seqx;\parametersImag)$,
the maximization in \eqref{eq:mlNew} can be re-expressed as
\begin{align}
\max_{\parametersRealg,\parametersRealf,\parametersImagg,\parametersImagf}  
 \Re\left\{ \sum_{\varMonomial=0}^{2^{\numberOfIterations-1}\shortMinus 1}  \exponentialBase^{\funcgForFinalAmplitudeDerivateDec[\referenceDerivative][\varMonomial;\parametersRealg]-\constantj \funcgForFinalPhaseDerivateDec[\referenceDerivative][\varMonomial;\parametersImagg] } \weightedElementAfterSum[\varMonomial][\referenceDerivative]  \shortMinus  \exponentialBase^{2\funcgForFinalAmplitudeDerivateDec[\referenceDerivative][\varMonomial;\parametersRealg]}\channelPowerElementAfterSum[\varMonomial][\referenceDerivative]\right\}~,
\label{eq:mlNeww}
\end{align}
where
\begin{align}
\weightedElementAfterSum[\varMonomial][\referenceDerivative] &= \weightedElement[{\mappingFunction[\varMonomial]}] + \weightedElement[{\mappingFunction[\varMonomial]}+2^{\numberOfIterations-\referenceDerivative}]\exponentialBase^{\funcfForFinalAmplitudeDerivateDec[\referenceDerivative][\varMonomial;\parametersRealf]-\constantj \funcfForFinalPhaseDerivateDec[\referenceDerivative][\varMonomial;\parametersImagf] }~,
\label{eq:newSeq}
\\
\channelPowerElementAfterSum[\varMonomial][\referenceDerivative] &= \channelPowerElement[{\mappingFunction[\varMonomial]}] + \channelPowerElement[{\mappingFunction[\varMonomial]}+2^{\numberOfIterations-\referenceDerivative}]\exponentialBase^{2\funcfForFinalAmplitudeDerivateDec[\referenceDerivative][\varMonomial;\parametersRealf]}~,
\label{eq:newChn}
\end{align}
and $\mappingFunction[\varMonomial]$ maps the integers from $0$ to $2^{\numberOfIterations-1}-1$ to the values of $ \sum_{\indexFirstOrderMonomial=1}^{\numberOfIterations}\monomial[\indexFirstOrderMonomial]2^{\numberOfIterations-\indexFirstOrderMonomial}$    in ascending order for $\monomial[\referenceDerivative]=0$. The variables $\weightedElementAfterSum[\varMonomial][\referenceDerivative]$ and $\channelPowerElementAfterSum[\varMonomial][\referenceDerivative] $ can be considered as the element of a new received sequence and the half of absolute square of channel gain as in the original expression in \eqref{eq:mlNew}, where the original sequence length is halved.

Theorem~\ref{th:reduced}  provides  $\funcfForFinalAmplitude(\seqx)$ and $\funcfForFinalPhase(\seqx)$. By using \eqref{eq:identity2}, $\funcfForFinalAmplitude(\seqx)$ in \eqref{eq:realPartReduced} can be re-written as
\begin{align}
	\funcfForFinalAmplitude(\seqx)&=\arbitraryScaleE\shortPlus\scaleEexp[1]\monomial[{\permutationMono[1]}]\shortPlus\sum_{\indexIteration=2}^{\numberOfIterations}(\scaleEexp[\indexIteration]\shortPlus\scaleEexp[\indexIteration-1])\monomial[{\permutationMono[\indexIteration]}]\shortMinus 2 \sum_{\indexIteration=1}^{\numberOfIterations-1}\scaleEexp[\indexIteration]  \monomial[{\permutationMono[\indexIteration]}] \monomial[{\permutationMono[\indexIteration+1]}]. \nonumber
\end{align} Therefore, $\funcfForFinalAmplitudeDerivate[\referenceDerivative][\seqx^\referenceDerivative]$ can be calculated as
\begin{align}
	\funcfForFinalAmplitudeDerivate[\referenceDerivative][\seqx^\referenceDerivative] = 
	\begin{cases}
		\scaleEexp[\numberOfIterations-1] \shortMinus 2\scaleEexp[\numberOfIterations-1]\monomial[{\permutationMono[\numberOfIterations-1]}]\shortPlus  \scaleEexp[\numberOfIterations] , & {\permutationMono[\numberOfIterations]}=\referenceDerivative\\
		\scaleEexp[\indexIteration-1] \shortMinus 2\scaleEexp[\indexIteration-1]\monomial[{\permutationMono[\indexIteration-1]}]\shortPlus \scaleEexp[\indexIteration]\shortMinus 2\scaleEexp[\indexIteration]\monomial[{\permutationMono[\indexIteration+1]}], & {\permutationMono[\indexIteration]}=\referenceDerivative\\
		\scaleEexp[1] \shortMinus 2\scaleEexp[1]\monomial[{\permutationMono[2]}], & {\permutationMono[1]}=\referenceDerivative
	\end{cases}~. \nonumber
\end{align} 
Similarly,   we can express $\funcfForFinalPhase(\seqx)=\funcgForFinalPhaseDerivate[\referenceDerivative][\seqx^\referenceDerivative]+\monomial[\referenceDerivative]\funcfForFinalPhaseDerivate[\referenceDerivative][\seqx^\referenceDerivative]$, where
\begin{align}
	\funcfForFinalPhaseDerivate[\referenceDerivative][\seqx^\referenceDerivative] = 
	\begin{cases}
		\angleexpAll[\numberOfIterations]+ \frac{\numberOfPointsForPSK}{2}\monomial[{\permutationMono[\numberOfIterations-1]}], & {\permutationMono[\numberOfIterations]}=\referenceDerivative\\
		\angleexpAll[\indexIteration]+  \frac{\numberOfPointsForPSK}{2}\monomial[{\permutationMono[\indexIteration-1]}]+ \frac{\numberOfPointsForPSK}{2}\monomial[{\permutationMono[\indexIteration+1]}], & {\permutationMono[\indexIteration]}=\referenceDerivative\\
		\angleexpAll[1]+\frac{\numberOfPointsForPSK}{2}\monomial[{\permutationMono[2]}], & {\permutationMono[1]}=\referenceDerivative
	\end{cases}~. \nonumber
\end{align} 
%Since $\funcfForFinalAmplitudeDerivate[\referenceDerivative][\seqx^\referenceDerivative]$ and $\funcfForFinalPhaseDerivate[\referenceDerivative][\seqx^\referenceDerivative] $ are functions of $\seqPermutationCompShift$, the enumerations are not tractable. On the other hand, 
As  $\permutationMono[1]$ is fixed and available at the receiver, $\funcfForFinalAmplitudeDerivate[\referenceDerivative][\seqx^\referenceDerivative] =\scaleEexp[1] - 2\scaleEexp[1]\monomial[{\permutationMono[2]}]$ and $\funcfForFinalPhaseDerivate[\referenceDerivative][\seqx^\referenceDerivative] = \angleexpAll[1]+\frac{\numberOfPointsForPSK}{2}\monomial[{\permutationMono[2]}]$ for 
$\referenceDerivative=\permutationMono[1]$. Hence, 
$\parametersImagf=\{\angleexpAll[1],\permutationMono[2]\}$ and $\parametersRealf=\{\scaleEexp[1],\permutationMono[2]\}$, where $\angleexpAll[1]=\angleexpAllBit[1]+\phaseOffsetl[1]$ and $\scaleEexp[1]$ depend on $\Indexuvl$. Hence, for a hypothesized $\Indexuvl$, we can calculate \eqref{eq:newSeq} and \eqref{eq:newChn} for different $\angleexpAllBit[1]$ and $\permutationMono[2]$. This procedure can done continuously, where the sequence length is  halved for each step. By backtracking the branch that maximizes \eqref{eq:mlNeww}, a  recursive \ac{ML} detector  can be achieved. We adopt this principle and address the exponential growth by terminating unpromising branches early.

\if\IEEEsubmission0
\else
\renewcommand{\baselinestretch}{1}
\fi
\begin{algorithm}[t]
	\scriptsize
	\caption{\small ML Decoder for Fading Channel \& Pruning }\label{alg:decoder}
	\SetKwInput{KwInput}{Input}                % Set the Input
	\SetKwInput{KwOutput}{Output}              % set the Output
	\DontPrintSemicolon
	
	\KwInput{$\permutationMono[1]$,  $\numberOfIterations$, $\numberOfPointsForPSK$,  $(\channel[\varMonomial])_{\varMonomial=0}^{2^{\numberOfIterations}-1}$, $(\receivedElement[\varMonomial])_{\varMonomial=0}^{2^{\numberOfIterations}-1}$}
	\KwOutput{$\IndexuvlpiD$,   $(\angleexpAllBitD[1],\mydots,\angleexpAllBitD[\numberOfIterations], \arbitraryPhaseKBitD)$}
	
	% Set Function Names
	\SetKwFunction{FMain}{main}
	\SetKwFunction{Fdecoder}{dec}
	\SetKwFunction{Fesc}{offsets}
	
	\SetKwProg{Fn}{Function}{}{}
	\Fn{\FMain}{
		Prepare $\setOfweightedSequence$, $\setOfweightedChannel$, $\setOfpermutations$, $\setOfOffsets$ for $\Nuvl$ sequences\;
		Run ($\setScores$, $\setPi$, $\setPhase$)=\Fdecoder{$\setOfweightedSequence$, $\setOfweightedChannel$, $\setOfpermutations$, $\setOfOffsets$}\;
		Calculate  $\IndexuvlpiD$ and  $(\angleexpAllBitD[1],\mydots,\angleexpAllBitD[\numberOfIterations], \arbitraryPhaseKBitD)$ from $\setPi$ and $\setPhase$ for the index that maximizes $\setScores$\;
		\KwRet $\IndexuvlpiD$,  $(\angleexpAllBitD[1],\mydots,\angleexpAllBitD[\numberOfIterations], \arbitraryPhaseKBitD)$\;
	}\;
	
	\SetKwProg{Fn}{Function}{}{}
	\Fn{($\setScores$, $\setPi$, $\setPhase$) =  \Fdecoder{$\setOfweightedSequence$, $\setOfweightedChannel$, $\setOfpermutations$, $\setOfOffsets$}}{
			
		Enumerate $\setOfweightedSequenceSub$, $\setOfweightedChannelSub$, $\setOfpermutationsSub$, $\setOfOffsetSub$ based on \eqref{eq:newSeq} and \eqref{eq:newChn}\;

		\eIf{$\numberOfIterations=1$}{
				Obtain $\arbitraryPhaseKBitD,\angleexpAllBitD[1]$ for each of $\numberOfSequences$ sequences  based in \eqref{eq:mlNeww}\;
				Populate $\setPi|_\indexSequence\leftarrow1$,
				$\setPhase|_\indexSequence\leftarrow(\angleexpAllBitD[1],\arbitraryPhaseKBitD)$, 
				$\setScores|_\indexSequence\leftarrow \eqref{eq:mlNeww}$
		%	}
		}{
		
		Populate the indices $\numberOfGoodSeqences$ best sequences in $\setOfgoodSequenceIndex$\; 
		Prune $\setOfweightedSequenceSub$, $\setOfweightedChannelSub$, $\setOfpermutationsSub$, $\setOfOffsetSub$ based on $\setOfgoodSequenceIndex$\;

			Run ($\setScoresSub$, $\setPiSub$, $\setPhaseSub$) = \Fdecoder{$\setOfweightedSequenceSub$, $\setOfweightedChannelSub$, $\setOfpermutationsSub$, $\setOfOffsetSub$}\;

			Calculate $\setScores$, $\setPi$, $\setPhase$ for each sequence indexed in   $\setOfgoodSequenceIndex$\;
			Set the scores in $\setScores$ for the prunned sequences  to $-\infty$ \;
		}
	\KwRet $\setScores$, $\setPi$, $\setPhase$\;
	}
\end{algorithm}
\renewcommand{\baselinestretch}{\baselineSize}

\subsubsection{Algorithm} The proposed algorithm has four different phases given as follows:

\paragraph{Preparation} We calculate $\Nuvl$ offset parameters, i.e., $\phaseOffset, \phaseOffsetll[1], \mydots, \phaseOffsetll[\numberOfIterations],\arbitraryScaleE,\scaleEexp[1],\mydots,\scaleEexp[\numberOfIterations]$ based on the \ac{ESC} for all $\{(\indexU,\indexV),\indexChoosen\}$ combinations. Since $\arbitraryScaleE$ and $\phaseOffset$ alter the amplitude and phase of all elements in the sequence, we combine them with the channel coefficients as $\channel[\varMonomial]\leftarrow\channel[\varMonomial]\exponentialBase^{\arbitraryScaleE+\constantj\phaseOffset}$. We then calculate  $\weightedElement[\varMonomial] \leftarrow \channel[\varMonomial]^* \receivedElement[\varMonomial]$ and
$\channelPowerElement[\varMonomial]\leftarrow{|\channel[\varMonomial]|^2}/{2}$ for each offset and
store $\Nuvl$ resulting sequences in  $\setOfweightedSequence$ and $\setOfweightedChannel$, respectively. %The offset parameters $(\phaseOffsetll[1], \mydots, \phaseOffsetll[\numberOfIterations])$ and $(\scaleEexp[1],\mydots,\scaleEexp[\numberOfIterations])$ are populated in $\setScaleOffset$ and $\setPhaseOffset$, respectively. 
The parameter $\referenceDerivative$ (i.e., the fixed $\permutationMono[1]$ for the preparation) and $\Indexuvl$ are stored in  $\setOfpermutations$ and $\setOfOffsets$, respectively, for  identifying the offsets in the following steps.% and backtracking. 

\paragraph{Enumeration} There are $\numberOfIterations-1$ and $\numberOfPointsForPSK$ options for $\permutationMono[2]$ and  $\angleexpAll[1]$ for each sequence in $\setOfweightedSequence$, respectively. 
By taking $\scaleEexp[1]$ and $\phaseOffsetll[1]$ into account, we calculate \eqref{eq:newSeq} and \eqref{eq:newChn} for all combinations of $\permutationMono[2]$ and  $\angleexpAllBit[1]$. We populate $(\numberOfIterations-1)\numberOfPointsForPSK\numberOfSequences$  resulting sequences in $\setOfweightedSequenceSub$ and $\setOfweightedChannelSub$, respectively, where $\numberOfSequences$ is the number of sequences given to the algorithm. We populate the corresponding $\Indexuvl$ and the information related to $\referenceDerivative$ for the next recursion (i.e., the enumerated values for $\permutationMono[2]$)  in $\setOfOffsetSub$ and $\setOfpermutationsSub$, respectively. %The first element of all sequences in $\setScaleOffset$ and $\setPhaseOffset$ are removed and the remaining parts are stored $\setScaleOffsetSub$ and $\setPhaseOffsetSub$, respectively. 

\paragraph{Pruning}
\label{subsubsec:prun}
%Algorithm~\ref{alg:decoder} systematically solves  \eqref{eq:mlNew}  based on the polynomials on Theorem~\ref{th:reduced}. However,  it can still be very complex since $\Nuvl\numberOfPointsForPSK^i\prod_{j=1}^{i}(\numberOfIterations-j)$ sequences of length $2^{\numberOfIterations-\indexRecursion}$ are generated  for the $(\indexRecursion+1)$th recursion step, i.e., 
To address the exponential growth of the enumerated sequences, we keep only $\numberOfGoodSeqences$ enumarated sequences for the next  step. By using the high \ac{SNR} approximations of $\exponentialBase^{\funcfForFinalAmplitude(\seqx;\parametersReal)+\constantj \funcfForFinalPhase(\seqx; \parametersImag)} \approxeq \weightedElement[\varMonomial]/{2\channelPowerElement[\varMonomial]}$ and $\exponentialBase^{2\funcfForFinalAmplitude(\seqx;\parametersReal)} \approxeq |\weightedElement[\varMonomial]|^2/4\channelPowerElement[\varMonomial]^2$ in \eqref{eq:mlNew}, we calculate an estimate of \eqref{eq:mlNew} and choose $\numberOfGoodSeqences$ sequences based on the resulting metric, i.e., $\sum_{\varMonomial=1}^{2^\numberOfIterations-1}  {|\weightedElement[\varMonomial]|^2}/{4\channelPowerElement[\varMonomial]}$. The indices of the best sequences are populated in $\setOfgoodSequenceIndex$ for backtracking.
\begin{figure}
	\centering
	\includegraphics[width =2.8in]{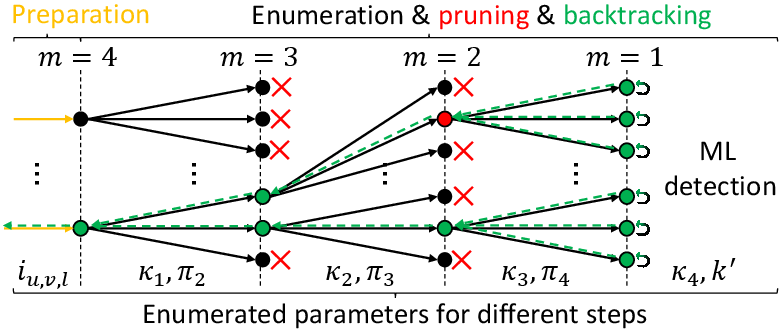}~
	\label{fig:dec}
	\if\IEEEsubmission\vspace{-2mm}\fi
	\caption{An example for preparation, enumeration, pruning, backtracking and the  parameters for $\numberOfIterations=4$. While $\arbitraryScaleE$ and $\phaseOffset$ are used for the preparation,  $\phaseOffsetll[\indexIteration]$ and $\scaleEexp[\indexIteration]$ are used only at the $\indexIteration$th step for calculating \eqref{eq:newSeq} and \eqref{eq:newChn}.}
	\if\IEEEsubmission\vspace{-5mm}\fi
\end{figure}

\paragraph{Backtracking} 
We recursively run the enumeration and pruning phases till $\numberOfIterations=1$. For $\numberOfIterations=1$, we perform an \ac{ML} detection based on \eqref{eq:mlNeww}. The detected parameters, i.e., $(\angleexpAllBitD[1],\arbitraryPhaseKBitD)$, $\seqPermutationCompShift=(1)$, and the value of \eqref{eq:mlNeww}  for all sequences  are stored in	$\setPi$,
$\setPhase$, and $\setScores$, where $\angleexpAllBitD[1]$ and $\arbitraryPhaseKBitD$ correspond to the detected $\angleexpAllBit[\numberOfIterations]$ and $\arbitraryPhaseKBit$, respectively. The algorithm then returns $\setPi$,
$\setPhase$, and $\setScores$ for the step above.
For $\numberOfIterations\ge2$, the detected parameters from the previous finalized step are utilized to calculate \eqref{eq:mlNeww} for each enumerated sequence. The best $\angleexpAllBitD[1]$ and $\permutationMonoD[1]$ for the sequences indicated in  $\setOfgoodSequenceIndex$ are obtained by using $\setOfOffsets$ and $\setOfpermutations$  and the results are combined with the parameters from the previous step in $\setPi$ and $\setPhase$. The scores with \eqref{eq:mlNeww} for different sequences are also populated in $\setScores$. 

After the recursions are finalized, the  sequence index that returns the maximum  \eqref{eq:mlNeww} among $\Nuvl$  sequences is assigned to $\IndexuvlD$. The parameters $\seqPermutationCompShiftD$ and $(\angleexpAllBitD[1],\mydots,\angleexpAllBitD[\numberOfIterations], \arbitraryPhaseKBitD)$ are detected from ${\setPi}$ and ${\setPhase}$. After  mapping  $\seqPermutationCompShiftD$ to an integer $\IndexpiD$, $\IndexuvlpiD=\IndexpiD\Nuvl+\IndexuvlD$ is calculated. The binary representation of $\IndexuvlpiD$ corresponds to the received bits over $\seqPermutationCompShift$ and $\{(\indexU,\indexV),\indexChoosen\}$. The pseudocode and an example for the phases are given in Algorithm~\ref{alg:decoder} and \figurename~\ref{fig:dec}, respectively.

\subsubsection{Complexity Reduction with Pruning}
With pruning, the remaining complexity is due to the generation of $\max(\numberOfIterations-\indexRecursion,1)\numberOfPointsForPSK\min(\numberOfSequences,\numberOfGoodSeqences)$  sequences of length $2^{\numberOfIterations-\indexRecursion}$ at the $\indexRecursion$th  step, the process of choosing $\numberOfGoodSeqences$ sequences, and the calculation of \eqref{eq:mlNeww} for $\numberOfGoodSeqences$ sequences. For instance, for $\numberOfGoodSeqences=400$, maximum $400\numberOfPointsForPSK(i-1)$ sequences of length $2^{\numberOfIterations-\indexRecursion}$ are enumerated  and  $400$ of them survive for the $(\indexRecursion+1)$th  step, which corresponds to a substantially reduced search space.

\subsubsection{Comparisons with Other Decoders for CSs}
Although there are many studies that focus on the enumeration of \acp{CS}, the literature is {\em not} rich with decoding algorithms for the codes related to \acp{CS}. 
Particularly, there are only few studies that consider the coset term changing based on $\seqPermutationCompShift$.  In \cite{davis_1999}, a supercode decoder that subtracts each possible coset representative and decodes the remaining sequences with a \ac{FHT} were proposed. In \cite{Grant_1998}, \ac{FHT} was extended for decoding the first-order generalized \ac{RM} codes. In \cite{Schmidt2005}, for a given coset term, a recursive decoding  which can reduce the complexity of generalized \ac{FHT} further is proposed. In  \cite{Paterson2000decode}, $\funcfForANF^\referenceDerivative(\seqx^\referenceDerivative)$ is defined as $\funcfForANF(\seqx|\monomial[{\referenceDerivative}]=1)-\funcfForANF(\seqx|\monomial[{\referenceDerivative}]=0)$ and utilized for developing efficient decoding algorithms. Although Algorithm~8 in \cite{Paterson2000decode} considers the coset term, it consists of a step (i.e., Step 2) that multiplies two noisy terms, which would cause noise enhancement. %In addition, the available decoders in the literature were investigated only for \ac{AWGN} channel. % To the best of our knowledge, the systematic decoding of codes based on \acp{CS} with various constellations in fading channels has not been studied. 

The proposed decoder can be considered as an extension of the decoders in \cite{Schmidt2005} and \cite{Paterson2000decode}. The recursion steps in our decoder follow the same rationale discussed in \cite{Schmidt2005} and exploit the order reduction proposed in \cite{Paterson2000decode}. However, it differs from \cite{Schmidt2005} as we consider the coset term. As opposed to the decoder in \cite{Paterson2000decode}, we utilize $\funcfForANF^\referenceDerivative(\seqx^\referenceDerivative)$ such that the decoder does not cause noise enhancement. We also use the derivatives of both phase and amplitude functions in  Theorem~\ref{th:reduced}, which allow us to develop the decoder for different constellations and selective fading channels.

\section{Numerical Results}
\label{sec:numerical}

\begin{figure*}
	\centering
	\subfloat[Non-contiguous resource allocation.]{\includegraphics[width =3.2in]{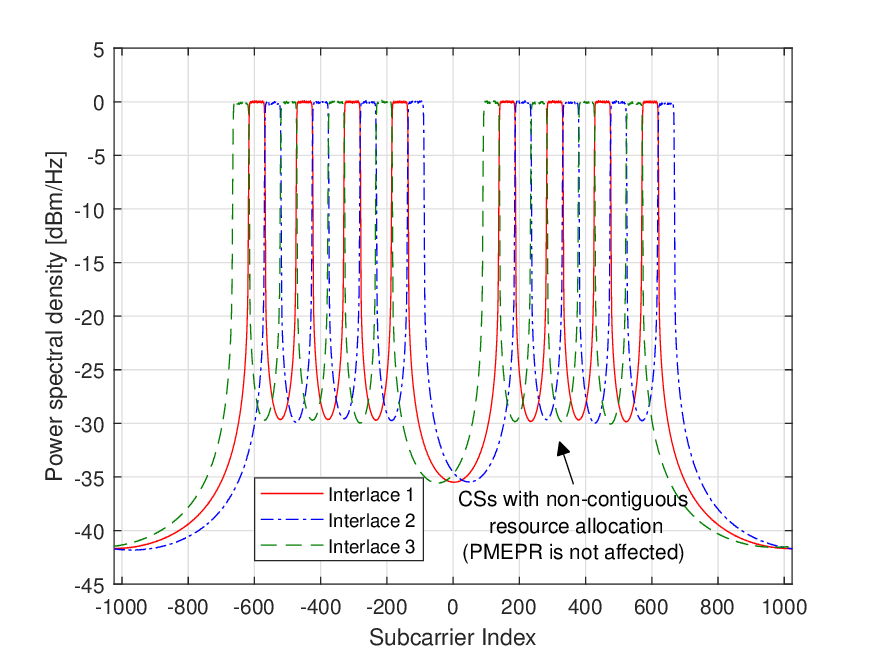}
		\label{subfig:spectrum}}~
	\subfloat[PMEPR distribution.]{\includegraphics[width =3.2in]{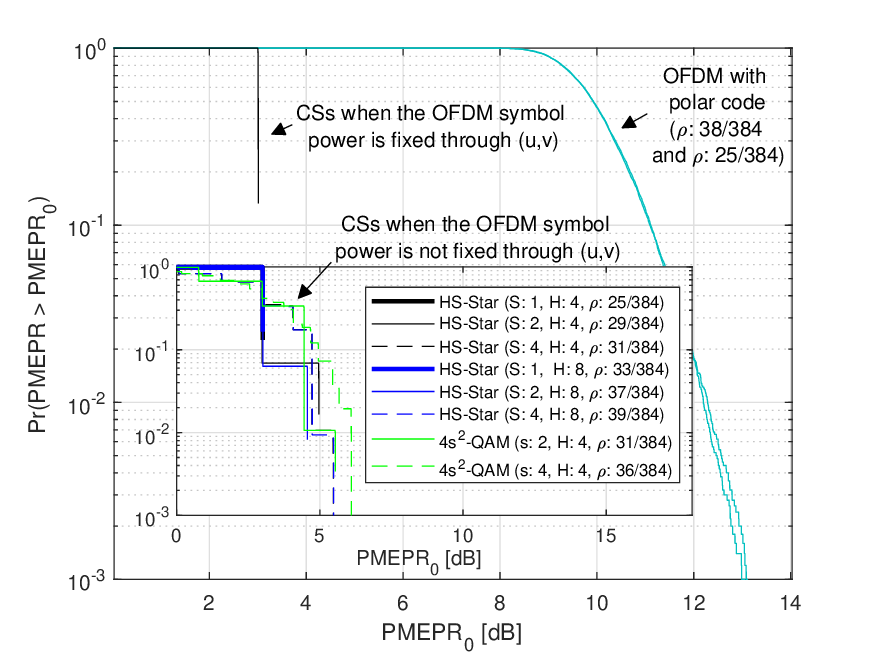}
		\label{subfig:pmepr}}
	\vspace{-3mm}\\
	\subfloat[BER versus Eb/N0 in AWGN channel.]{\includegraphics[width =3.2in]{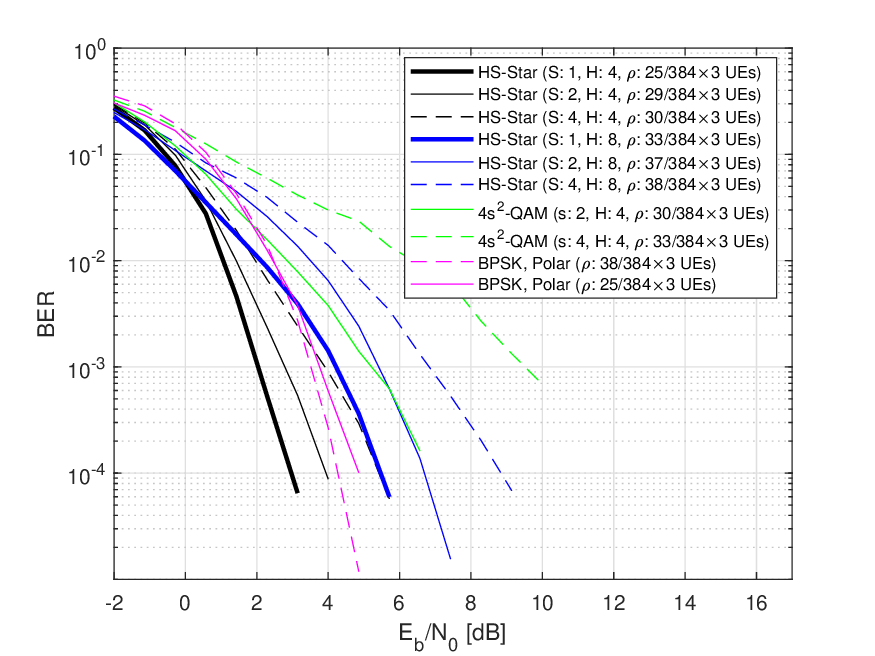}
		\label{subfig:berawgn}}~	
	\subfloat[BLER versus SNR in  fading channel.]{\includegraphics[width =3.2in]{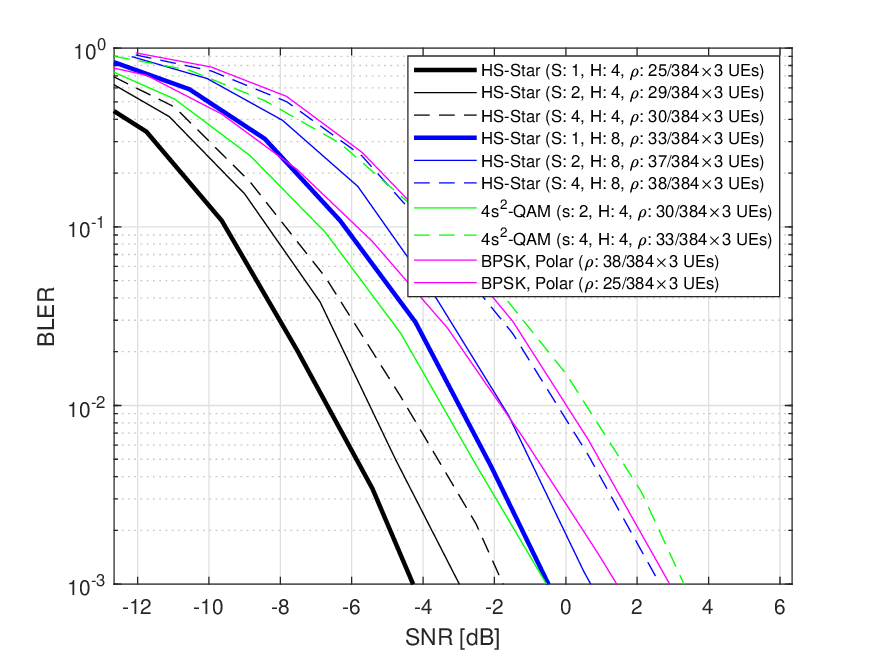}
		\label{subfig:blerfading}}
	\caption{Resource allocation, PMEPR distribution, BER, and BLER  for different schemes ($\numberOfGoodSeqences=400$, $\numberOfIterations=7$, $\lengthGaGb=3$).}
	\label{fig:performance}
	\vspace{-2mm}
\end{figure*}

In this section, we evaluate the proposed encoder and decoder for an uplink scenario numerically.  For comparison, we also generate results with a polar code and \ac{BPSK}  under similar \ac{SE} conditions. 
We assume that multiple users access the communication medium with a single \ac{OFDM} symbol synchronously over a non-contiguous resource allocation in the frequency domain to exploit the frequency diversity. Considering the compatibility with 3GPP \ac{5G} \ac{NR} waveform parameters, we assume that each user utilizes $384$ subcarriers distributed over $8$ clusters in the frequency domain. Each cluster consists of 48 subcarriers. We assume that $132$ subcarriers at the center of the band are allocated for random access channel and left unoccupied as shown in \figurename~\ref{fig:performance}\subref{subfig:spectrum}. The spacing between the clusters on the left and right portions of the band is set to $96$ subcarriers. Hence, three interlaced groups are constructed. We support multiple access on each group by exploiting orthogonal seed sequences of length $\lengthGaGb=3$. Therefore, $9$ users can be multiplexed  in the uplink within a single \ac{OFDM} symbol. 
%In this scenario, it is not trivial to reduce the \ac{PMEPR} due to the non-contiguous resource allocation in the frequency domain.
 To align with the resource allocation, we set $\numberOfIterations=7$ for the proposed encoder and polar code. We choose $\permutationMono[1]=7$.
For \acp{ESC}, we consider  $\numberOfPointsForPSK\numberOfPointsOnConstellation$-Star for $\numberOfPointsForPSK=\{4,8\}$ and $\numberOfPointsOnConstellation=\{1,2,4\}$ and $4\sqrtofNumPointsInQuad^2$-QAM  for $\sqrtofNumPointsInQuad=\{2,4\}$. For  $\numberOfPointsForPSK\numberOfPointsOnConstellation$-Star,  $(\ratioBetweenDistanceAndInner[1]^2,\ratioBetweenDistanceAndInner[2]^2)$ and $(\ratioBetweenDistanceAndInner[1]^2,\ratioBetweenDistanceAndInner[2]^2,\ratioBetweenDistanceAndInner[3]^2,\ratioBetweenDistanceAndInner[4]^2)$ are chosen as $(2-a^2,a^2)$ where $a=1.256$ for $\numberOfPointsOnConstellation=2$  and  $(2-b^2,2-a^2,a^2,b^2)$ where $a=1.1$ and $b=1.33$ for $\numberOfPointsOnConstellation=4$, respectively. For the proposed decoder, we set $\numberOfGoodSeqences=400$. For polar decoder, we consider  soft decoding with \ac{SIC} and the equalizer is \ac{MMSE} \ac{FDE}. The design \ac{SNR} for the polar code is set to $3$~dB.  The demodulation and equalization are combined in the proposed decoder for the \acp{CS} (see Algorithm~\ref{alg:decoder}). The subcarrier spacing is set to $30~$kHz.

In Figure~\ref{fig:performance}\subref{subfig:pmepr}, we compare \ac{PMEPR} distributions of different schemes. While \ac{PMEPR} of \ac{OFDM} symbols with the polar code reaches to $12$~dB at the rate of $\expnumber{1}{-2}$,  the maximum \ac{PMEPR} is $6$~dB for all cases of \acp{CS}. When the \ac{OFDM} symbol power is fixed by limiting the $(\indexU,\indexV)$ combinations, the maximum \acp{PMEPR} of the schemes based on \acp{CS} reduce to $3$~dB, which leaves a $9$~dB extra  room for transmit power as compared to \ac{OFDM} with the polar code. It is worth noting that the \ac{PMEPR} of the \ac{OFDM} symbols generated with the \acp{CS} is not affected by the non-contiguous resource allocation as the support can be achieved via $\funcfForCommonShift(\seqx)$ given in \eqref{eq:shift}, i.e., $\separationGolay[{\permutationMono[{\indexIteration}]}=1]=288$, $\separationGolay[{\permutationMono[{\indexIteration}]}=2]=96$, $\separationGolay[{\permutationMono[{\indexIteration}]}=3]=48$, and $\separationGolay[{\permutationMono[{\indexIteration}]}\neq{1,2,3}]=0$. The \acp{SE} of the schemes  are  provided in  Figure~\ref{fig:performance}\subref{subfig:pmepr} for the fixed \ac{OFDM} symbol power.  While the number of bits sacrificed for fixed \ac{OFDM} symbol power is only $1$ bit for $16$-QAM and $\numberOfPointsForPSK\numberOfPointsOnConstellation$-Star for $\numberOfPointsOnConstellation=4$ and  $\numberOfPointsForPSK=\{4,8\}$, it is $3$ bits for $64$-QAM. No bit is sacrificed with $\numberOfPointsForPSK\numberOfPointsOnConstellation$-Star for $\numberOfPointsOnConstellation=2$ and  $\numberOfPointsForPSK=\{4,8\}$.

In \figurename~\ref{fig:performance}\subref{subfig:berawgn} and \figurename~\ref{fig:performance}\subref{subfig:blerfading}, we provide \ac{BER} and \ac{BLER} curves  in \ac{AWGN} and fading channel, respectively. We consider ITU Vehicular A for the multi-path channel model  without mobility. We perform the evaluations with constant-modulus \acp{CS} by restricting $(\indexU,\indexV)$ permutations. For $\numberOfPointsOnConstellation=1$, $\numberOfPointsForPSK\numberOfPointsOnConstellation$-Star is equivalent to $\numberOfPointsForPSK$-\ac{PSK} constellation and our encoder reduces to the one in \cite{davis_1999}. % The proposed decoder decodes the encoded bits in both \ac{AWGN} and fading channel. 
The \acp{BER} are at $\expnumber{1}{-3}$ when $\EbNO=2$~dB and $\EbNO=4$~dB for $\numberOfPointsForPSK=4$ and $\numberOfPointsForPSK=8$ in \ac{AWGN} channel as shown in  \figurename~\ref{fig:performance}\subref{subfig:berawgn}. In the fading channel, $\expnumber{1}{-2}$ \ac{BLER} is achieved  when \acp{SNR} are $-7$~dB and  $-3$~dB for $\numberOfPointsForPSK=4$ and $\numberOfPointsForPSK=8$, respectively, as in  \figurename~\ref{fig:performance}\subref{subfig:blerfading}. For $\numberOfPointsOnConstellation=\{2,4\}$, the \ac{SE} increases gradually as more distinct \acp{CS} are synthesized without sacrificing $3$~dB \ac{PMEPR}.   For $\numberOfPointsForPSK=4$, $\expnumber{1}{-2}$ \ac{BLER}  is achieved at $-5.6$~dB and $-4.5$~dB when $\numberOfPointsOnConstellation=2$ and $\numberOfPointsOnConstellation=4$, respectively. For $\numberOfPointsForPSK=8$, the number of information bits reach up to $37$ and $38$ bits per user for $\numberOfPointsOnConstellation=2$ and $\numberOfPointsOnConstellation=4$, respectively, where $\expnumber{1}{-2}$ \ac{BLER} is achieved at $-2$~dB and $0$~dB. The results indicate that $\numberOfPointsForPSK\numberOfPointsOnConstellation$-Star performs better than \ac{QAM} in average \ac{BLER}. While both $\numberOfPointsForPSK\numberOfPointsOnConstellation$-Star for $\numberOfPointsForPSK=4$ and $\numberOfPointsOnConstellation=4$ and $16$-\ac{QAM} result in the same \ac{SE} under \ac{PMEPR} constraint, $\numberOfPointsForPSK\numberOfPointsOnConstellation$-Star provides $1$~dB improvement at $\expnumber{1}{-2}$ \ac{BLER}. The SNR gap between $\numberOfPointsForPSK\numberOfPointsOnConstellation$-Star for $\numberOfPointsForPSK=1$ and $\numberOfPointsOnConstellation=8$ and $64$-\ac{QAM} reaches to $3$~dB. The \ac{OFDM} with polar code provides similar \ac{BER} versus $\EbNO$ curves for both $\spectralEfficient=25/384$ bit/s/Hz and $\spectralEfficient=38/384$ bit/s/Hz (i.e., the code rates are 25/128 and 38/128, respectively) as shown in  \figurename~\ref{fig:performance}\subref{subfig:berawgn}. It performs $2$~dB worse than the \ac{QPSK}  \acp{CS} as the decoder for polar code relies on \ac{SIC}.
%(the degradation is severe in the fading channel  as in \figurename~\ref{fig:performance}\subref{subfig:blerfading}). 
The polar code is more reliable than the \acp{CS} with high-order constellations in \ac{AWGN}. It outperforms \acp{CS} with $\numberOfPointsForPSK\numberOfPointsOnConstellation$-Star for $\numberOfPointsOnConstellation=\{2,4\}$ for $\numberOfPointsForPSK=8$ with $2$~dB and $4$~dB, respectively. Nevertheless, the proposed encoder under a similar \ac{SE} provides a $9$~dB room for transmit power and can offset the degradation due to the \ac{SNR} loss in practice.

\section{Conclusion}\label{sec:conclusion}

In this study, we introduce a method that allows one to re-express a polynomial generated with a recursion consisting of linear operators by indicating how the operators are distributed to coefficients of the polynomial via Boolean functions.
By applying this method to a  recursive \ac{GCP} constructions, we show how the seed sequences, phase rotations, signs, real scalars, and the shifting factors applied at each step of the recursion alter the elements of the  sequences in the \ac{GCP}. As a result, we obtain a generic \ac{CS} construction that extends David and Jedwab's construction in \cite{davis_1999} by deriving separate pseudo-Boolean functions for the amplitude, the phase, the seed sequences, and the support of the \ac{CS}. We show that distinct \acp{CS} with a wide range of  constellations can be enumerated systematically. We also present an encoder  and an \ac{ML} decoder. We show that the proposed encoder can achieve low-\ac{PMEPR} \ac{OFDM} symbols without utilizing any optimization technique while supporting multiple users,  different constellations, and non-contiguous resource allocation. The proposed decoder also decodes the information bits on the second-order coset in the phase function while being compatible with different constellations. 

We demonstrate the performance of the encoder through an uplink simulation. The error rate is a strong function of the constellation. Under a maximum $3$~dB \ac{PMEPR} constraint, the  encoder with $\numberOfPointsForPSK\numberOfPointsOnConstellation$-Star outperforms the one with \ac{QAM}  in terms of average \ac{BLER}. The polar code is superior to the proposed encoder with high-order constellations in \ac{AWGN} while the performance gap diminishes in a fading channel. Also, the proposed encoder offers a $9$~dB room for increasing the transmit power without any clipping, which can offset the performance difference in practice. %This result indicates that the \acp{CS} can be effective for achieving reliable low-data rate communication links where the \ac{PA} non-linearity and/or power efficiency are major concerns.

\appendices

\section{Proof of Lemma~\ref{th:framework}}
\begin{proof}
\label{app:algConstruction}
The sequence ${\bf 1}\triangleq(1)_{i=0}^{2^\numberOfIterations-1}$ can be decomposed as
\if \IEEEsubmission0
\begin{align}
&\polySeq[{1}][\polyVariableIt]  =
\prod_{\indexIteration=1}^{\numberOfIterations}(1 + \polyVariableIt^{2^\permutationShift[{\indexIteration}]})\nonumber  \\&= \underbrace{\prod_{\substack{\indexIteration=1 \\ \indexIteration\neq\indexIterationANF}}^{\numberOfIterations}(1 + \polyVariableIt^{2^\permutationShift[{\indexIteration}]}}_{\polySeq[{\seqSubP[1]}][\polyVariableIt]})  + \underbrace{\polyVariableIt^{2^\permutationShift[{\indexIterationANF}]} \prod_{\substack{\indexIteration=1 \\ \indexIteration\neq\indexIterationANF}}^{\numberOfIterations}(1 + \polyVariableIt^{2^\permutationShift[{\indexIteration}]}}_{\polySeq[{\seqSubP[2]}][\polyVariableIt]}) \nonumber
\\&= \underbrace{\prod_{\substack{\indexIteration=1 \\ \indexIteration\neq\indexIterationANF,\indexIterationANF+1}}^{\numberOfIterations}(1 + \polyVariableIt^{2^\permutationShift[{\indexIteration}]}}_{\polySeq[{\seqSubP[11]}][\polyVariableIt]})  + \underbrace{\polyVariableIt^{2^\permutationShift[{\indexIterationANF+1}]} \prod_{\substack{\indexIteration=1 \\ \indexIteration\neq\indexIterationANF,\indexIterationANF+1}}^{\numberOfIterations}(1 + \polyVariableIt^{2^\permutationShift[{\indexIteration}]}}_{\polySeq[{\seqSubP[21]}][\polyVariableIt]})~\nonumber
\\&+ \underbrace{\polyVariableIt^{2^\permutationShift[{\indexIterationANF}]}\prod_{\substack{\indexIteration=1 \\ \indexIteration\neq\indexIterationANF,\indexIterationANF+1}}^{\numberOfIterations}(1 + \polyVariableIt^{2^\permutationShift[{\indexIteration}]}}_{\polySeq[{\seqSubP[12]}][\polyVariableIt]})  + \underbrace{\polyVariableIt^{2^\permutationShift[{\indexIterationANF}]+2^\permutationShift[{\indexIterationANF+1}]} \prod_{\substack{\indexIteration=1 \\ \indexIteration\neq\indexIterationANF,\indexIterationANF+1}}^{\numberOfIterations}(1 + \polyVariableIt^{2^\permutationShift[{\indexIteration}]}}_{\polySeq[{\seqSubP[22]}][\polyVariableIt]})~.\nonumber
\end{align}
\else
{
	\small
\begin{align}
&\polySeq[{1}][\polyVariableIt]  =
\prod_{\indexIteration=1}^{\numberOfIterations}(1 + \polyVariableIt^{2^\permutationShift[{\indexIteration}]})= \underbrace{\prod_{\substack{\indexIteration=1 \\ \indexIteration\neq\indexIterationANF}}^{\numberOfIterations}(1 + \polyVariableIt^{2^\permutationShift[{\indexIteration}]}}_{\polySeq[{\seqSubP[1]}][\polyVariableIt]})  + \underbrace{\polyVariableIt^{2^\permutationShift[{\indexIterationANF}]} \prod_{\substack{\indexIteration=1 \\ \indexIteration\neq\indexIterationANF}}^{\numberOfIterations}(1 + \polyVariableIt^{2^\permutationShift[{\indexIteration}]}}_{\polySeq[{\seqSubP[2]}][\polyVariableIt]}) \nonumber
\\&= \underbrace{\prod_{\substack{\indexIteration=1 \\ \indexIteration\neq\indexIterationANF,\indexIterationANF+1}}^{\numberOfIterations}(1 + \polyVariableIt^{2^\permutationShift[{\indexIteration}]}}_{\polySeq[{\seqSubP[11]}][\polyVariableIt]})  + \underbrace{\polyVariableIt^{2^\permutationShift[{\indexIterationANF+1}]} \prod_{\substack{\indexIteration=1 \\ \indexIteration\neq\indexIterationANF,\indexIterationANF+1}}^{\numberOfIterations}(1 + \polyVariableIt^{2^\permutationShift[{\indexIteration}]}}_{\polySeq[{\seqSubP[21]}][\polyVariableIt]})~\nonumber
+ \underbrace{\polyVariableIt^{2^\permutationShift[{\indexIterationANF}]}\prod_{\substack{\indexIteration=1 \\ \indexIteration\neq\indexIterationANF,\indexIterationANF+1}}^{\numberOfIterations}(1 + \polyVariableIt^{2^\permutationShift[{\indexIteration}]}}_{\polySeq[{\seqSubP[12]}][\polyVariableIt]})  + \underbrace{\polyVariableIt^{2^\permutationShift[{\indexIterationANF}]+2^\permutationShift[{\indexIterationANF+1}]} \prod_{\substack{\indexIteration=1 \\ \indexIteration\neq\indexIterationANF,\indexIterationANF+1}}^{\numberOfIterations}(1 + \polyVariableIt^{2^\permutationShift[{\indexIteration}]}}_{\polySeq[{\seqSubP[22]}][\polyVariableIt]})~.\nonumber
\end{align}
}
\fi
By the definitions,  $\polySeq[{\seqSubP[1]}][\polyVariableIt]$ and $\polySeq[{\seqSubP[2]}][\polyVariableIt]$ should have no common monomials, i.e., $\seqSub[1]+\seqSub[2]={\bf 1}$. 
The polynomial $\polySeq[{\seqSubP[2]}][\polyVariableIt]$ corresponds to a polynomial where the degree of each monomial in $ \polySeq[{\seqSubP[1]}][\polyVariableIt]$ is increased by $2^\permutationShift[{\indexIterationANF}]$ by their definitions. Hence, the sequence $\seqSub[2]$ should be the shifted version of the sequence $\seqSub[1]$ by $2^\permutationShift[{\indexIterationANF}]$. Since $\seqSub[1]+\seqSub[2]={\bf 1}$ and  $\seqSub[2]$ is the shifted version of $\seqSub[1]$ by $2^\permutationShift[{\indexIterationANF}]$, the $u$th element of the sequence $\seqSub[1]$ should be $1$ for $u=s2^\permutationShift[{\indexIterationANF}]+(1,2,\dots,2^\permutationShift[{\indexIterationANF}])$  for even $s$ and $0$ for odd $s$ where $0\le s < 2^{\numberOfIterations - \permutationShift[{\indexIterationANF}]}$. 
Thus, the Boolean functions generating  $\seqSub[1]$  and  $\seqSub[2]$ can be obtained as $\monomial[{\permutationMono[\indexIterationANF]}]$ and $1-\monomial[{\permutationMono[\indexIterationANF]}]$, respectively. The polynomials $\polySeq[{\seqSubP[1]}][\polyVariableIt]$ and $\polySeq[{\seqSubP[2]}][\polyVariableIt]$ can also be decomposed as $\polySeq[{\seqSubP[11]}][\polyVariableIt]$ + $\polySeq[{\seqSubP[21]}][\polyVariableIt]$ and  $\polySeq[{\seqSubP[12]}][\polyVariableIt]$ + $\polySeq[{\seqSubP[22]}][\polyVariableIt]$, respectively, i.e., $\seqSub[11]+\seqSub[21]=\seqSub[1]$ and $\seqSub[12]+\seqSub[22]=\seqSub[2]$. The sequences $\seqSub[21]$ and $\seqSub[22]$ are the shifted version of the sequence $\seqSub[11]$ and $\seqSub[12]$ by $2^\permutationShift[{\indexIterationANF+1}]$, respectively. Thus, the Boolean functions generating $\seqSub[11]$, $\seqSub[21]$, $\seqSub[12]$, and $\seqSub[22]$  can then be expressed as
$
({ 1} - {\permutationMono[{\indexIterationANF}]})({ 1} - {\permutationMono[{\indexIterationANF+1}]})
$,
$
({ 1} - {\permutationMono[{\indexIterationANF}]}){\permutationMono[{\indexIterationANF+1}]}
$,
$
{{\permutationMono[{\indexIterationANF}]}}  ({ 1} - {\permutationMono[{\indexIterationANF+1}]})
$, and
$
{{\permutationMono[{\indexIterationANF}]}}{\permutationMono[{\indexIterationANF+1}]}
$, respectively.

Now, consider a new recursion given by 
$\functionfdot[\indexIterationdot] $ is  $\binaryAsignment[11][{\indexIterationANF}]{\functionfdot[\indexIterationdot-1]} + \binaryAsignment[12][{\indexIterationANF}]{\functiongdot[\indexIterationdot-1]}\polyVariableIt^{2^\permutationShift[{\indexIterationdot}]}$ for $\indexIterationdot=\indexIterationANF$, otherwise ${\functionfdot[\indexIterationdot-1]} + {\functiongdot[\indexIterationdot-1]}\polyVariableIt^{2^\permutationShift[{\indexIterationdot}]}$ and $\functiongdot[\indexIterationdot] $ is  $\binaryAsignment[21][{\indexIterationANF}]{\functionfdot[\indexIterationdot-1]} + \binaryAsignment[22][{\indexIterationANF}]{\functiongdot[\indexIterationdot-1]}\polyVariableIt^{2^\permutationShift[{\indexIterationdot}]}$ for $\indexIterationdot=\indexIterationANF$, otherwise $	{\functionfdot[\indexIterationdot-1]} + {\functiongdot[\indexIterationdot-1]}\polyVariableIt^{2^\permutationShift[{\indexIterationdot}]}$,
	where $\functionfdot[0]=\functiongdot[0]=1$. By the definition, $\functionfdot[\numberOfIterations]$ and $\functiongdot[\numberOfIterations]$ correspond to the associated polynomials for the  sequences $\seqGf[\indexIterationANF]$ and $\seqGg[\indexIterationANF]$, respectively. By evaluating the new recursion,  $\functionfdot[\numberOfIterations]$ and $\functiongdot[\numberOfIterations]$ can be expressed as
\if\IEEEsubmission0
	\begin{align}
	\polySeq[{\seqGfP[\indexIterationANF]}][\polyVariableIt]  =  \functionfdot[\numberOfIterations] = 
	\begin{cases}
	\displaystyle
	\binaryAsignment[11][{\indexIterationANF}]\polySeq[{\seqSubP[1]}][\polyVariableIt]+\binaryAsignment[12][{\indexIterationANF}]\polySeq[{\seqSubP[2]}][\polyVariableIt]~,
	& \indexIterationANF=\numberOfIterations \\
	\displaystyle
	\binaryAsignment[11][{\indexIterationANF}]\polySeq[{\seqSubP[11]}][\polyVariableIt]+\binaryAsignment[12][{\indexIterationANF}]\polySeq[{\seqSubP[12]}][\polyVariableIt]\\+\binaryAsignment[21][{\indexIterationANF}]\polySeq[{\seqSubP[21]}][\polyVariableIt]+\binaryAsignment[22][{\indexIterationANF}]\polySeq[{\seqSubP[22]}][\polyVariableIt]~,
	& \indexIterationANF<\numberOfIterations \\
	\end{cases}~,\nonumber\\
	\polySeq[{\seqGgP[\indexIterationANF]}][\polyVariableIt] = \functiongdot[\numberOfIterations]=
	\begin{cases}
	\displaystyle
	\binaryAsignment[21][{\indexIterationANF}]\polySeq[{\seqSubP[1]}][\polyVariableIt]+\binaryAsignment[22][{\indexIterationANF}]\polySeq[{\seqSubP[2]}][\polyVariableIt]~,
	& \indexIterationANF=\numberOfIterations \\
	\displaystyle
	\binaryAsignment[11][{\indexIterationANF}]\polySeq[{\seqSubP[11]}][\polyVariableIt]+\binaryAsignment[12][{\indexIterationANF}]\polySeq[{\seqSubP[12]}][\polyVariableIt]\\+\binaryAsignment[21][{\indexIterationANF}]\polySeq[{\seqSubP[21]}][\polyVariableIt]+\binaryAsignment[22][{\indexIterationANF}]\polySeq[{\seqSubP[22]}][\polyVariableIt]~,
	& \indexIterationANF<\numberOfIterations \\
	\end{cases}~.
	\nonumber
	\end{align}
	\else
	\begin{align}
	\polySeq[{\seqGfP[\indexIterationANF]}][\polyVariableIt]  =  \functionfdot[\numberOfIterations] = 
	\begin{cases}
	\displaystyle
	\binaryAsignment[11][{\indexIterationANF}]\polySeq[{\seqSubP[1]}][\polyVariableIt]+\binaryAsignment[12][{\indexIterationANF}]\polySeq[{\seqSubP[2]}][\polyVariableIt]~,
	& \indexIterationANF=\numberOfIterations \\
	\displaystyle
	\binaryAsignment[11][{\indexIterationANF}]\polySeq[{\seqSubP[11]}][\polyVariableIt]+\binaryAsignment[12][{\indexIterationANF}]\polySeq[{\seqSubP[12]}][\polyVariableIt]+\binaryAsignment[21][{\indexIterationANF}]\polySeq[{\seqSubP[21]}][\polyVariableIt]+\binaryAsignment[22][{\indexIterationANF}]\polySeq[{\seqSubP[22]}][\polyVariableIt]~,
	& \indexIterationANF<\numberOfIterations \\
	\end{cases}~,\nonumber\\
	\polySeq[{\seqGgP[\indexIterationANF]}][\polyVariableIt] = \functiongdot[\numberOfIterations]=
	\begin{cases}
	\displaystyle
	\binaryAsignment[21][{\indexIterationANF}]\polySeq[{\seqSubP[1]}][\polyVariableIt]+\binaryAsignment[22][{\indexIterationANF}]\polySeq[{\seqSubP[2]}][\polyVariableIt]~,
	& \indexIterationANF=\numberOfIterations \\
	\displaystyle
	\binaryAsignment[11][{\indexIterationANF}]\polySeq[{\seqSubP[11]}][\polyVariableIt]+\binaryAsignment[12][{\indexIterationANF}]\polySeq[{\seqSubP[12]}][\polyVariableIt]+\binaryAsignment[21][{\indexIterationANF}]\polySeq[{\seqSubP[21]}][\polyVariableIt]+\binaryAsignment[22][{\indexIterationANF}]\polySeq[{\seqSubP[22]}][\polyVariableIt]~,
	& \indexIterationANF<\numberOfIterations \\
	\end{cases}~.
	\nonumber
	\end{align}
	\fi
\end{proof}

\if\IEEEsubmission0
\renewcommand{\baselinestretch}{1}
\else
\renewcommand{\baselinestretch}{0.8}
\fi
\begin{table*}[t]
	\caption{The operator definitions  and the Boolean functions.}
	\centering
	\resizebox{\textwidth}{!}{%
		\begin{tabular}{l|c|c|c|c|c|}
			& $\operatorBinary[0][\indexIteration]\{\functionh\}$                   & $\operatorBinary[1][\indexIteration]\{\functionh\}$  & $\vecArrangement[\indexIteration]^{\rm T}$ & $\funcGfForANF[\indexIteration](\seqx)$ & $\funcGgForANF[\indexIteration](\seqx)$  \\ \hline

			\hspace{-3mm}\begin{tabular}{l}Scalar \\ $\scaleA[\indexIteration]$  \end{tabular} 													& $\operatorScaleA[0][{\indexIteration}]\{\functionh\}\triangleq\exponentialBase^0 \functionh $	& $\operatorScaleA[1][{\indexIteration}]\{\functionh\}\triangleq\exponentialBase^{\scaleAexp[\indexIteration]} \functionh $   & $[1~0~0~1]$   & $ 
			\begin{cases}
			1-\monomial[{\permutationMono[{\numberOfIterations}]}], 		& \indexIteration = \numberOfIterations\\
			1-\monomial[{\permutationMono[{\indexIteration}]}]-\monomial[{\permutationMono[{\indexIteration+1}]}],   	& \indexIteration < \numberOfIterations
			\end{cases}$  &  $ 
			\begin{cases}
			\monomial[{\permutationMono[{\numberOfIterations}]}], 		& \indexIteration = \numberOfIterations\\
			1-\monomial[{\permutationMono[{\indexIteration}]}]-\monomial[{\permutationMono[{\indexIteration+1}]}],   	& \indexIteration < \numberOfIterations
			\end{cases}$ \\ \hline
			
			\hspace{-3mm}\begin{tabular}{l}Scalar \\ $\scaleB[\indexIteration]$  \end{tabular} 													& $\operatorScaleB[0][{\indexIteration}]\{\functionh\}\triangleq\exponentialBase^0 \functionh $	& $\operatorScaleB[1][{\indexIteration}]\{\functionh\}\triangleq\exponentialBase^{\scaleBexp[\indexIteration]} \functionh $   & $[0~1~1~0]$   & $ 
			\begin{cases}
			\monomial[{\permutationMono[{\numberOfIterations}]}], 		& \indexIteration = \numberOfIterations\\
			\monomial[{\permutationMono[{\indexIteration}]}]+\monomial[{\permutationMono[{\indexIteration+1}]}],   	& \indexIteration < \numberOfIterations
			\end{cases}$  &  $ 
			\begin{cases}
			1-\monomial[{\permutationMono[{\numberOfIterations}]}], 		& \indexIteration = \numberOfIterations\\
			\monomial[{\permutationMono[{\indexIteration}]}]+\monomial[{\permutationMono[{\indexIteration+1}]}],   	& \indexIteration < \numberOfIterations
			\end{cases}$ \\ \hline
			
			\hspace{-3mm}\begin{tabular}{l}Phase \\ $\angleScaleA[\indexIteration][]$  \end{tabular} 													& $\operatorAngleScaleA[0][{\indexIteration}]\{\functionh\}\triangleq\exponentialBase^0 \functionh $	& $\operatorAngleScaleA	[1][{\indexIteration}]\{\functionh\}\triangleq\exponentialBase^{\constantj\angleScaleAexp[\indexIteration]} \functionh $   & $[1~0~0~0]$   & $ 
			\begin{cases}
			1-\monomial[{\permutationMono[{\numberOfIterations}]}], 		& \indexIteration = \numberOfIterations\\
			(1-\monomial[{\permutationMono[{\indexIteration}]}])(1-\monomial[{\permutationMono[{\indexIteration+1}]}]),   	& \indexIteration < \numberOfIterations
			\end{cases}$  &  $ 
			\begin{cases}
			0, 		& \indexIteration = \numberOfIterations\\
			(1-\monomial[{\permutationMono[{\indexIteration}]}])(1-\monomial[{\permutationMono[{\indexIteration+1}]}]),   	& \indexIteration < \numberOfIterations
			\end{cases}$ \\ \hline
			
			\hspace{-3mm}\begin{tabular}{l}Phase \\ $\angleScaleA[\indexIteration][*]$  \end{tabular} 													& $\operatorAngleScaleA[0][{\indexIteration}]\{\functionh\}\triangleq\exponentialBase^0 \functionh $	& $\operatorAngleScaleA	[1][{\indexIteration}]\{\functionh\}\triangleq\exponentialBase^{-\constantj\angleScaleAexp[\indexIteration]} \functionh $   & $[0~0~0~1]$   & $ 
			\begin{cases}
			0, 		& \indexIteration = \numberOfIterations\\
			\monomial[{\permutationMono[{\indexIteration}]}]\monomial[{\permutationMono[{\indexIteration+1}]}],   	& \indexIteration < \numberOfIterations
			\end{cases}$  &  $ 
			\begin{cases}
			\monomial[{\permutationMono[{\numberOfIterations}]}], 		& \indexIteration = \numberOfIterations\\
			\monomial[{\permutationMono[{\indexIteration}]}]\monomial[{\permutationMono[{\indexIteration+1}]}],   	& \indexIteration < \numberOfIterations
			\end{cases}$ \\ \hline

			\hspace{-3mm}\begin{tabular}{l}Phase \\ $\angleScaleB[\indexIteration][]$  \end{tabular} 													& $\operatorAngleScaleB[0][{\indexIteration}]\{\functionh\}\triangleq\exponentialBase^0 \functionh $	& $\operatorAngleScaleB	[1][{\indexIteration}]\{\functionh\}\triangleq\exponentialBase^{\constantj\angleScaleBexp[\indexIteration]} \functionh $   & $[0~1~0~0]$   & $ 
			\begin{cases}
			\monomial[{\permutationMono[{\numberOfIterations}]}], 		& \indexIteration = \numberOfIterations\\
			\monomial[{\permutationMono[{\indexIteration}]}](1-\monomial[{\permutationMono[{\indexIteration+1}]}]),   	& \indexIteration < \numberOfIterations
			\end{cases}$  &  $ 
			\begin{cases}
			0, 		& \indexIteration = \numberOfIterations\\
			\monomial[{\permutationMono[{\indexIteration}]}](1-\monomial[{\permutationMono[{\indexIteration+1}]}]),   	& \indexIteration < \numberOfIterations
			\end{cases}$ \\ \hline
			
			\hspace{-3mm}\begin{tabular}{l}Phase \\ $\angleScaleB[\indexIteration][*]$  \end{tabular} 													& $\operatorAngleScaleB[0][{\indexIteration}]\{\functionh\}\triangleq\exponentialBase^0 \functionh $	& $\operatorAngleScaleB	[1][{\indexIteration}]\{\functionh\}\triangleq\exponentialBase^{-\constantj\angleScaleBexp[\indexIteration]} \functionh $   & $[0~0~1~0]$   & $ 
			\begin{cases}
			0, 		& \indexIteration = \numberOfIterations\\
			(1-\monomial[{\permutationMono[{\indexIteration}]}])\monomial[{\permutationMono[{\indexIteration+1}]}],   	& \indexIteration < \numberOfIterations
			\end{cases}$  &  $ 
			\begin{cases}
			1-\monomial[{\permutationMono[{\numberOfIterations}]}], 		& \indexIteration = \numberOfIterations\\
			(1-\monomial[{\permutationMono[{\indexIteration}]}])\monomial[{\permutationMono[{\indexIteration+1}]}],   	& \indexIteration < \numberOfIterations
			\end{cases}$ \\ \hline     			
			
			\hspace{-3mm}\begin{tabular}{l}Sign \\  \end{tabular} 													& $\operatorSign[0][{\indexIteration}]\{\functionh\}\triangleq\exponentialBase^0 \functionh $	& $\operatorSign[1][{\indexIteration}]\{\functionh\}\triangleq\exponentialBase^ {\constantj\frac{\numberOfPointsForPSK}{2}}\functionh $   & $[0~0~0~1]$   & $ 
			\begin{cases}
			0, 		& \indexIteration = \numberOfIterations\\
			\monomial[{\permutationMono[{\indexIteration}]}]\monomial[{\permutationMono[{\indexIteration+1}]}],   	& \indexIteration < \numberOfIterations
			\end{cases}$  &  $ 
			\begin{cases}
			\monomial[{\permutationMono[{\numberOfIterations}]}], 		& \indexIteration = \numberOfIterations\\
			\monomial[{\permutationMono[{\indexIteration}]}]\monomial[{\permutationMono[{\indexIteration+1}]}],  	& \indexIteration < \numberOfIterations
			\end{cases}$ \\ \hline        
			
			\hspace{-3mm}\begin{tabular}{l}Phase \\ $\angleGolay[\indexIteration]$   \end{tabular}  													& $\operatorAngle[0][{\indexIteration}]\{\functionh\}\triangleq\exponentialBase^0  \functionh $	& $\operatorAngle[1][{\indexIteration}]\{\functionh\}\triangleq\exponentialBase^{\constantj\angleexp[\indexIteration]} \functionh $   & $[0~1~0~1]$   & $ \monomial[{\permutationMono[{\indexIteration}]}]$  &  $ 
			\monomial[{\permutationMono[{\indexIteration}]}]$ \\ \hline
			
			\hspace{-3mm}\begin{tabular}{l}Shift \\  $\separationIterative[\indexIteration]$  \end{tabular} 													& $\operatorSeparation[0][{\indexIteration}]\{\functionh\}\triangleq\polyVariable^0\functionh $	& $\operatorSeparation[1][{\indexIteration}]\{\functionh\}\triangleq\polyVariable^{\separationGolay[\indexIteration]}  \functionh $   & $[0~1~0~1]$   & $ \monomial[{\permutationMono[{\indexIteration}]}]$  &  $ 
			\monomial[{\permutationMono[{\indexIteration}]}]$ \\ \hline

			\hspace{-3mm}\begin{tabular}{l} $\polySeq[{\seqGaP}][\polyVariable]$  \end{tabular}  														& $\operatorOrderA[0][{\indexIteration}]\{\functionh\}\triangleq\exponentialBase^0\functionh $	& $\operatorOrderA[1][{\indexIteration}]\{\functionh\}\triangleq\begin{cases}
			\exponentialBase^0\functionh~, 		& \indexIteration \neq 1\\
			\polySeq[{\seqGaP}][\polyVariable]~, 	& \indexIteration = 1
			\end{cases} $   & $[1~0~1~0]$   & 
			%         $ 
			% 		\begin{cases}
			% 		0, 		& \indexIteration \neq 1\\
			% 		1-\monomial[{\permutationMono[{1}]}]   	& \indexIteration = 1
			% 		\end{cases}$ 
			$ 1-\monomial[{\permutationMono[{\indexIteration}]}]$
			&  
			%         $ 
			% 		\begin{cases}
			% 		0, 		& \indexIteration \neq 1\\
			% 		1-\monomial[{\permutationMono[{1}]}]   	& \indexIteration = 1
			% 		\end{cases}$
			$ 1-\monomial[{\permutationMono[{\indexIteration}]}]$
			\\ \hline

			\hspace{-3mm}\begin{tabular}{l}  $\polySeq[{\seqGbP}][\polyVariable]$  \end{tabular} 														& $\operatorOrderB[0][{\indexIteration}]\{\functionh\}\triangleq\exponentialBase^0\functionh $	& $\operatorOrderB[1][{\indexIteration}]\{\functionh\}\triangleq\begin{cases}
			\exponentialBase^0\functionh~, 		& \indexIteration \neq 1\\
			\polySeq[{\seqGbP}][\polyVariable]~,  	& \indexIteration = 1
			\end{cases} $   & $[0~1~0~1]$   & $ \monomial[{\permutationMono[{\indexIteration}]}]$
			%         $ 
			% 		\begin{cases}
			% 		0, 		& \indexIteration \neq 1\\
			% 		\monomial[{\permutationMono[{1}]}]   	& \indexIteration = 1
			% 		\end{cases}$
			&  
			$ \monomial[{\permutationMono[{\indexIteration}]}]$
			%         $ 
			% 		\begin{cases}
			% 		0, 		& \indexIteration \neq 1\\
			% 		\monomial[{\permutationMono[{1}]}]   	& \indexIteration = 1
			% 		\end{cases}$
			\\ \hline	
		\end{tabular}
	}
	\label{table:ANFgolay}
	\if\IEEEsubmission1	\vspace{-5mm} 	\fi
\end{table*} 
\renewcommand{\baselinestretch}{\baselineSize}

\if\IEEEsubmission1	\vspace{-10mm} 	\fi
\section{Proof of Theorem~\ref{th:reduced}}
	\label{app:red}
\begin{proof}
We re-write  $\polySeq[{\seqGaItP[\indexIteration]}][\polyVariable]$ and $\polySeq[{\seqGbItP[\indexIteration]}][\polyVariable]$ in Theorem~\ref{th:golayIterative} as 
\if\IEEEsubmission0
	\begin{align}
	\functionf[\indexIteration] &=
	\operatorAngleScaleA[1][{\indexIteration}]
	\operatorAngleConjScaleA[0][{\indexIteration}]
	\operatorAngleScaleB[0][{\indexIteration}]
	\operatorAngleConjScaleB[0][{\indexIteration}]
	\nonumber\\&
	\operatorSign[0][{\indexIteration}]
	\operatorScaleA[1][{\indexIteration}]
	\operatorScaleB[0][{\indexIteration}]
	\operatorAngle[0][{\indexIteration}]
	\operatorSeparation[0][{\indexIteration}]
	\operatorOrderA[1][{\indexIteration}]
	\operatorOrderB[0][{\indexIteration}]
	\{\functionf[\indexIteration-1]\}\nonumber\\
	&+ 
	\operatorAngleScaleA[0][{\indexIteration}]
	\operatorAngleConjScaleA[0][{\indexIteration}]
	\operatorAngleScaleB[1][{\indexIteration}]
	\operatorAngleConjScaleB[0][{\indexIteration}]
	\nonumber\\&
	\operatorSign[0][{\indexIteration}]
	\operatorScaleA[0][{\indexIteration}]
	\operatorScaleB[1][{\indexIteration}]
	\operatorAngle[1][{\indexIteration}]
	\operatorSeparation[0][{\indexIteration}]
	\operatorOrderA[0][{\indexIteration}]
	\operatorOrderB[1][{\indexIteration}]
	\{\functiong[\indexIteration-1]\}
	\polyVariableIt^{\varUpsample2^\permutationShift[{\indexIteration}]}~, \nonumber
\\
	\functiong[\indexIteration] &= 
	\operatorAngleScaleA[0][{\indexIteration}]
	\operatorAngleConjScaleA[0][{\indexIteration}]
	\operatorAngleScaleB[0][{\indexIteration}]
	\operatorAngleConjScaleB[1][{\indexIteration}]
	\nonumber\\&
	\operatorSign[0][{\indexIteration}]
	\operatorScaleA[0][{\indexIteration}]
	\operatorScaleB[1][{\indexIteration}]
	\operatorAngle[0][{\indexIteration}]
	\operatorSeparation[1][{\indexIteration}]
	\operatorOrderA[1][{\indexIteration}]
	\operatorOrderB[0][{\indexIteration}]
	\{\functionf[\indexIteration-1]\}\nonumber\\
	&+ 
	\operatorAngleScaleA[0][{\indexIteration}]
	\operatorAngleConjScaleA[1][{\indexIteration}]
	\operatorAngleScaleB[0][{\indexIteration}]
	\operatorAngleConjScaleB[0][{\indexIteration}]
	\nonumber\\&
	\operatorSign[1][{\indexIteration}]
	\operatorScaleA[1][{\indexIteration}]
	\operatorScaleB[0][{\indexIteration}]
	\operatorAngle[1][{\indexIteration}]
	\operatorSeparation[1][{\indexIteration}]
	\operatorOrderA[0][{\indexIteration}]
	\operatorOrderB[1][{\indexIteration}]
	\{\functiong[\indexIteration-1]\}
	\polyVariableIt^{\varUpsample2^\permutationShift[{\indexIteration}]}~,
\nonumber%	\label{eq:iterationGolayWithOperators}
	\end{align}
\else
$
	\functionf[\indexIteration] =
	\operatorAngleScaleA[1][{\indexIteration}]
	\operatorAngleConjScaleA[0][{\indexIteration}]
	\operatorAngleScaleB[0][{\indexIteration}]
	\operatorAngleConjScaleB[0][{\indexIteration}]
	\operatorSign[0][{\indexIteration}]
	\operatorScaleA[1][{\indexIteration}]
	\operatorScaleB[0][{\indexIteration}]
	\operatorAngle[0][{\indexIteration}]\\
	\operatorSeparation[0][{\indexIteration}]
	\operatorOrderA[1][{\indexIteration}]
	\operatorOrderB[0][{\indexIteration}]
	\{\functionf[\indexIteration-1]\}
	+ 
	\operatorAngleScaleA[0][{\indexIteration}]
	\operatorAngleConjScaleA[0][{\indexIteration}]
	\operatorAngleScaleB[1][{\indexIteration}]
	\operatorAngleConjScaleB[0][{\indexIteration}]
	\operatorSign[0][{\indexIteration}]
	\operatorScaleA[0][{\indexIteration}]
	\operatorScaleB[1][{\indexIteration}]
	\operatorAngle[1][{\indexIteration}]
	\operatorSeparation[0][{\indexIteration}]
	\operatorOrderA[0][{\indexIteration}]
	\operatorOrderB[1][{\indexIteration}]
	\{\functiong[\indexIteration-1]\}
	\polyVariableIt^{\varUpsample2^\permutationShift[{\indexIteration}]} 
	$ and $
	\functiong[\indexIteration] = 
	\operatorAngleScaleA[0][{\indexIteration}]
	\operatorAngleConjScaleA[0][{\indexIteration}]
	\operatorAngleScaleB[0][{\indexIteration}]
	\operatorAngleConjScaleB[1][{\indexIteration}]
	%\nonumber\\&
	\operatorSign[0][{\indexIteration}]
	\operatorScaleA[0][{\indexIteration}]
	\operatorScaleB[1][{\indexIteration}]
	\operatorAngle[0][{\indexIteration}]
	\operatorSeparation[1][{\indexIteration}]
	\operatorOrderA[1][{\indexIteration}]
	\operatorOrderB[0][{\indexIteration}]
	({\functionf[\indexIteration-1]})+ 
	\operatorAngleScaleA[0][{\indexIteration}]
	\operatorAngleConjScaleA[1][{\indexIteration}]
	\operatorAngleScaleB[0][{\indexIteration}]
	\operatorAngleConjScaleB[0][{\indexIteration}]
	%\nonumber\\&
	\operatorSign[1][{\indexIteration}]
	\operatorScaleA[1][{\indexIteration}]
	\operatorScaleB[0][{\indexIteration}]
	\operatorAngle[1][{\indexIteration}]
	\operatorSeparation[1][{\indexIteration}]\\
	\operatorOrderA[0][{\indexIteration}]
	\operatorOrderB[1][{\indexIteration}]
	\{\functiong[\indexIteration-1]\}
	\polyVariableIt^{\varUpsample2^\permutationShift[{\indexIteration}]}
%	\label{eq:iterationGolayWithOperators}
$,
%	\end{align}
\fi
by using the operators defined in \tablename~\ref{table:ANFgolay}. The operators related to the scalars, i.e.,  $\scaleA[\indexIteration]$ and $\scaleB[\indexIteration]$, and the phase rotations, i.e., $\angleScaleA[\indexIteration][]$, $\angleScaleA[\indexIteration][*]$, $\angleScaleB[\indexIteration][]$, $\angleScaleB[\indexIteration][*]$, and $\angleGolay[\indexIteration]$, are denoted by $\operatorScaleA[\{0,1\}][{\indexIteration}]$, 
$\operatorAngleScaleA[\{0,1\}][{\indexIteration}]$,
$\operatorAngleConjScaleA[\{0,1\}][{\indexIteration}]$,
$\operatorAngleScaleB[\{0,1\}][{\indexIteration}]$,
$\operatorAngleConjScaleB[\{0,1\}][{\indexIteration}]$,
and $\operatorAngle[\{0,1\}][{\indexIteration}]$ respectively. 
The signs for $\polySeq[{\seqGaItP[\indexIteration-1]}][\polyVariable]$ and $\polySeq[{\seqGbItP[\indexIteration-1]}][\polyVariable]$ in Theorem~\ref{th:golayIterative} are generated with the operators $\operatorSign[\{0,1\}][{\indexIteration}]$. While the operators $\operatorSeparation[\{0,1\}][{\indexIteration}]$ apply a shift to their arguments,  the operators $\operatorOrderA[\{0,1\}][{\indexIteration}]$ and $\operatorOrderB[\{0,1\}][{\indexIteration}]$ generate the polynomials $\polySeq[{\seqGaP}][\polyVariable]$ and $\polySeq[{\seqGbP}][\polyVariable]$ for $\indexIteration=1$, respectively. For the sake of unifying the operator formats, we express the  operators by using $\exponentialBase$ as   $\scaleA[\indexIteration]\triangleq\exponentialBase^{\scaleAexp[\indexIteration]}$, $\scaleB[\indexIteration]\triangleq\exponentialBase^{\scaleBexp[\indexIteration]}$, 
$\angleScaleA[\indexIteration][]\triangleq\exponentialBase^{\constantj\angleScaleAexp[\indexIteration]}$,
$\angleScaleB[\indexIteration][]\triangleq\exponentialBase^{\constantj\angleScaleBexp[\indexIteration]}$,
$\angleGolay[\indexIteration]\triangleq\exponentialBase^{\constantj\angleexp[\indexIteration]}$,
and $\exponentialBase^{\constantj\separationIterative[\indexIteration]}\triangleq\exponentialBase^{\constantj\separationGolay[\indexIteration]}$,  and exploit the identity $\exponentialBase^{\constantj\frac{\numberOfPointsForPSK}{2}}=-1$ in  \tablename~\ref{table:ANFgolay}. The  configuration vectors $\vecArrangement[\indexIteration]$ for the sub-recursions are also given in  \tablename~\ref{table:ANFgolay}. By using \eqref{eq:closedformANFsF} and  \eqref{eq:closedformANFsG},  $\funcGfForANF[\indexIteration](\seqx)$ and $\funcGgForANF[\indexIteration](\seqx)$ are obtained for each sub-recursion as in \tablename~\ref{table:ANFgolay}. 
Finally, by composing all operators based on $\funcGfForANF[\indexIteration](\seqx)$ and $\funcGgForANF[\indexIteration](\seqx)$, the polynomials $\functionf[\numberOfIterations]$ and $\functiong[\numberOfIterations]$  can be calculated as 
\begin{align}
\functionf[\numberOfIterations]  &= \sum_{\varMonomial=0}^{2^\numberOfIterations-1} 
\funcfForCommonOrderDec(\varMonomial,\polyVariable)
\exponentialBase^{\funcfForFinalAmplitudeDec(\varMonomial)   + \constantj \funcfForFinalAmplitudeDec(\varMonomial)}
\polyVariableIt^{\funcfForCommonShiftDec(\varMonomial)+\varMonomial\varUpsample}~,
\label{eq:encodedCSa}
\\
\functiong[\numberOfIterations]  &= \sum_{\varMonomial=0}^{2^\numberOfIterations-1}
\funcfForCommonOrderDec(\varMonomial,\polyVariable)
\exponentialBase^{\funcgForFinalAmplitudeDec(\varMonomial)   + \constantj \funcgForFinalAmplitudeDec(\varMonomial)}
\polyVariableIt^{\funcfForCommonShiftDec(\varMonomial)+\varMonomial\varUpsample}~,
\label{eq:encodedCSb}
\end{align}
where $\varMonomial = \sum_{\indexFirstOrderMonomial=1}^{\numberOfIterations}\monomial[\indexFirstOrderMonomial]2^{\numberOfIterations-\indexFirstOrderMonomial}$,
$ \funcfForCommonShift(\seqx) =\sum_{\indexIteration=1}^\numberOfIterations\separationGolay[\indexIteration]\monomial[{\permutationMono[{\indexIteration}]}]$, $\funcfForCommonOrder(\seqx,\polyVariable)=\polySeq[{\seqGaP}][\polyVariable](1-\monomial[{\permutationMono[{1}]}])_2+\polySeq[{\seqGbP}][\polyVariable]\monomial[{\permutationMono[{1}]}]
$, 	$\funcfForPartAAmplitude(\seqx)	= \funcfForCommonAmplitude(\seqx)+\scaleAexp[\numberOfIterations](1-\monomial[{\permutationMono[{\numberOfIterations}]}])_2+\scaleBexp[\numberOfIterations]\monomial[{\permutationMono[{\numberOfIterations}]}]$, 
$
\funcfForPartBAmplitude(\seqx)
= \funcfForCommonAmplitude(\seqx) +\scaleBexp[\numberOfIterations](1-\monomial[{\permutationMono[{\numberOfIterations}]}])_2 +\scaleAexp[\numberOfIterations]\monomial[{\permutationMono[{\numberOfIterations}]}]
$, $\funcfForCommonPhaseA(\seqx)= \funcfForCommonPhase(\seqx)+\angleScaleAexp[{\numberOfIterations}] (1-\monomial[{\permutationMono[{\numberOfIterations}]}])_2 +\angleScaleBexp[{\numberOfIterations}] \monomial[{\permutationMono[{\numberOfIterations}]}]$, 	$\funcfForCommonPhaseB(\seqx)= \funcfForCommonPhase(\seqx) - \angleScaleAexp[{\numberOfIterations}] \monomial[{\permutationMono[{\numberOfIterations}]}] -\angleScaleBexp[{\numberOfIterations}] (1-\monomial[{\permutationMono[{\numberOfIterations}]}])_2+ \frac{\numberOfPointsForPSK}{2}\monomial[{\permutationMono[{\numberOfIterations}]}]$, where
$ \funcfForCommonAmplitude(\seqx) \triangleq  \sum_{\indexIteration=1}^{\numberOfIterations-1}\scaleAexp[\indexIteration](1-\monomial[{\permutationMono[{\indexIteration}]}] -\monomial[{\permutationMono[{\indexIteration+1}]}])_2+\scaleBexp[\indexIteration](\monomial[{\permutationMono[{\indexIteration}]}] +\monomial[{\permutationMono[{\indexIteration+1}]}])_2$  and	
\if\IEEEsubmission0
	\begin{align}
	\funcfForCommonPhase(\seqx)
	\triangleq& {\frac{\numberOfPointsForPSK}{2}\sum_{\indexIteration=1}^{\numberOfIterations-1}\monomial[{\permutationMono[{\indexIteration}]}]\monomial[{\permutationMono[{\indexIteration+1}]}]}+\sum_{\indexIteration=1}^\numberOfIterations \angleexp[\indexIteration]\monomial[{\permutationMono[{\indexIteration}]}]\nonumber
	\\
	&+\sum_{\indexIteration=1}^{\numberOfIterations-1}\angleScaleAexp[{\indexIteration}] ((1-\monomial[{\permutationMono[{\indexIteration}]}])(1-\monomial[{\permutationMono[{\indexIteration+1}]}]))_2 
	%\nonumber
	%\\
	%&-	\sum_{\indexIteration=1}^{\numberOfIterations-1}
	-\angleScaleAexp[{\indexIteration}](\monomial[{\permutationMono[{\indexIteration}]}]\monomial[{\permutationMono[{\indexIteration+1}]}])_2\nonumber
	\\
	& +\sum_{\indexIteration=1}^{\numberOfIterations-1}\angleScaleBexp[{\indexIteration}] (\monomial[{\permutationMono[{\indexIteration}]}](1-\monomial[{\permutationMono[{\indexIteration+1}]}]))_2
	%\nonumber
	%\\
	%& -\sum_{\indexIteration=1}^{\numberOfIterations-1}
	-\angleScaleBexp[{\indexIteration}]((1-\monomial[{\permutationMono[{\indexIteration}]}])\monomial[{\permutationMono[{\indexIteration+1}]}])_2\nonumber
	~,
	\end{align}
\else
	%\begin{align}
	$
	\funcfForCommonPhase(\seqx)
	\triangleq {\frac{\numberOfPointsForPSK}{2}\sum_{\indexIteration=1}^{\numberOfIterations-1}\monomial[{\permutationMono[{\indexIteration}]}]\monomial[{\permutationMono[{\indexIteration+1}]}]}+\sum_{\indexIteration=1}^\numberOfIterations \angleexp[\indexIteration]\monomial[{\permutationMono[{\indexIteration}]}]\nonumber
	+\sum_{\indexIteration=1}^{\numberOfIterations-1}\angleScaleAexp[{\indexIteration}] ((1-\monomial[{\permutationMono[{\indexIteration}]}])(1-\monomial[{\permutationMono[{\indexIteration+1}]}]))_2 
	%\nonumber
	%\\
	%&-	\sum_{\indexIteration=1}^{\numberOfIterations-1}
	-\angleScaleAexp[{\indexIteration}](\monomial[{\permutationMono[{\indexIteration}]}]\monomial[{\permutationMono[{\indexIteration+1}]}])_2
	 +\sum_{\indexIteration=1}^{\numberOfIterations-1}\angleScaleBexp[{\indexIteration}] (\monomial[{\permutationMono[{\indexIteration}]}](1-\monomial[{\permutationMono[{\indexIteration+1}]}]))_2
	%\nonumber
	%\\
	%& -\sum_{\indexIteration=1}^{\numberOfIterations-1}
	-\angleScaleBexp[{\indexIteration}]((1-\monomial[{\permutationMono[{\indexIteration}]}])\monomial[{\permutationMono[{\indexIteration+1}]}])_2$
	.
	%\end{align}
\fi

Based on \eqref{eq:identity2} in Lemma~\ref{lemma:identity}, 	$\scaleAexp[\indexIteration](1-\monomial[{\permutationMono[{\indexIteration}]}] -\monomial[{\permutationMono[{\indexIteration+1}]}])_2 = \scaleAexp[\indexIteration] - \scaleAexp[\indexIteration](\monomial[{\permutationMono[{\indexIteration}]}] +\monomial[{\permutationMono[{\indexIteration+1}]}])_2 $, 		$\scaleAexp[\numberOfIterations](1-\monomial[{\permutationMono[{\numberOfIterations}]}])_2=\scaleAexp[\numberOfIterations]-\scaleAexp[\numberOfIterations]\monomial[{\permutationMono[{\numberOfIterations}]}]$,		and $\scaleBexp[\numberOfIterations](1-\monomial[{\permutationMono[{\numberOfIterations}]}])_2=\scaleBexp[\numberOfIterations]-\scaleBexp[\numberOfIterations]\monomial[{\permutationMono[{\numberOfIterations}]}]$,	Hence, $\funcfForFinalAmplitude(\seqx)$ and $\funcgForFinalAmplitude(\seqx)$ can be re-written as in \eqref{eq:realPartReduced} and \eqref{eq:realPartReducedG}, respectively, where $\scaleEexp[\indexIteration] = (\scaleBexp[\indexIteration]-\scaleAexp[\indexIteration])$ for $\indexIteration = 1,2,\mydots,\numberOfIterations$, and $\arbitraryScaleE =   \sum_{\indexIteration=1}^{\numberOfIterations}\scaleAexp[\indexIteration]$. By using the identities given in \eqref{eq:identity2} and \eqref{eq:identity4},	$\angleScaleAexp[{\numberOfIterations}] (1-\monomial[{\permutationMono[{\numberOfIterations}]}])_2 +\angleScaleBexp[{\numberOfIterations}] \monomial[{\permutationMono[{\numberOfIterations}]}] =    \angleScaleAexp[{\numberOfIterations}]+(\angleScaleBexp[{\numberOfIterations}]-\angleScaleAexp[{\numberOfIterations}])\monomial[{\permutationMono[{\numberOfIterations}]}]	$,	$	 \angleScaleAexp[{\numberOfIterations}] \monomial[{\permutationMono[{\numberOfIterations}]}] + \angleScaleBexp[{\numberOfIterations}] (1-\monomial[{\permutationMono[{\numberOfIterations}]}])_2 =     \angleScaleBexp[{\numberOfIterations}]+(\angleScaleAexp[{\numberOfIterations}]-\angleScaleBexp[{\numberOfIterations}])\monomial[{\permutationMono[{\numberOfIterations}]}]	$,	$		\angleScaleAexp[{\indexIteration}] ((1-\monomial[{\permutationMono[{\indexIteration}]}])(1-\monomial[{\permutationMono[{\indexIteration+1}]}]))_2-\angleScaleAexp[{\indexIteration}](\monomial[{\permutationMono[{\indexIteration}]}]\monomial[{\permutationMono[{\indexIteration+1}]}])_2=	\angleScaleAexp[{\indexIteration}] -	\angleScaleAexp[{\indexIteration}] \monomial[{\permutationMono[{\indexIteration}]}] -\angleScaleAexp[{\indexIteration}]\monomial[{\permutationMono[{\indexIteration+1}]}]$,	and	$
		\angleScaleBexp[{\indexIteration}] (\monomial[{\permutationMono[{\indexIteration}]}](1-\monomial[{\permutationMono[{\indexIteration+1}]}]))_2
		-\angleScaleBexp[{\indexIteration}]((1-\monomial[{\permutationMono[{\indexIteration}]}])\monomial[{\permutationMono[{\indexIteration+1}]}])_2 =
		\angleScaleBexp[{\indexIteration}] \monomial[{\permutationMono[{\indexIteration}]}]
		-\angleScaleBexp[{\indexIteration}]\monomial[{\permutationMono[{\indexIteration+1}]}]
	$.
		Hence, $\funcfForFinalPhase(\seqx)$ and $\funcfForCommonPhaseB(\seqx)$ can be re-expressed as in \eqref{eq:imagPartReduced} and \eqref{eq:imagPartReducedG}, respectively, where 	$\angleexpAll[1] =\angleexp[1]+\angleScaleBexp[1]-\angleScaleAexp[1]$, $ 
		\angleexpAll[\indexIteration|\indexIteration=2,3,\mydots,\numberOfIterations]=\angleexp[\indexIteration]+\angleScaleBexp[\indexIteration]-\angleScaleAexp[\indexIteration]-\angleScaleBexp[\indexIteration-1]-\angleScaleAexp[\indexIteration-1]$, $\arbitraryPhaseK =  \sum_{\indexIteration=1}^{\numberOfIterations}\angleScaleAexp[\indexIteration]$, and $\arbitraryPhaseKPP = -\angleScaleBexp[{\numberOfIterations}] + \sum_{\indexIteration=1}^{\numberOfIterations-1}\angleScaleAexp[{\indexIteration}]$, respectively.		
	\end{proof}

\section{Identities}\label{app:iden}
{
\begin{lemma}
	\label{lemma:identity}
	For  $\funcfForANF_1:{\integers_2^\numberOfIterations}\rightarrow\integers_2$, $\funcfForANF_2:{\integers_2^\numberOfIterations}\rightarrow\integers_2$  and $\coefficientArbitrary[1]\in\realNumbers$,
	\if\IEEEsubmission0
	\begin{align}
		&\coefficientArbitrary[1](\funcaArbitrary[\seqx]+\funcbArbitrary[\seqx])_2 \nonumber \\& ~~~~~=   \coefficientArbitrary[1]\funcaArbitrary[\seqx]+\coefficientArbitrary[1]\funcbArbitrary[\seqx] - 2\coefficientArbitrary[1](\funcaArbitrary[\seqx]\funcbArbitrary[\seqx])_2~,
		\label{eq:identity2}\\
		&\coefficientArbitrary[1](\funcaArbitrary[\seqx](1+\funcbArbitrary[\seqx]))_2-\coefficientArbitrary[1]((1+\funcaArbitrary[\seqx])\funcbArbitrary[\seqx])_2 \nonumber\\&~~~~~=   \coefficientArbitrary[1]\funcaArbitrary[\seqx]-\coefficientArbitrary[1]\funcbArbitrary[\seqx]~. 
		\label{eq:identity4}	
	\end{align}
	\else
	\begin{align}
		& \coefficientArbitrary[1](\funcaArbitrary[\seqx]\pm\funcbArbitrary[\seqx])_2  =  \coefficientArbitrary[1]\funcaArbitrary[\seqx]+\coefficientArbitrary[1]\funcbArbitrary[\seqx] - 2\coefficientArbitrary[1](\funcaArbitrary[\seqx]\funcbArbitrary[\seqx])_2~, 
		\label{eq:identity2}\\
		& \coefficientArbitrary[1](\funcaArbitrary[\seqx](1\pm\funcbArbitrary[\seqx]))_2-\coefficientArbitrary[1]((1\pm\funcaArbitrary[\seqx])\funcbArbitrary[\seqx])_2 =   \coefficientArbitrary[1]\funcaArbitrary[\seqx]-\coefficientArbitrary[1]\funcbArbitrary[\seqx]~. 	\label{eq:identity4}
	\end{align}
	\fi
\end{lemma}	
	
\begin{proof}
	The right hand side of \eqref{eq:identity2} is $0$ for $\funcaArbitrary[\seqx]=\funcbArbitrary[\seqx]$ and $\coefficientArbitrary[1]$ for $\funcaArbitrary[\seqx]\neq\funcbArbitrary[\seqx]$. For  \eqref{eq:identity4}, after the arguments of $(\cdot)_2$ are arranged as the summations of two Boolean functions,  \eqref{eq:identity2} is applied. 
\end{proof}
}

\bibliographystyle{IEEEtran}
\if\IEEEsubmission1
\renewcommand{\baselinestretch}{1.451}
\fi
\bibliography{golayModulator}

\end{document}